\documentclass[
  journal=pasa,
  manuscript=research-paper, 
  year=2022,
  volume=00,
]{cup-journal}

\usepackage{orcidlink}

\usepackage{microtype,siunitx,booktabs}
\title{The Rapid ASKAP Continuum Survey III: Spectra and Polarisation In Cutouts of Extragalactic Sources (SPICE-RACS) First Data Release}

\author{Alec J.M. Thomson \orcidlink{0000-0001-9472-041X}}
\affiliation{ATNF, CSIRO Space \& Astronomy, PO Box 1130, Bentley, WA 6102, Australia}
\email[Alec J.M. Thomson]{\url{alec.thomson@csiro.au}}

\author{David McConnell \orcidlink{0000-0002-2819-9977}}
\affiliation{ATNF, CSIRO Space \& Astronomy, PO Box 76, Epping, NSW 1710, Australia}

\author{Emil Lenc \orcidlink{0000-0002-9994-1593}}
\affiliation{ATNF, CSIRO Space \& Astronomy, PO Box 76, Epping, NSW 1710, Australia}

\author{Timothy J Galvin \orcidlink{0000-0002-2801-766X}}
\affiliation{ATNF, CSIRO Space \& Astronomy, PO Box 1130, Bentley, WA 6102, Australia}
\alsoaffiliation{International Centre for Radio Astronomy Research, Curtin University, Kent St, Bentley WA 6102, Australia}

\author{Lawrence Rudnick \orcidlink{0000-0001-5636-7213}}
\affiliation{Minnesota Institute for Astrophysics, University of Minnesota, 116 Church St. SE, Minneapolis, MN, 55455, USA}

\author{George Heald \orcidlink{0000-0002-2155-6054}}
\affiliation{ATNF, CSIRO Space \& Astronomy, PO Box 1130, Bentley, WA 6102, Australia}

\author{Catherine L. Hale \orcidlink{0000-0001-6279-4772}}
\affiliation{Institute for Astronomy, School of Physics and Astronomy, University of Edinburgh, Royal Observatory Edinburgh, Blackford Hill, Edinburgh, EH9 3HJ, UK}

\author{Stefan W. Duchesne \orcidlink{0000-0002-3846-0315}}
\affiliation{ATNF, CSIRO Space \& Astronomy, PO Box 1130, Bentley, WA 6102, Australia}

\author{Craig S. Anderson \orcidlink{0000-0002-6243-7879}}
\affiliation{Research School of Astronomy \& Astrophysics, Australian National University, Canberra, ACT~2611, Australia}

\author{Ettore	Carretti \orcidlink{0000-0002-3973-8403}}
\affiliation{INAF, Istituto di Radioastronomia, Via Gobetti 101, 40129, Italy}

\author{Christoph Federrath \orcidlink{0000-0002-0706-2306}}
\affiliation{Research School of Astronomy \& Astrophysics, Australian National University, Canberra, ACT~2611, Australia}
\alsoaffiliation{Australian Research Council Centre of Excellence in All Sky Astrophysics (ASTRO3D), Canberra, ACT~2611, Australia}

\author{B. M. Gaensler \orcidlink{0000-0002-3382-9558}}
\affiliation{Dunlap Institute for Astronomy and Astrophysics, University of Toronto, 50 St. George Street, Toronto, ON M5S 2H4, Canada}
\alsoaffiliation{David A. Dunlap Department of Astronomy and Astrophysics, University of Toronto, 50 St. George Street, Toronto, ON M5S 2H4, Canada}

\author{Lisa Harvey-Smith \orcidlink{0000-0001-9179-4560}}
\affiliation{Faculty of Science, UNSW, Sydney, NSW, 2052, Australia}
\alsoaffiliation{Western Sydney University, Locked Bag 1797, Penrith, NSW 2751, Australia}

\author{Marijke Haverkorn \orcidlink{0000-0002-5288-312X}}
\affiliation{Department of Astrophysics/IMAPP, Radboud University, PO Box 9010, 6500 GL Nijmegen, The Netherlands}

\author{Aidan W. Hotan \orcidlink{0000-0001-7464-8801}}
\affiliation{ATNF, CSIRO Space \& Astronomy, PO Box 1130, Bentley, WA 6102, Australia}

\author{Yik Ki Ma \orcidlink{0000-0003-0742-2006}}
\affiliation{Research School of Astronomy \& Astrophysics, Australian National University, Canberra, ACT~2611, Australia}

\author{Tara Murphy \orcidlink{0000-0002-2686-438X}}
\affiliation{Sydney Institute for Astronomy, School of Physics, University of Sydney, Sydney, NSW 2006, Australia}

\author{N.\ M.\	McClure-Griffiths\orcidlink{0000-0003-2730-957X}}
\affiliation{Research School of Astronomy \& Astrophysics, Australian National University, Canberra, ACT~2611, Australia}

\author{Vanessa	A. Moss \orcidlink{0000-0002-3005-9738}}
\affiliation{ATNF, CSIRO Space \& Astronomy, PO Box 76, Epping, NSW 1710, Australia}
\alsoaffiliation{Sydney Institute for Astronomy, School of Physics, The University of Sydney, NSW 2006, Australia}

\author{Shane P. O'Sullivan  \orcidlink{0000-0002-3968-3051}}
\affiliation{School of Physical Sciences and Centre for Astrophysics \& Relativity, Dublin City University, Glasnevin, D09 W6Y4, Ireland}

\author{Wasim	Raja}
\affiliation{ATNF, CSIRO Space \& Astronomy, PO Box 76, Epping, NSW 1710, Australia}

\author{Amit	Seta \orcidlink{0000-0001-9708-0286}}
\affiliation{Research School of Astronomy \& Astrophysics, Australian National University, Canberra, ACT~2611, Australia}

\author{Cameron L.	Van Eck \orcidlink{0000-0002-7641-9946}} 
\affiliation{Research School of Astronomy \& Astrophysics, Australian National University, Canberra, ACT~2611, Australia}

\author{Jennifer L. West \orcidlink{0000-0001-7722-8458}} 
\affiliation{National Research Council Canada, Herzberg Research Centre for Astronomy and Astrophysics, Dominion Radio Astrophysical Observatory, PO Box 248, Penticton, V2A 6J9, Canada}

\author{Matthew T.	Whiting \orcidlink{0000-0003-1160-2077}}
\affiliation{ATNF, CSIRO Space \& Astronomy, PO Box 76, Epping, NSW 1710, Australia}

\author{Mark H. Wieringa \orcidlink{0000-0002-7721-8660}}
\affiliation{ATNF, CSIRO Space \& Astronomy, PO Box 76, Epping, NSW 1710, Australia}

\doi{00.0000/pasa.0000.00}

\received {dd Mmm YYYY}
\revised  {dd Mmm YYYY}
\accepted {dd Mmm YYYY}
\published{00 Mmm YYY}

\usepackage{xcolor}
\usepackage[normalem]{ulem}

\usepackage{tikz}

\usepackage{longtable}
\usepackage{graphicx}
\usepackage{amsmath}
\usepackage{amssymb}
\usepackage{booktabs}
\usepackage{subcaption}
\usepackage{makecell}
\usepackage[htt]{hyphenat}
\usepackage{xurl}
\usepackage{hyperref}
\definecolor{dodgerblue}{RGB}{30, 144, 255}
\definecolor{crimson}{RGB}{220, 20, 60}
\definecolor{darkerblue}{RGB}{0, 0, 139}
\hypersetup{colorlinks,citecolor=dodgerblue,linkcolor=blue,urlcolor=darkerblue}

\usepackage{siunitx}

\usepackage[nolist,nohyperlinks]{acronym}

\usepackage{listings}
\newcommand\ion[2]{\text{#1\,\textsc{\lowercase{#2}}}}

\DeclareSIUnit\degreesq{deg\text{$^2$}}
\DeclareSIUnit\perdegreesq{deg\text{$^{-2}$}}

\DeclareMathOperator*{\argmax}{argmax}

\newcommand{\jansky}{\text{Jy}}

\graphicspath{{./}{figures/}}

\begin{document}

\def \ncomponents {105912}
\def \ngoodI {24680}
\def \npol {9092}
\def \nrms {5818}
\def \ncomplex {695}
\def \ncomplexsigma {692}
\def \ncomplexmcc {178}
\def \ncomplexsigmanotmcc {517}
\def \ncomplexmccnotsigma {3}

\def \medianabsRM {9}
\def \stdabsRM {13}
\def \absmaxRM {362}
\def \absmaxRMerr {4}
\def \maxRM {362}
\def \maxRMerr {4}
\def \minRM {-273}
\def \minRMerr {2}
\def \nrmgthundred {3}
\def \nlocalrmflag {81}
\def \nlocalrmflaggthundred {3}
\def \absmaxRMnolocal {-98.8}
\def \absmaxRMerrnolocal {0.9}
\def \maxRMnolocal {34.5}
\def \maxRMerrnolocal {0.3}
\def \flagstats{270.0 \substack{+8.0\\-46}}

\def \ncross {1553}
\def \ncrossbad {473}
\def \ncrossgood {1080}
        
\def \insigma {1.67 \substack{+1.7\\-0.51}}
\def \outsigma {3.7 \substack{+4.0\\-1.9}}
\def \inmcc {0.14 \substack{+0.32\\-0.13}}
\def \outmcc {0.72 \substack{+3.9\\-0.45}}
        
\def \inleak {1.173 \substack{+0.36\\-0.063}}
\def \outleak {1.188 \substack{+0.81\\-0.069}}
\def \inleakhi {2.3}
\def \outleakhi {3.4}
        
\def \medfracspicexnvssspice {$3.4 \substack{+3.2\\-1.7}$}
\def \medfracspicexnvssnvss {$7.3 \substack{+4.7\\-3.2}$}
\def \medsigmarm {$5.8 \substack{+1.5\\-1.6}$}
\def \sigmaaddcross {$1.67 \substack{+1.6\\-0.52}$}
\def \sigmaaddcrossgt {$3.1 \substack{+4.2\\-1.7}$}
\def \sigmaaddcrosslt {$1.64 \substack{+1.3\\-0.51}$}

\def \nalpha {18021}
\def \nflat {6675}
\def \avgalpha {$-0.8\pm0.4$}
        
\def \stokesLnoise {80 \substack{+28\\-16}}
\def \stokesIavnoise {340 \substack{+170\\-110}}
\def \stokesInoise {5.7 \substack{+2.9\\-1.8}}
\def \stokesQavnoise {78 \substack{+25\\-16}}
\def \stokesQnoise {1.33 \substack{+0.42\\-0.27}}
\def \stokesUavnoise {74 \substack{+25\\-16}}
\def \stokesUnoise {1.26 \substack{+0.42\\-0.27}}

\begin{abstract}
The Australian SKA Pathfinder (ASKAP) radio telescope has carried out a survey of the entire Southern Sky at \qty[]{887.5}{\mega\hertz}. The wide area, high angular resolution, and broad bandwidth provided by the low-band Rapid ASKAP Continuum Survey (RACS-low) allow the production of a next-generation rotation measure (RM) grid across the entire Southern Sky. Here we introduce this project as Spectral and Polarisation in Cutouts of Extragalactic sources from RACS (SPICE-RACS). In our first data release, we image 30 RACS-low fields in Stokes $I$, $Q$, $U$ at \ang{;;25} angular resolution, across \qtyrange[range-units = single]{744}{1032}{\MHz} with \qty[]{1}{\mega\hertz} spectral resolution. Using a bespoke, highly parallelised, software pipeline we are able to rapidly process wide-area spectro-polarimetric ASKAP observations. Notably, we use `postage stamp' cutouts to assess the polarisation properties of \ncomponents\ radio components detected in total intensity. We find that our Stokes $Q$ and $U$ images have an \textit{rms} noise of $\sim$\qty{80}{\micro\jansky\per PSF}, and our correction for instrumental polarisation leakage allows us to characterise components with $\gtrsim1\%$ polarisation fraction over most of the field of view. We produce a broadband polarised radio component catalogue that contains \nrms\ RM measurements over an area of $\sim$\qty[]{1300}{\degreesq} with an average error in RM of $1.6 \substack{+1.1\\-1.0}$\,\unit{\radian\per\metre\squared}, and an average linear polarisation fraction  $3.4 \substack{+3.0\\-1.6}$\,\%. We determine this subset of components using the conditions that the polarised signal-to-noise ratio is $>8$, the polarisation fraction is above our estimated polarised leakage, and the Stokes $I$ spectrum has a reliable model. Our catalogue provides an areal density of $4\pm2$\,\qty{}{RMs\,\perdegreesq}; an increase of $\sim4$ times over the previous state-of-the-art~\citep{Taylor2009}. Meaning that, having used just 3\% of the RACS-low sky area, we have produced the 3$^\text{rd}$ largest RM catalogue to date. This catalogue has broad applications for studying astrophysical magnetic fields; notably revealing remarkable structure in the Galactic RM sky. We will explore this Galactic structure in a follow-up paper. We will also apply the techniques described here to produce an all-Southern-sky RM catalogue from RACS observations. Finally, we make our catalogue, spectra, images, and processing pipeline publicly available.
\end{abstract}

\section{INTRODUCTION }\label{sec:intro}
\defcitealias{McConnell2020}{Paper~I}
\defcitealias{Hale2021}{Paper~II}
\defcitealias{Taylor2009}{NVSS}
\defcitealias{Schnitzeler2019}{S-PASS/ATCA}

Magnetic fields are pervasive throughout the cosmos. From planetary systems to the interstellar medium (ISM) of galaxies through to the cosmic web, magnetic fields are present and play a variety of roles in various processes\citep[see e.g.][and references therein for a review]{Klein2015, Han2017}. From the scales of stars through to galaxies, magnetic fields are important drivers and regulators of numerous astrophysical phenomena~\citep[e.g.][]{Crutcher2010,Padoan2011,Harvey-Smith2011a,Federrath2012, Beck2016}. On larger scales magnetic fields play a more passive role, but can be a powerful tool to illuminate the tenuous intergalactic medium~\citep[e.g.][]{Anderson2021, Carretti2023, Vernstrom2023}. Further, the origin of magnetic fields in the universe is still unknown. Since magnetic fields are not directly observable, we must rely on observable proxies to infer their strength and structure.

By measuring the Faraday rotation of the polarised emission to distant sources, we are able to probe the magneto-ionic medium along the line-of-sight to the source of the emission. In this way, cosmic magnetic fields can be studied through observations of linearly polarised emission at radio frequencies. The extragalactic radio sky is dominated by synchrotron emission, mostly from distant active galactic nuclei (AGN)~\citep{Condon1998, Condon1992}. Synchrotron radiation theory predicts that sources can be intrinsically highly linearly polarised \citep[up to $\sim70\%$, ][]{Rybicki1986}, depending on the energy distribution of the relativistic particles. Due to various depolarisation effects \citep[see e.g.][]{Sokoloff1998}, however, extragalactic sources typically exhibit linear polarisation fractions between \qtyrange[range-units = single]{0}{10}{\%}~\citep{Taylor2009}.

Using collections of background polarised sources we can constrain the multi-scale structure of foreground magneto-ionic material; constructing what is known as the `rotation measure (RM) grid'~\citep{Gaensler2004, Heald2020}. The angular scale of foreground features that can be probed is limited by both the density of observed polarised sources and the contiguous area of the observations. Further, intrinsic properties of the polarised sources themselves can be studied, both individually and en masse, if sufficient angular resolution and bandwidth can be obtained. These requirements demand a radio survey that is simultaneously sensitive, high-resolution, broad in bandwidth, and wide in area.

To date, the largest catalogue of polarised sources is from \citet{Taylor2009}, derived from the NRAO VLA Sky Survey \citep[NVSS,][]{Condon1998}. Even over a decade later, the \citetalias{Taylor2009} catalogue remains the largest by an order of magnitude, with a total of $\sim\num{3.7e4}$ RMs. The \citetalias{Taylor2009} catalogue was produced from images at two narrowly-spaced frequencies around \qty{1.4}{\giga\hertz}, with a Southern-most declination limit of $\delta=-\ang{40}$ providing an average RM density of $\sim\qty{1}{\perdegreesq}$ over the Northern sky. The next largest RM catalogue\footnote{as consolidated by \citet{VanEck2023} \url{https://github.com/CIRADA-Tools/RMTable} v1.1.0} is \citetalias{Schnitzeler2019}~\citep{Schnitzeler2019}, which characterised $\sim\num{6.9e3}$ polarised sources with $\sim\qty{2}{\giga\hertz}$ of bandwidth, centred on $\sim\qty{2}{\GHz}$, over the Southern sky up to a declination of $\delta=\ang{0}$. Despite the significantly better characterisation of each source in \citetalias{Schnitzeler2019}, the average source density is only $\qty{0.2}{\perdegreesq}$. As such, large-area RM grids, and derived products such as \citet{Oppermann2012, Oppermann2015} and \citet{Hutschenreuter2020, Hutschenreuter2021}, are both dominated, and fundamentally limited, by the properties of the \citetalias{Taylor2009} catalogue~\citep[see e.g.][]{Ma2019}. We note that many bespoke RM grids have been carried out towards particular regions and objects of interest \citep[e.g.][]{Mao2008, Mao2012, Kaczmarek2017, Betti2019, Livingston2021, Livingston2022}. The targeted nature of these surveys, however, limits their scope and applicability to a broad range range of science cases. 

It is at this stage that we require surveys from new instruments such as the Square Kilometre Array (SKA) and its precursors and pathfinders. Recently, there have been a number of polarisation survey releases from LoTSS~\citep{OSullivan2023}, Apertif~\citep{Adebahr2022}, and the Murchison Widefield Array (MWA)~\citep{Riseley2018,Riseley2020}, with plans to produce polarisation products from VLASS~\citep{Lacy2020}. This calls for a $\sim$\qty{1}{\giga\hertz} polarisation survey in the Southern sky to maximise both areal and frequency coverage.

The Rapid ASKAP Continuum Survey \citep[RACS,][`\citetalias{McConnell2020}']{McConnell2020} is the first all-Southern-sky survey undertaken using the Australian SKA Pathfinder \citep[ASKAP;][]{Hotan2021} radio telescope. ASKAP was designed as a purpose-built survey telescope. It therefore enables rapid observations of the radio sky with the properties (high sensitivity and angular resolution, broad bandwidth, and wide areal coverage) we desire for a spectro-polarimetric survey. ASKAP has recently begun a multi-year campaign to observe the entire Southern sky. The primary continuum surveys that form part of this campaign are the Evolutionary Map of the Universe \citep[EMU,][]{Norris2011, Norris2021} and the Polarisation Sky Survey of the Universe's Magnetism \citep[POSSUM,][Gaensler et al. in prep.]{Gaensler2010}. RACS has been conducted in preparation for these surveys. The first, low-frequency, component of RACS (RACS-low) covers the entire sky from the South celestial pole to a declination of $\delta=+\ang{41}$. RACS-low has a central frequency of \qty{887.5}{\mega\hertz}, with a bandwidth of \qty{288}{\mega\hertz}. The all-Southern-sky catalogue in Stokes $I$ was presented by \citet[hereafter \protect{\citetalias{Hale2021}}]{Hale2021}, who analysed the survey area from $\delta=+\ang{30}$ down to $\delta=-\ang{80}$ across 799 tiles with a common angular resolution of \ang{;;25}. Excluding the Galactic plane (where $|b|>\ang{5}$), we found $\sim2.5$ million radio components in total intensity, with an estimated completeness of 95\% for point sources at $\sim\qty{5}{\milli\jansky}$ and an overall completeness of 95\% at $\sim\qty{3}{\milli\jansky}$ when integrated across the full catalogue, assuming a typical sky model.

Here we present the first data release (DR1) of Spectra and Polarisation In Cutouts of Extra-galactic sources from RACS (SPICE-RACS). SPICE-RACS is the linearly polarised counterpart to the Stokes $I$ catalogue presented in \citetalias{Hale2021}. This paper is organised as follows: In \S\ref{sec:observations}, we provide an overview of the RACS-low observations, and our selection of fields for this data release. In \S~\ref{sec:data}, we describe our data reduction and processing techniques, and describe the resulting polarised component catalogue. We assess the quality of our data and describe their overall properties in \S\ref{sec:analysis}. We describe our final data products, and where to find them, in \S\ref{sec:access}. Finally, we provide our outlook for future data releases in \S\ref{sec:discussion} and conclude in \S\ref{sec:conclusions}.

\section{OBSERVATIONS}\label{sec:observations}

\subsection{Rapid ASKAP Continuum Survey}
RACS-low was observed between April 2019 and June 2020, as we described in \citetalias{McConnell2020}. RACS-low tiled the observable sky into 903 overlapping fields spanning declinations $-\ang{90}<\delta<+\ang{41}$, each receiving \qty{15}{\minute} of integration time. During beam-forming, RACS-low configured ASKAP's 36 beams\footnote{Throughout we use the term `beam' to refer to the 36 formed beams on the sky. Each formed beam has its own `primary beam', describing the response of the antenna to a source of emission. The angular resolution of the beam is described by the synthesised beam which is then deconvolved and restored with a Gaussian model. We refer to the resolution element of our restored images as the `point-spread function' (PSF).} into the \texttt{square\_6x6} configuration, with a pitch of \ang{1.05} separating each beam centre for each field. Each field is uniquely identified by its field name. Further, each field also has corresponding target and calibration scheduling-block identifier codes~\citepalias[SBID, see][]{McConnell2020}. We note that the majority of RACS-low was observed in ASKAP's `multi-field' mode, whereby multiple fields or calibrators are observed in a single scheduling block. The ASKAP Observatory now prefers a single SBID per field. In future RACS releases, therefore, an SBID will also uniquely identify an observation.

In \citetalias{Hale2021}, we defined two subsets of RACS-low sources. Sources with Galactic latitude $|b|<\ang{5}$ were classed as `Galactic' with the remainder classed as `non-Galactic'. This split was motivated by the imaging performance of RACS-low. The snapshot imaging of RACS provides a sensitivity to a maximum angular scale of \ang{;25}--\ang{;50} over most of the sky. RACS is therefore suited to imaging of compact sources, with a loss of sensitivity and image fidelity along the Galactic plane. Source-finding and characterisation of highly extended emission is also a unique challenge unto itself that we leave for future work. By a simple count from the catalogue, we have 704 `non-Galactic' fields and 113 `Galactic' fields lying along the Galactic plane. We note that some fields fall into both categories, as some portions may lie within $|b|<\ang{5}$.

Here we select 30 representative RACS-low fields for analysis in linear polarisation. In our selection of these fields we weigh several factors. First, we choose a contiguous subset of the 704 non-Galactic fields. Given the significant computational requirements for processing the entire RACS-low survey in full polarisation, we present our processing methodology here along with a catalogue produced from a small subset of the full RACS-low survey. Further, we desire a subset at an intermediate declination, allowing for coverage between, and comparison with, previous large-area surveys in both hemispheres; namely \citetalias{Taylor2009} and \citetalias{Schnitzeler2019}. We also select a region of interest within the Galaxy with a large angular extent, allowing the quality of wide-area ASKAP observations to be highlighted. Using the same methods that are presented here, we will produce and publish an all-sky compact source polarisation catalogue in a subsequent release.

\begin{figure*}
	\begin{center}
		\includegraphics[width=\linewidth]{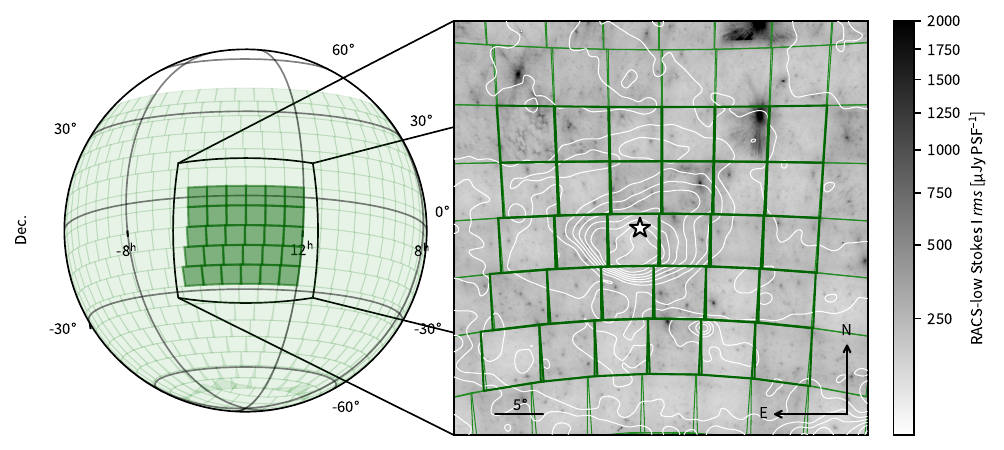}
		\caption{Sky coverage of SPICE-RACS DR1. We show the tiling of fields for the entirety of RACS-low in light green, and the 30 fields selected for this data release in dark green. In the inset panel we show the Stokes $I$ $rms$ noise~\citepalias[from][]{McConnell2020} in the region surrounding the Spica nebula. In white contours, we show emission from the nebula itself in H$\alpha$ from WHAM \citep{Haffner2003}, and we show the position of the Spica star with a white star.}\label{fig:tiles}
	\end{center}
\end{figure*}

For our first data release, we have selected fields towards the Spica Nebula; a Galactic \ion{H}{II} region ionised by the nearby star Spica /$\alpha$Vir~\citep{Reynolds1985_spica}. We list our RACS-low fields in Table~\ref{tab:fields}, and show their distribution on the sky in Figure~\ref{fig:tiles}. We will provide detailed analysis of these data, and the magneto-ionic properties of the Spica Nebula, in a forthcoming paper.

 \begin{table*}
 \sisetup{
 angle-symbol-degree = deg,
 angle-symbol-minute = arcmin,
 angle-symbol-second = arcsec
 }
 \caption{RACS fields selected for SPICE-RACS DR1. The columns (from left to right) are: the RACS-low field name, the field centre J2000 right ascension (RA), the field centre J2000 declination (Dec.), the field centre Galactic longitude (GLON), the field centre Galactic latitude (GLAT), the date of the observation, the integration time $T_\text{int}$, the scheduling block identifier (SBID) of the bandpass calibration, the SBID of the target observation, the SBID of the beam-forming observation, and the number of Gaussian components in total intensity from \citetalias{Hale2021}$^a$ ($N_\text{Gauss}$).}
  \begin{tabular}{ccccccccccc}
  	\hline
  	     Field name      &     RA      &     Dec.     &        GLON        &        GLAT        &     Date     & $T_\text{int}$ & Calibrator & Target & Beam weight & $N_\text{Gauss}$ \\
  	                     &   [h:m:s]   &   [d:m:s]    & [$\unit{\degree}$] & [$\unit{\degree}$] & [YYYY-MM-DD] &      [s]       &    SBID    &  SBID  &    SBID     &                  \\ \hline
  	\verb|RACS_1213-25A| & 12:13:35.09 & -25:07:47.24 &       292.20       &       36.95        &  2019-04-25  &      985       &    8577    &  8576  &    8247     &       3522       \\
  	\verb|RACS_1424-18A| & 14:24:00.00 & -18:51:45.43 &       331.45       &       38.83        &  2019-04-27  &      926       &    8585    &  8584  &    8247     &       3348       \\
  	\verb|RACS_1213-18A| & 12:13:05.45 & -18:51:45.43 &       290.47       &       43.08        &  2019-04-27  &      926       &    8585    &  8584  &    8247     &       3470       \\
  	\verb|RACS_1305-18A| & 13:05:27.27 & -18:51:45.43 &       307.53       &       43.88        &  2019-04-27  &      916       &    8585    &  8584  &    8247     &       3030       \\
  	\verb|RACS_1402-25A| & 14:02:15.85 & -25:07:47.24 &       322.58       &       35.03        &  2019-04-27  &      926       &    8585    &  8584  &    8247     &       3577       \\
  	\verb|RACS_1357-18A| & 13:57:49.09 & -18:51:45.43 &       324.01       &       41.28        &  2019-04-27  &      936       &    8585    &  8584  &    8247     &       3621       \\
  	\verb|RACS_1331-18A| & 13:31:38.18 & -18:51:45.43 &       315.98       &       42.99        &  2019-04-27  &      916       &    8585    &  8584  &    8247     &       3383       \\
  	\verb|RACS_1307-25A| & 13:07:55.47 & -25:07:47.24 &       307.64       &       37.59        &  2019-04-27  &      926       &    8585    &  8584  &    8247     &       3424       \\
  	\verb|RACS_1351+00A| & 13:51:43.45 &  0:00:00.00  &       333.50       &       59.25        &  2019-04-28  &      926       &    8590    &  8589  &    8247     &       2786       \\
  	\verb|RACS_1302-06A| & 13:02:04.14 & -6:17:58.73  &       307.72       &       56.47        &  2019-04-28  &      936       &    8590    &  8589  &    8247     &       2479       \\
  	\verb|RACS_1326-06A| & 13:26:53.79 & -6:17:58.73  &       318.62       &       55.49        &  2019-04-29  &      926       &    8594    &  8593  &    8247     &       3078       \\
  	\verb|RACS_1418-12A| & 14:18:56.84 & -12:35:08.47 &       333.86       &       44.97        &  2019-05-06  &      956       &    8675    &  8674  &    8669     &       3523       \\
  	\verb|RACS_1212-06A| & 12:12:24.83 & -6:17:58.73  &       285.74       &       55.26        &  2020-03-26  &      906       &   12455    & 12422  &    11371    &       3665       \\
  	\verb|RACS_1237-06A| & 12:37:14.48 & -6:17:58.73  &       296.55       &       56.40        &  2020-03-26  &      906       &   12455    & 12423  &    11371    &       3202       \\
  	\verb|RACS_1328-12A| & 13:28:25.26 & -12:35:08.47 &       316.84       &       49.28        &  2020-03-26  &      906       &   12455    & 12426  &    11371    &       3932       \\
  	\verb|RACS_1351-06A| & 13:51:43.45 & -6:17:58.73  &       328.70       &       53.53        &  2020-03-26  &      906       &   12455    & 12428  &    11371    &       3430       \\
  	\verb|RACS_1416-06A| & 14:16:33.10 & -6:17:58.73  &       337.68       &       50.74        &  2020-03-26  &      906       &   12455    & 12429  &    11371    &       3262       \\
  	\verb|RACS_1212-12A| & 12:12:37.89 & -12:35:08.47 &       288.36       &       49.18        &  2020-03-27  &      906       &   12509    & 12494  &    11371    &       3291       \\
  	\verb|RACS_1237-12A| & 12:37:53.68 & -12:35:08.47 &       297.77       &       50.15        &  2020-03-27  &      906       &   12509    & 12496  &    11371    &       3548       \\
  	\verb|RACS_1303-12A| & 13:03:09.47 & -12:35:08.47 &       307.40       &       50.18        &  2020-03-27  &      906       &   12509    & 12497  &    11371    &       3234       \\
  	\verb|RACS_1353-12A| & 13:53:41.05 & -12:35:08.47 &       325.74       &       47.51        &  2020-03-27  &      906       &   12509    & 12500  &    11371    &       3088       \\
  	\verb|RACS_1326+00A| & 13:26:53.79 &  0:00:00.00  &       321.81       &       61.56        &  2020-04-30  &      906       &   13615    & 13591  &    11371    &       3404       \\
  	\verb|RACS_1416+00A| & 14:16:33.10 &  0:00:00.00  &       343.43       &       56.03        &  2020-04-30  &      906       &   13615    & 13595  &    11371    &       2867       \\
  	\verb|RACS_1237+00A| & 12:37:14.48 &  0:00:00.00  &       295.19       &       62.66        &  2020-05-01  &      906       &   13708    & 13671  &    13624    &       1718       \\
  	\verb|RACS_1302+00A| & 13:02:04.14 &  0:00:00.00  &       308.74       &       62.75        &  2020-05-01  &      906       &   13708    & 13672  &    13624    &       3475       \\
  	\verb|RACS_1212+00A| & 12:12:24.83 &  0:00:00.00  &       282.27       &       61.30        &  2020-05-01  &      906       &   13708    & 13673  &    13624    &       3190       \\
  	\verb|RACS_1429-25A| & 14:29:26.04 & -25:07:47.24 &       329.41       &       32.65        &  2020-05-01  &      906       &   13708    & 13678  &    13624    &       4020       \\
  	\verb|RACS_1240-25A| & 12:40:45.28 & -25:07:47.24 &       299.88       &       37.68        &  2020-05-02  &      896       &   13708    & 13746  &    13624    &       4173       \\
  	\verb|RACS_1239-18A| & 12:39:16.36 & -18:51:45.43 &       298.94       &       43.91        &  2020-05-02  &      906       &   13708    & 13747  &    13624    &       3741       \\
  	\verb|RACS_1335-25A| & 13:35:05.66 & -25:07:47.24 &       315.28       &       36.69        &  2020-05-02  &      906       &   13708    & 13749  &    13624    &       4175       \\ \hline
  \end{tabular}
  \label{tab:fields}
 \medskip
 \begin{tablenotes}
 \item[a] Here we have only used the observations as listed above. The de-duplication process in \citetalias{Hale2021}, however, results in some components as being listed in adjacent fields in the final catalogue. In our SPICE-RACS-DR1 catalogue we modify the \verb|tile_id| and other metadata columns so only the 30 fields listed in the table above will appear. In the RACS-low catalogue the adjacent fields are: \verb|RACS_1351-06A|, \verb|RACS_1328-12A|, \verb|RACS_1237-12A|, \verb|RACS_1213-25A|, \verb|RACS_1335-25A|, \verb|RACS_1424-18A|, \verb|RACS_1240-25A|, \verb|RACS_1331-18A|, \verb|RACS_1302+00A|, \verb|RACS_1303-12A|, \verb|RACS_1351+00A|, \verb|RACS_1326+00A|, \verb|RACS_1213-18A|, \verb|RACS_1212-06A|, \verb|RACS_1353-12A|, \verb|RACS_1416+00A|, \verb|RACS_1302-06A|, \verb|RACS_1305-18A|, \verb|RACS_1326-06A|, \verb|RACS_1239-18A|, \verb|RACS_1416-06A|, \verb|RACS_1212+00A|, \verb|RACS_1307-25A|, \verb|RACS_1237-06A|, \verb|RACS_1237+00A|, \verb|RACS_1357-18A|, \verb|RACS_1418-12A|, \verb|RACS_1429-25A|, \verb|RACS_1402-25A|, and \verb|RACS_1212-12A|.
 \end{tablenotes}
 \end{table*}

\section{DATA}\label{sec:data}
We process our data in two primary stages. First, calibration of the visibilities and production of the beam images is done using the \textsc{ASKAPsoft} pipeline~\citep{Guzman2019}. Second, the image products then flow through our own pipeline `\textsc{Arrakis}', which we make publicly available\footnote{\url{https://github.com/AlecThomson/arrakis}}. This pipeline is a modular and parallelised Python framework, built using the \textsc{Prefect} and \textsc{Dask}~\citep{dask} libraries for flow management and parallelisation. Currently, the \textsc{Arrakis} pipeline performs the following tasks, which we detail below:
\begin{itemize}
	\item Cutouts of sources from the beam image cubes (\S\ref{sec:cutout}).
	\item Mosaicking with primary beam and widefield leakage correction (see \S\ref{sec:off_leak}).
	\item Rotation measure determination (\S\ref{sec:rm_determine}).
	\item Catalogue production (\S\ref{sec:catalogue}).
\end{itemize}
In a future release of the pipeline, we will also include an imaging module, which will allow end-to-end processing of calibrated visibilities produced by the ASKAP Observatory.

It is important to note that the POSSUM collaboration is also developing a pipeline (Van Eck et al. in prep.) to produce value-added data products from the images provided by the Observatory. This pipeline has several practical and conceptual differences from the pipeline presented here. These differences are primarily driven by the atomic unit of data for each survey; being full fields and cutouts for POSSUM and SPICE-RACS, respectively. We have developed the `\textsc{Arrakis} pipeline to rapidly produce an RM catalogue from compact sources detected in shallow ASKAP observations. Further, we already have the full all-sky dataset in-hand. This presents a significant computational challenge to process and reduce in reasonable timeframes. By contrast, POSSUM observations will be made over a period of $\sim5\,$years, and will be sensitive to both compact and diffuse emission. We note, however, that the \textsc{RM-Tools}~\citep{Purcell2020} library provides many of the core routines for both pipelines.

\subsection{Calibration}\label{sec:calibration}

Our primary calibration procedures, including beam-forming, phase, and gain calibration, are all detailed in \citetalias{McConnell2020}. Here our starting data are the calibrated visibilities as described in \citetalias{McConnell2020}, each with a spectral resolution of \qty{1}{\mega\hertz} across the \qty{288}{\mega\hertz} bandwidth. For the purpose of full spectro-polarimetric analysis, we apply the following additional calibration steps.

\subsubsection{On-axis leakage}

We derive the on-axis leakage solutions using the \texttt{cbpcalibrator} tool in \textsc{ASKAPsoft}. This tool finds the leakages sequentially following the bandpass solution from observations of our primary flux calibrator PKS~B1934-638~\citep{Reynolds1994}. After leakage solutions are derived \citep[e.g.][]{Hamaker1996,Sault1996,Hales2017}, we apply them directly to the bandpass-and-self-calibrated visibilities. At the time of writing only a subset of publicly available RACS-low data has had on-axis leakage calibration applied, which required us to re-derive and apply this calibration. For future RACS releases (included RACS-mid, high, and low2) this calibration will be part of the standard Observatory processing.

\subsubsection{Off-axis leakage}\label{sec:off_leak}

After the correction of the on-axis leakage, images will still suffer from instrumental polarisation. This off-axis leakage is direction-dependent and will increase in magnitude with separation from the pointing centre of each beam. The sky is dominated primarily by emission in Stokes $I$, and therefore leakage from $I$ into $Q$, $U$, and $V$ will be the dominant instrumental error.

The ASKAP dishes feature a unique mount design that allows for a third `roll' axis, in addition to the altitude and azimuth axes~\citep{Hotan2021}. This allows the reference frame of the antenna to remain fixed with respect to the sky, and therefore the parallactic angle remains fixed in time throughout a given track. Field sources, therefore, remain at fixed points in each beam throughout a given observation.

To characterise the wide-field leakage itself, we make use of the field sources to probe the primary beam behaviour, under the assumption that most are intrinsically unpolarised. This technique is also being developed for the full POSSUM survey (Anderson et al. in prep.), and draws on previous work such as \citet{Farnsworth2011} and \citet{Lenc2018}. We begin by selecting high signal-to-noise ($\geq100\sigma$, full-band) total intensity components. We then extract Stokes $I$, $Q$ and $U$ spectra following the same procedure that we outline below in \S\ref{sec:cutout} and \S\ref{sec:spectra}. For each component, we fit a power-law polynomial to Stokes $I$ of the same functional form described in \S\ref{sec:spectra}, and a third-order polynomial to the fractional $q=Q/I_\text{model}$ and $u=U/I_\text{model}$. Doing this initial fitting is required to suppress the effects of large noise spikes. In addition to residual RFI, there are further two likely sources of these spikes. First, the sensitivity per \qty{1}{\mega\hertz} channel is relatively low (by a factor of $1/\sqrt{288}$ compared to the full band). Even with pure Gaussian noise, large value spikes are possible but less common. Second, since \textsc{ASKAPsoft} images each channel independently, each channel only receives a relatively shallow \textsc{clean}. This leave can leave residual sidelobes, along with other artefacts, which can appear as spikes in a given spectrum.

For each beam at each \qty{1}{\mega\hertz} channel we fit a Zernike polynomial~\citep{Zernike1934} as function of separation from the beam centre in the instrument frame (see \ref{app:leakage}). We use the Zernike polynomials implemented in \textsc{GalSim}~\citep{Rowe2015}, who follow the `Noll index' ($j_\text{max}$) convention~\citep{Noll1976}. For our surface fitting we select $j_\text{max}=10$, corresponding to the third radial order and meridional frequency of the Zernike polynomials. For all our model-fitting we use the least-squares approach, implemented in \textsc{NumPy}'s \texttt{linalg.lstsq} routine~\citep{Harris2020}.

ASKAP determines beamformer weights using a maximum sensitivity algorithm, taking the Sun as a reference source~\citep{Hotan2021}. The beam-forming system reuses the beam-forming weights for a period of several months, using the on-dish-calibration (ODC) system to correct for any instrumental drifts that may have occurred since the beam-forming observation. We can therefore expect the primary beams, including the wide-field leakage response, to potentially change significantly after each major beam-forming weight update. We explore and quantify these changes in greater depth in the RACS-mid description paper~\citep[][Paper~IV]{Duchesne2023}.

For the 30 RACS-low fields we have selected, there were four independent sets of beam weights used in the observations (see Table~\ref{tab:fields}). We choose to fit frequency dependent, two-dimensional leakage surfaces independently to components from each set of beam-forming weights for each beam. In doing so, we sacrifice the number of components available to characterise the beam, but gain greater accuracy in our derived surface. For three of the four weights, we were able to obtain robust fits to the leakage with around $\sim200$ components per beam on average. There was only a single field using beams formed with SB8669, giving us an average of $\sim20$ components per beam. For this field we reduce the order of Zernike polynomial to $j_\text{max}=6$ to avoid over-fitting.

We find that fitted surfaces are very sensitive to outlying points, which presumably include some truly polarised components. This becomes particularly problematic when noise spikes produce very high fractional polarisation in a given channel. We therefore exclude components with $q$ or $u$ greater than 100\%. We do this both before fitting our third-order polynomials along the frequency axis, and again before fitting a Zernike polynomial surface. We provide further detailed analysis of the fitted and residual leakage in \ref{app:leakage}.

Having characterised the leakage patterns in $Q$ and $U$, we perform a correction for the effects of wide-field leakage in the image domain during linear mosaicking of each beam. We produce corrected images of Stokes $Q$ and $U$ ($\{Q,U\}_\text{cor}$) using the Stokes $I$ images and the leakage maps ($\{Q,U\}_\text{leakage}$) via
\begin{equation}
 \{Q,U\}_\text{cor} = \{Q,U\} - I \times \{Q,U\}_\text{leakage},
\end{equation}
as implemented in the \textsc{YandaSoft}\footnote{\url{https://github.com/ATNF/yandasoft}} tool \texttt{linmos}. The linear mosaicking process itself also suppresses the instrumental leakage. After mosaicking beams together in a field, subtracting the leakage model, and mosaicking adjacent fields into the full `patch', we conclude that instrumental polarisation in our images is of the order 1\% across the centre of a given field, but increases to a few percent from about \ang{3} separation from a field's centre (see Figure~\ref{fig:leakage_res}).

\subsubsection{Ionospheric Faraday rotation}

We perform a correction for ionospheric Faraday rotation in the image domain using a time-integrated correction. We make use of the \textsc{FRion}\footnote{\url{https://github.com/CIRADA-Tools/FRion}}~(Van Eck in prep.) framework to both derive and apply the ionosphere correction. Following \citet{VanEck2021}, our observed complex polarisation ($P_\text{meas}(\lambda^2)$) relates to the polarisation from the sky ($P_\text{sky}(\lambda^2)$) by
\begin{equation}
 P_\text{meas}(\lambda^2) = P_\text{sky}(\lambda^2) \Theta(\lambda^2),
\end{equation}
where $\Theta$ is the integrated effect of the ionosphere, $\lambda$ is the observed wavelength, and $P=Q+iU$. We can then compute $\Theta$ as
\begin{equation}
 \Theta(\lambda^2) = \frac{1}{t_s - t_f} \int_{t_s}^{t_f} e^{2i\lambda^2\phi_{\text{ion}}(t)} dt,
\end{equation}
where $t_s$ and $t_f$ are the starting and ending times of an observation, and $\phi_{\text{ion}}(t)$ is the ionospheric Faraday rotation as a function of time $t$. Note that the amplitude ($|\Theta|$) and the phase ($\arg\Theta$) of $\Theta$ (in the range $-\pi/2<\arg\Theta<+\pi/2$) can be interpreted as the depolarisation and change in polarisation angle, respectively, due to the ionosphere.

\begin{figure}
	\begin{center}
		\includegraphics[width=\linewidth]{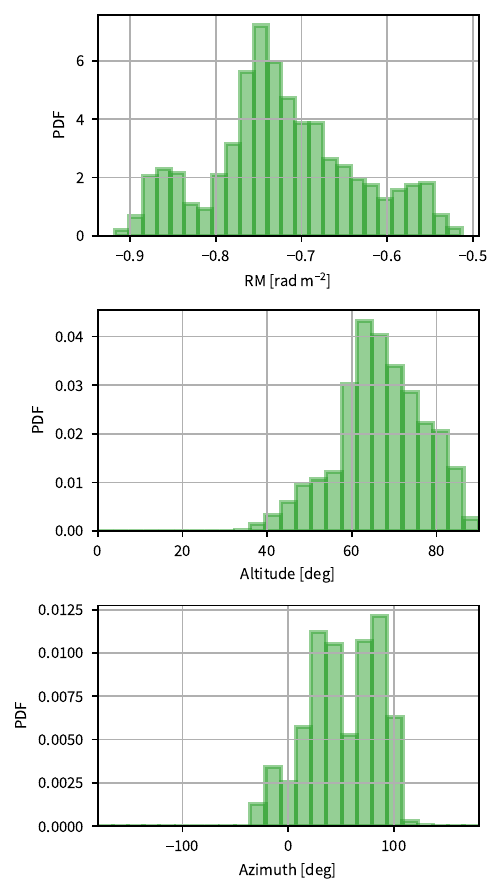}
		\caption{Impact of the ionosphere in SPICE-RACS-DR1. In the top panel we show the probably distribution function (PDF) of rotation measure (RM) over time, in the middle we show the PDF of the telescope altitude angle, and in the lower panel we show the PDF of the telescope azimuth angle.}
		\label{fig:ion}
	\end{center}
\end{figure}

\textsc{FRion} derives the time-dependent Faraday rotation using \textsc{RMExtract}~\citep{Mevius2018} before computing and applying the integrated correction. For our observations, \textsc{FRion} and \textsc{RMExtract} derive 5 time samples of the ionosphere. Other than stochastic variations, we find no large trends in the extracted ionospheric Faraday rotation as a function of time for our RACS observations. We show the distribution of the ionospheric RM in Figure~\ref{fig:ion}, along with the distribution of each field's altitude and azimuth angle, with respect to ASKAP. For all our 30 fields, the magnitude of the ionospheric Faraday rotation was less than \qty{1}{\radian\per\metre\squared}, with a median value of $-0.73 \substack{+0.10\\-0.07}$\,\unit{\radian\per\metre\squared}. Within each day of observing the standard deviation of the ionospheric Faraday rotation is \qty{0.016(0.002)}{\radian\per\metre\squared}. We also see that the distribution of observing altitude and azimuth is relatively narrow. We can conclude that the small fluctuations ionospheric RM in SPICE-RACS-DR1 driven primarily by variations in the ionospheric total electron column and we are mostly looking along the same local geomagnetic field.

\begin{figure}
	\begin{center}
		\includegraphics[width=\linewidth]{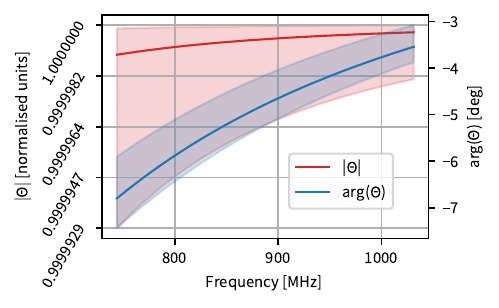}
		\caption{Integrated effect of ionospheric Faraday rotation as a function of frequency ($\Theta$). We can interpret the absolute value $|\Theta|$ the depolarisation from the ionosphere, and the phase $\arg(\Theta)$ as the change in polarisation angle.}
		\label{fig:theta}
	\end{center}
\end{figure}

Applying a time-integrated correction comes with a few caveats. First, in the case of a highly disturbed ionosphere over the period of integration, the value of $|\Theta(\lambda^2)|$ can go to 0, resulting in catastrophic depolarisation. Even in less dire cases, however, uncertainties in $\Theta(\lambda^2)$ will be propagated into our estimated $P_\text{sky}(\lambda^2)$, and will also amplify the errors in $P_\text{meas}(\lambda^2)$. Each of these effects become worse as $|\Theta(\lambda^2)|\rightarrow0$, but are negligible as $|\Theta(\lambda^2)|\rightarrow1$. In Figure~\ref{fig:theta} we show the distribution of the absolute value and angle of $\Theta(\lambda^2)$ as function of frequency across our observations. We find that $0.9999929<|\Theta(\lambda^2)|\leq1$ for all our fields. This is due to both our short (\qty{15}{\minute}) integration time, as well as a relatively stable ionosphere during our observations. A time-integrated ionospheric Faraday rotation correction is well suited to short, RACS-style, observations.

\subsection{Imaging}
For each field we perform imaging of the calibrated visibility data using the \textsc{ASKAPsoft} software applications and pipeline \citep{Guzman2019} on the Pawsey Supercomputing Research Centre\footnote{\url{https://pawsey.org.au}}, producing an image cube for each of ASKAP's 36 beams in Stokes $I$, $Q$, and $U$. At all stages, the pipeline treats each beam as an independent observation of the sky. Each cube is channelised at \qty{1}{\mega\hertz} spectral resolution, spanning \SIrange[range-units = single]{744}{1032}{\mega\hertz} in 288 channels.

Imaging and deconvolution are handled by the \texttt{cimager} application. We outline the key parameters that we specify for \texttt{cimager} in Table~\ref{tab:imaging}. Our approach is similar to that described in \citetalias{McConnell2020}. As we are producing image cubes, however, we make some different choices in imaging parameters. We produce images $\sim\ang{2.8}$ in size, with a pixel grid allowing 10 samples (\ang{;;2.5} pixels) across the final point-spread function (PSF, \ang{;;25} full width at half maximum). Wide-field effects are corrected using the $w$-projection algorithm~\citep{Cornwell2008}. We use the \texttt{BasisfunctionMFS} algorithm for multi-scale deconvolution, allowing 3 major cycles with a target residual threshold of \qty{0.06}{\milli\jansky\per PSF}.

RACS-low was observed with the third axis rotated to maintain a constant feed rotation of \ang{-45} from the North celestial pole in the plane of the sky. We enable the parallactic angle correction, with a rotation of \ang{+90}, in \texttt{cimager} which produces correct polarisation angles in the frame of the sky. For future observations (with SBID$\geq 16000$), parallactic angle correction is applied to the visibilities on ingest from the telescope.

For spectro-polarimetric analysis, we require a uniform resolution across all of our image planes and along the frequency axis. As described in \citetalias{McConnell2020}, the combination of wide-field and snapshot imaging results in variation in the PSF from beam-to-beam as a function of position on the sky. In \citetalias{Hale2021}, a common angular resolution of \ang{;;25} was selected to maximise uniform sky coverage. To satisfy our resolution requirement we first utilise the Wiener filter pre-conditioning of \textsc{ASKAPsoft} with robust weighting~\citep{Briggs1995} $-0.5$ to ensure our lowest frequency channels can meet our required PSF size. Similar to \citetalias{Hale2021}, we then convolve each restored image plane to a common, circular \ang{;;25} using a Gaussian kernel. From here we break from `standard' ASKAP processing, and the use of the ASKAP pipeline.

\begin{table*}
\caption{Key imaging parameters used in ASKAPsoft.}\label{tab:imaging}
 \begin{tabular}{p{0.35\linewidth} p{0.35 \linewidth} c}
  \hline
  Description & ASKAPsoft \texttt{cimager} parameter & Value \\ \hline
  Minimum $uv$-distance in metres. & \texttt{MinUV} & 100 \\
  Polarisation planes to be produced for the image. & \texttt{Images.polarisation} & ["I","Q","U"] \\
  Enable parallactic angle rotation based on feed angle recorded in the feed table. & \texttt{gridder.parotation} & true \\
  Rotation to apply in addition to value in feed table. & \texttt{gridder.parotation.angle} & 90 \\
  Size of the images in pixels. & \texttt{Images.shape} & [4096, 4096] \\
  Size of each pixel in RA/Dec. & \texttt{Images.cellsize} & [2.5arcsec, 2.5arcsec] \\
  Use $w$-projection to account for the $w$-term in the Fourier transform. & \texttt{gridder} & WProject \\
  Largest allowed absolute value of the $w$-term in wavelengths. & \texttt{gridder.WProject.wmax} & 26000 \\
  Number of $w$-planes. & \texttt{gridder.WProject.nwplanes} & 513 \\
  \textsc{clean} deconvolution algorithm. & \texttt{solver.Clean.algorithm} & BasisfunctionMFS \\
  Number of minor cycles. & \texttt{solver.Clean.niter} & 600 \\
  \textsc{clean} loop gain. & \texttt{solver.Clean.gain} & 0.2 \\
  Scales to be solved (defined in pixels). & \texttt{solver.Clean.scales} & [0,3,10] \\
  Minor cycle stopping threshold. & \texttt{threshold.minorcycle} & [40\%, 0.5mJy, 0.05mJy] \\
  The target peak residual. & \texttt{threshold.majorcycle} & 0.06mJy \\
  Number of major cycles. & \texttt{ncycles} & 3 \\
  ASKAPsoft precondtioner(s) to apply. & \texttt{preconditioner.Names} & [Wiener] \\
  Equivalent of Briggs robustness. & \texttt{preconditioner.Wiener.robustness} & $-0.5$ \\
  Restore the residual image with the model (after convolving with the given 2D gaussian). & \texttt{restore} & true \\
  Fit a 2D gaussian to the PSF image. & \texttt{restore.beam} & fit \\
  \hline
 \end{tabular}
\end{table*}

\subsection{Cutout Procedure}\label{sec:cutout}

The image cubes for each Stokes parameter, each with 288 \qty{1}{\mega\hertz} channels, require large amounts of disk space to store. In contrast to previous treatments of ASKAP images~\citep[e.g.][]{Anderson2021}, we do not produce mosaics of each observed field. The snapshot imaging of RACS provides a sensitivity to a maximum angular scale of \ang{;25}--\ang{;50} over most of the sky. \citetalias{Hale2021} finds that $\sim40\%$ of RACS sources are unresolved at \ang{;;25}. As such, images from RACS contain a large portion of empty sky. We take advantage of this fact, and produce cutouts around each source for Stokes $I$, $Q$, and $U$, for each beam cube in which they appear across all fields in total intensity. We use the criterion that a source must be within \ang{1} of a beam centre to be associated with that beam. We define the upper (lower) cutout boundary for a given source as the maximum (minimum) in RA/Dec of all Gaussian components comprising that source in the total-intensity catalogue, offset by the major axis of that component. We further pad the size of each cutout by three times the size of the PSF (\ang{;;75}). In this way, each cutout contains all the components we require for our later catalogue construction, and enough data to estimate the local $rms$ noise.

The trade-off we make here is significantly reduced data size in exchange for many more data products to manage. We track these data products, as well as their metadata, in a \textsc{mongoDB}\footnote{\url{https://www.mongodb.com}} database. Within this database, we create four `collections' \footnote{Equivalent to `tables' in SQL.}:
\begin{itemize}
 \item \texttt{beams}: One `document'\footnote{Equivalent to `rows' in SQL.} per source. Lists which RACS-low field, and beams within that field, a given source appears in.
 \item \texttt{fields}: One `document' per field. Contains meta-data on each field as described in \citetalias{McConnell2020}. Built directly from the RACS database\footnote{\url{https://bitbucket.csiro.au/projects/ASKAP_SURVEYS/repos/racs}}.
 \item \texttt{islands}: One `document' per source. Contains data and meta-data on a given RACS-low source. Initially populated by the source catalogue from \citetalias{Hale2021}.
 \item \texttt{components}: One `document' per Gaussian component. Contains data and meta-data on a given RACS-low Gaussian component. Initially populated by the Gaussian catalogue from \citetalias{Hale2021}.
\end{itemize}


We now apply our image-based corrections for ionospheric Faraday rotation, as described in \S\ref{sec:calibration}. Using the \textsc{YandaSoft} tool \texttt{linmos}, we simultaneously correct for the primary beam and wide-field leakage, and mosaic the images of each source across different beams within a field. At this stage we have a single cutout, corrected for the primary beam and wide-field leakage, for each source within each field. We cutout and then mosaic, equivalent to mosaicking first and then cutting out, in order to make it more efficient.

Finally, we can mosaic adjacent fields together to correct for the loss of sensitivity towards the edge of a field. We search for all sources that appear in more than one field. Due to the RACS-low tiling scheme, a source can appear in four fields at most. Having found all repeated sources, we again use \texttt{linmos} to perform a weighted co-addition of the images. This now provides us with a single image cube of each source in Stokes $I$, $Q$, and $U$, corrected for the primary beam, wide-field leakage, and ionospheric Faraday rotation across our 30 selected fields.

\subsection{Determination of Rotation Measure}\label{sec:rm_determine}

\subsubsection{Extraction of spectra}\label{sec:spectra}

In this first data release, we provide a polarised component catalogue, complementing the total intensity Gaussian catalogue described in \citetalias{Hale2021}. As such, we leave both direct source-finding on the polarisation images as well as further treatment of the image cubes (e.g. 3D RM-synthesis) to a future release.

Here we use our prior knowledge from the total intensity catalogue, and extract single-pixel spectra in Stokes $I$, $Q$, and $U$ from the location of peak total intensity. It is important to note that since we do not use source-finding, which can decompose a larger island of emission into overlapping Gaussians, our pencil-beam extractions may contain contributions from multiple components which are listed in the \citetalias{Hale2021} catalogue. In lieu of complete component fitting and separation, we provide a flag if given component is blended with another (see~\S\ref{sec:flags}) and additional columns to allow users to easily examine the relative flux of components that are blended with a component of interest (see~\S\ref{sec:catalogue}).

We also extract a `noise' spectrum for each component; estimating the \textit{rms} noise ($\sigma_{QU}$) from each source cutout cube per channel. We define an annulus with an origin at the centre of the image, an inner radius of 10 pixels ($=\ang{;;25}=1\,$PSF) and an outer radius of either 31 pixels ($=\ang{;;77.5}=3.1\,$PSF) or the largest radius allowable within the image if the cutout is less than 62 pixels across. For the ensemble of pixels lying within this annulus, we compute the median absolute deviation from the median (MADFM) of this ensemble for each channel and Stokes parameter. The MADFM is more robust to outlying values than the direct standard deviation. We then correct the MADFM value by a factor of $1/0.6745$ which allows it to approximate the value of the standard deviation. This estimator of the noise is potentially sensitive to a component falling inside of our annulus, which would result in an over-estimation. Such a scenario could, for example, arise in the case of a cutout around a double radio source. We perform a brief exploration of this effect in \ref{app:noise}, and we find that even in the case of a \qty{1}{\jansky\per beam} component in the annulus the error on our noise estimation remains below 50\%.

Before performing further analysis we apply flagging to the 1D spectra. We apply this to remove the effects of narrow-band radio-frequency interference (RFI) and channels for which the \textsc{clean} process diverged. For each Stokes parameter we apply iterative sigma-clipping to a threshold of $5\sigma$ about the median, as measured by the robust MADFM. We mask any channel flagged across all Stokes parameters. Across all of the \ncomponents\ components, the average number of channels in our spectra after flagging is $\flagstats$ out of 288.

To mitigate the effect of the Stokes $I$ spectral index, we perform RM-synthesis~\citep{Burn1966, Brentjens2005} on the fractional Stokes parameters ($q=Q/I$, $u=U/I$). We fit a power-law model as function of frequency ($\nu$) to each extracted Stokes $I$ spectrum of the form:
\begin{align}
 I(\nu_c) &= A \nu_c^{\alpha},\label{eqn:power}\\
 \nu_c &= \frac{\nu}{\nu_0},
\end{align}
where $\nu_0$ is the reference frequency, and all other terms are our model parameters. The reference frequency $\nu_0$ is determined independently for each spectrum as the mean frequency after flagging out bad channels. We fit using the \textsc{SciPy}'s \texttt{optimize.curve\_fit} routine, which also provides the uncertainties on each fitted parameter. We iterate through the number of fitted spectral terms; first fitting just $A$, and then $A,\alpha$. We select the `best' model as the one with minimum Akaike information criterion~\citep[AIC,][]{Akaike1974}, so long as there is not a simpler model within an AIC value of 2~\citep[following][]{Kass1995}. We also assess the quality of the fit, and apply a flag if we determine that the model is unreliable. We detail these quality checks in \S\ref{sec:flags}.

\subsubsection{Rotation measure synthesis}

We use RM-synthesis~\citep{Burn1966,Brentjens2005} in combination with \textsc{RM-clean}~\citep{Heald2009} to quantify the Faraday rotation of our spectra, implemented in \textsc{RM-Tools}\footnote{\url{https://github.com/CIRADA-Tools/RM-Tools}}~\citep{Purcell2020}. For a recent review, and in-depth derivation, of Faraday rotation in astrophysical contexts we refer the reader to \citet{Ferriere2021}.

In brief, as linearly polarised emission propagates through a magneto-ionised medium, the polarisation angle ($\psi$) will change in proportion to both the `Faraday depth' ($\phi$) and the wavelength squared. Faraday depth is a physical quantity, defined at every point in the interstellar medium along the line of sight~\citep{Ferriere2016}
\begin{equation}
 \phi(r) = \mathcal{C} \int_0^r n_e(r') B_\parallel (r') dr',
\end{equation}
where $r$ is the distance from the observer to the source along the line-of-sight (LOS), $n_e$ is the thermal electron density, $B_\parallel$ is the magnetic field projected along the LOS (taken to be positive towards the observer), and $\mathcal{C}\equiv\frac{e^3}{2\pi m_e^2 c^4}\approx0.812$ for $B_\parallel$ in \unit{\micro G}, $n_e$ in \unit{\per\centi\metre\cubed}, and $r$ in \unit{pc}.

In the most simplistic case, where a single source of polarised emission is behind a uniform rotating volume along the line of sight, the Faraday depth will collapse to a single value referred to as the rotation measure (RM). In this case the change in $\psi$ is linearly proportional to $\lambda^2$, where RM is the constant of proportionality:
\begin{align}
 \psi &= \frac{1}{2}\arctan\left(\frac{U}{Q}\right) \\
 \Delta\psi &= \phi\lambda^2 = \text{RM}\lambda^2.
\end{align}
The RM is a observational quantity which, if the simplistic case above holds, we can relate to the physical ISM via~\citep{Ferriere2021}
\begin{equation}
 \text{RM} = \mathcal{C} \int_0^d n_e B_\parallel ds,
\end{equation}
where $d$ is the distance from the source to the observer.

RM-synthesis is a Fourier-transform-like process, with many analogies to aperture synthesis in one dimension. The result of RM-synthesis is a spectrum of complex polarised emission ($pI=P=Q+iU=Le^{2i\psi}$, where $L$ is the linearly polarised intensity and $\psi$ the polarisation angle) as a function of Faraday depth~\citep[$F(\phi)$,][]{Burn1966}:
\begin{equation}
 F(\phi) = \frac{1}{\pi} \int_{-\infty}^{+\infty}p(\lambda^2) e^{-2i\phi(\lambda^2 - \lambda^2_0)} d\lambda^2,\label{eqn:fourier_depth}
\end{equation}
where $\lambda^2_0$ is the reference wavelength-squared to which all polarisation vectors are de-rotated. Following \citet{Brentjens2005} $\lambda_0^2$ is often set to be the weighted mean of the observed $\lambda^2$ range, but it can be set to an arbitrary real value such that $\lambda^2>0$ with consequences in the conjugate domain, as per the Fourier shift theorem. For our purposes, we choose to set $\lambda^2_0$ as the weighted average of the observed band.

Here we refer to $F(\phi)$ as the `Faraday spectrum'. \textsc{RM-Tools} computes the direct, discrete transform of the complex polarisation. We can take the RM of this emission to be the point of maximum polarised intensity ($L=\sqrt{Q^2+U^2}$) in this spectrum:
\begin{equation}
 \text{RM} \equiv \argmax{|F(\phi)|}.
\end{equation}

In RM-synthesis, we weight by the inverse of the measured variance per channel to maximise sensitivity. Noise bias in the polarised intensity is corrected by \textsc{RM-Tools} for all sources with polarised signal-to-noise-ratio (SNR) greater than 5 following \citet{George2012}:
\begin{equation}
 L_\text{corr} = \sqrt{L^2 - 2.3\sigma_{QU}^2}.
\end{equation}

Observationally, we sample a discrete and finite range in $\lambda^2$, causing the Faraday spectrum to be convolved with a transfer function referred to as the RM spread function (RMSF). The properties of the RMSF set the observational limits of the Faraday spectrum. We list these properties in Table~\ref{tab:properties}. Since we deconvolve our Faraday spectrum, we replace the observed RMSF with a smooth Gaussian function. \textsc{RM-Tools} selects the width of the Gaussian function by fitting to the amplitude of main lobe of the computed RMSF. The key properties, namely the resolution (FWHM) in Faraday depth ($\delta\phi$), the maximum Faraday depth ($\phi_\text{max}$), and the maximum Faraday depth scale ($\phi_\text{max-scale}$), are given by \citep{Brentjens2005,Dickey2018}:
\begin{align}
 \delta\phi &\approx \frac{3.8}{\Delta\lambda^2}, \label{eqn:delta_phi} \\
 \phi_\text{max} &\approx \frac{\sqrt{3}}{\delta\lambda^2}, \label{eqn:phi_max} \\
 \phi_\text{max-scale} &\approx \frac{\pi}{\lambda^2_\text{min}},\label{eqn:phi_max_scale}
\end{align}
where $\lambda^2_\text{min}$ is the smallest observed wavelength-squared, $\Delta\lambda^2$ is the total span in wavelength-squared $\Delta\lambda^2=\lambda^2_\text{max} - \lambda^2_\text{min}$, and $\delta\lambda^2$ is the width of each $\lambda^2$ channel (\textsc{RM-tools} takes the median channel width). Cotton \& Rudnick~\citetext{submitted} argue that $\phi_\text{max-scale}$ over-estimates the maximum Faraday depth scale. They provide the quantity
\begin{equation}
    \mathcal{W}_\text{max} = 0.67 \left( \frac{1}{\lambda^2_\text{min}} - \frac{1}{\lambda^2_\text{max}} \right),\label{eqn:wmax}
\end{equation}
which provides the Faraday depth scale at which the power over the entire band drops by a factor of 2. We list all these key values in Table~\ref{tab:properties}. Here we choose to use 100 samples across the main lobe of the RMSF, which provides us improved precision for \textsc{RM-clean}, and allow \textsc{RM-Tools} to automatically select a maximum Faraday depth based on the available channels for a given spectrum.

It is important to note that here our Faraday resolution ($\delta\phi$) is less than maximum Faraday depth scale ($\phi_\text{max-scale}$,$\mathcal{W}_\text{max}$). As such our observations lose sensitivity to polarised emission from Faraday depth structures with characteristic widths greater than these scales \citep[see discussions in e.g.][]{Schnitzeler2014, VanEck2017, Dickey2018}. The true total polarised emission from such Faraday thick emission cannot be recovered directly from our Faraday spectra, but instead must be carefully modelled \citep[e.g.][]{VanEck2017,Thomson2019}. We aim to lift this constraint in the future through the inclusion of higher frequency RACS observations.

\begin{figure}
	\begin{center}
		\includegraphics[width=\linewidth]{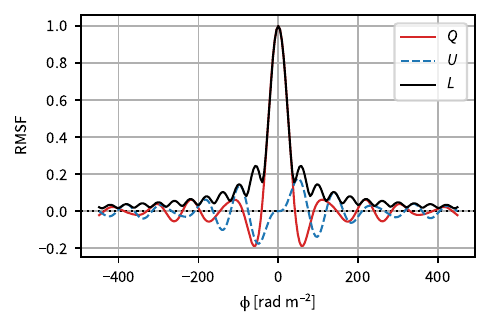}
		\caption{Ideal, uniformly weighted, RM spread function (RMSF).}
		\label{fig:rmsf}
	\end{center}
\end{figure}

\begin{table}
	\caption{Observational properties of SPICE-RACS DR1.}\label{tab:properties}
	\begin{tabular}{lcc}
		\toprule
		Property & Catalogue value \\
		\midrule
		Shortest baseline [\si{\m}] & 100 \\
		Longest baseline [\si{\km}] & 6.4 \\
		Angular resolution [\unit{arcsec}] & 25 \\
		J2000 declination [\unit{\degree}] & $-28.8$ to $+3.6$\\
		J2000 right ascension [\unit{\degree}] & $221.4$ to $179.3$ \\
		Areal coverage [\unit{\degreesq}] & $\sim1306$\\
		Bandwidth [\unit{\mega\hertz}] & 744 to 1032\\
		Channel width [\unit{\mega\hertz}] & 1\\
		$\lambda^2$ coverage [\unit{\metre\squared}] & 0.085 to 0.162\\
		$\lambda^2$ channel width [\unit{\metre\squared}] & $\sim 0.0003\pm0.0001$\\
        $\lambda_0^2$  [\unit{\metre\squared}] (see Eqn.~\ref{eqn:fourier_depth}) & $\sim 0.12\pm0.01$ \\
		Stokes $I$ \textit{rms} noise [\unit{\micro\jansky\per PSF}] & $\sim300$\\
		Stokes $Q$, $U$ \textit{rms} noise [\unit{\micro\jansky\per PSF}] & $\sim80$\\
		$\delta\phi$ [\unit{rad\,m^{-2}}] & $49\pm7$ \\
		$\phi_\text{max}$ [\unit{rad\,m^{-2}}] & $6700\pm300$ \\
		$\phi_\text{max-scale}$ [\unit{rad\,m^{-2}}] & $37.2\pm0.2$ \\
		$\phi$ range [\unit{rad\,m^{-2}}] & $\pm3976$\\
		$\phi$ sampling [\unit{rad\,m^{-2}}] & 0.445\\
        $\mathcal{W}_\text{max}$ [\unit{rad\,m^{-2}}] (see Eqn.~\ref{eqn:wmax}) & $\sim12.0\pm0.1$\\
		\bottomrule
	\end{tabular}
\end{table}

\subsubsection{RM-CLEAN}

The deconvolution of Faraday spectra, much like its two-dimensional counterpart in aperture synthesis imaging, is subject to several practical difficulties. In initially applying \textsc{RM-clean} to our dirty spectra, we found that side-lobes in high SNR spectra were being spuriously fitted and cleaned. We are therefore motivated to find algorithmic improvements to the \textsc{RM-clean} process. Here we take inspiration from the automatic masking feature in \textsc{WSClean}~\citep{offringa2017}. We call this improved algorithm `window' \textsc{RM-clean} and we have implemented it in \textsc{RM-Tools}\footnote{Implemented in \href{https://github.com/CIRADA-Tools/RM-Tools/pull/59}{PR\#59}, available since version 1.2.0.}.

We perform an initial round of deconvolution to some initial shallow threshold; we use $8\sigma$ in $L$ so that all of our spectra with reliable RMs have at least one \textsc{clean} component (see~\S\ref{sec:flags}). We then define a mask around each \textsc{clean} component along the Faraday depth axis at $\pm\delta\phi$ from the component's position. Within this mask we continue cleaning to a second, deeper threshold (we use $5\sigma$ in $L$). This process allows good modelling of strong and broad Faraday features, whilst avoiding the need to set a low global threshold. In addition, it allows us to avoid over-cleaning spurious side-lobe components.

\subsubsection{Faraday complexity}\label{sec:complex}

We wish to draw a distinction between two sets of classifications that are used in the literature: Faraday `thick' vs `thin' and Faraday `complex' vs `simple'. The terms `thin' and `thick' in Faraday space are analogous to `point' and `extended' source classifications in the image domain, respectively. Classifying in this way is therefore reliant on both the Faraday resolution ($\delta\phi$) and maximum scale ($\phi_\text{max-scale}$, $\mathcal{W}_\text{max}$). The simplest Faraday spectrum occurs in a `Faraday screen' scenario, where a polarised source illuminates a foreground Faraday rotating medium \citep{Brentjens2005}. In such a scenario, the fractional $q$ and $u$ spectra will be strictly sinusoidal as a function of $\lambda^2$. Identically, this will correspond to a delta-function in the Faraday spectrum. `Faraday complexity' (i.e. classified as `simple' or `complex'), as summarised in \citet{Alger2021}, is a deviation from this simplistic case. Determining whether a spectrum is Faraday thick or thin requires modelling and careful analysis of the spectrum \citep[e.g.][]{VanEck2017,OSullivan2018, Ma2019, Thomson2019}. For the purposes of our catalogue, we provide metrics to classify spectra as simple or complex, and leave further investigation of the spectral properties to future work.

We quantify the Faraday complexity in two ways. First, since we perform \textsc{RM-clean}, we can compute the second moment of the model components \citep[$m_{2,CC}$,][]{Anderson2015}. Following \citet{Dickey2018}, the zeroth, first, second, and simplified second moments of a Faraday spectrum are:
\begin{align}
 M_0 &= \sum_{i=1}^n F(\phi) d\phi, \label{eqn:M0} \\
 M_1 &= \frac{\sum_{i=1}^n F(\phi) \phi}{\sum_{i=1}^n F(\phi)}, \label{eqn:M1} \\
 M_2 &= \frac{\sum_{i=1}^n F(\phi) (\phi - M_1)^2}{\sum_{i=1}^n F(\phi)}, \label{eqn:M2} \\
 m_2 &= \sqrt{M_2} \label{eqn:m2}
\end{align}
respectively. We compute $m_{2,CC}$ by substituting the polarised intensity of the \textsc{RM-clean} components for $F(\phi)$ in Eqn.~\ref{eqn:M2}. Finally, we can produce a normalised complexity metric by dividing the second moment by the width of the RMSF ($\delta\phi$).

The second complexity parameter we provide is the $\sigma_\text{add}$ quantity provided by \textsc{RM-Tools}. A detailed description and investigation of this parameter is provided in \citet{Purcell2017} and will be published in Van Eck et al. (in prep.). In brief, $\sigma_\text{add}$ is computed by first subtracting a best-fit Faraday screen model from the spectrum. The distribution of residual spectrum is then normalised, and a standard normal distribution is fit. $\sigma_\text{add}$ is then computed for Stokes $q=Q/I$ and $u=U/I$ as the additional scatter beyond the standard normal distribution. Here we report $\sigma_\text{add}$ as:
\begin{equation}
  \sigma_\text{add} = \sqrt{\sigma_{\text{add},q}^2 + \sigma_{\text{add},u}^2}.\label{eqn:sigma_add}
\end{equation}
We also compute the uncertainty on $\sigma_\text{add}$ using a Monte-Carlo approach. \textsc{RM-Tools} reports the 16$^\text{th}$ and 84$^\text{th}$ percentile errors on $\sigma_{\text{add},q}$ and $\sigma_{\text{add},u}$. We take the probability distribution of these values to be log-normal, and draw a thousand random samples from these distributions. We then compute $\sigma_\text{add}$ for each sample and report the median value and error from the resulting distribution.

Finally, it is important to note that these complexity metrics are agnostic to the source of the complexity they are quantifying. In addition the distinctions between `thick' and `thin' that we describe above, these metrics are also unable to determine whether the complexity is intrinsic to the source or is instrumental. Instrumentally, there are many potential sources of spectral complexity including image artefacts, residual RFI, bandpass errors, and so on. Here we make no post hoc correction of the complexity metrics to try to nullify these effects. Instead, we prefer to correct such issues at their source and note that residual effects may still be present in our data.

\subsection{The Catalogue}\label{sec:catalogue}

Here we describe our Gaussian component catalogue. We describe the flags we use within this catalogue and our recommended subsets in \S\ref{sec:flags} and \S\ref{sec:subsets}, respectively. We have used the \textsc{RM-Table}\footnote{\url{https://github.com/CIRADA-Tools/RMTable}}~\citep{VanEck2023} format to construct our catalogue, which provides a minimum set of columns to include. We present the first two rows of our catalogue in Table~\ref{tab:catalogue}. Under this format, we reproduce some columns from the \citetalias{Hale2021} total intensity Gaussian catalogue. Where this is the case, we indicate the corresponding column name in the total intensity catalogue. Throughout we use quote position and polarisation angles in Celestial coordinates, measured North towards East following the IAU convention~\citep{IAU_1973}.

We provide the following \textsc{RM-Table} standard columns:
\begin{itemize}
	\item \texttt{ra} - Right ascension of the component. \texttt{RA} in \citetalias{Hale2021}.
	\item \texttt{dec} - Declination of the component. \texttt{Dec} in \citetalias{Hale2021}.
	\item \texttt{l} - Galactic longitude of the component.
	\item \texttt{b} - Galactic latitude of the component.
	\item \texttt{pos\_err} - Positional uncertainty of the component. Taken as the largest of either the error in RA or Dec. (see below).
	\item \texttt{rm} - Rotation measure.
	\item \texttt{rm\_err} - Error in RM.
	\item \texttt{rm\_width} - Width in Faraday depth. Taken as the second moment of the \textsc{RM-clean} components ($m_{2,CC}$)
	\item \texttt{complex\_flag} - Faraday complexity flag. See \S\ref{sec:flags} below.
	\item \texttt{rm\_method} - RM determination method. \texttt{RM Synthesis - Fractional polarisation} for all components.
	\item \texttt{ionosphere} - Ionospheric correction method, \texttt{FRion} for all components.
	\item \texttt{stokesI} - Stokes $I$ flux density at the reference frequency from the fitted model.
	\item \texttt{stokesI\_err} - Error in the Stokes $I$ flux density. \texttt{Noise} in \citetalias{Hale2021}.
	\item \texttt{spectral\_index} - Stokes $I$ spectral index. Taken as the fitted term $\alpha$ in Equation~\ref{eqn:power}.
	\item \texttt{spectral\_index\_err} - Error in the Stokes $I$ spectral index.
	\item \texttt{reffreq\_I} - Reference frequency for Stokes $I$.
	\item \texttt{polint} - Polarised intensity at the reference frequency.
	\item \texttt{polint\_err} - Error in polarised intensity.
	\item \texttt{pol\_bias} - Polarisation bias correction method. `\texttt{2012\-PASA...29..214G}' for all components, referring to \citet{George2012}.
	\item \texttt{flux\_type} - Spectrum extraction method. `\texttt{Peak}' for all components as we extract spectra from the peak in Stokes $I$.
	\item \texttt{aperture} - Integration aperture. \ang{0} for all components as we extract single-pixel spectra.
	\item \texttt{fracpol} - Fractional linear polarisation (in the range 0 to 1).
	\item \texttt{polangle} - Electric vector polarisation angle.
	\item \texttt{polangle\_err} - Error in the polarisation angle.
	\item \texttt{reffreq\_pol} - Reference frequency for polarisation.
	\item \texttt{stokesQ} - Stokes $Q$ flux density.
	\item \texttt{stokesU} - Stokes $U$ flux density.
	\item \texttt{derot\_polangle} - De-rotated polarisation angle.
	\item \texttt{derot\_polangle\_err} - Error in the de-rotated polarisation angle.
	\item \texttt{beam\_maj} - PSF major axis.
	\item \texttt{beam\_min} - PSF minor axis.
	\item \texttt{beam\_pa} - PSF position angle.
	\item \texttt{reffreq\_beam} - Reference frequency for the PSF.
	\item \texttt{minfreq} - Lowest frequency contributing to the RM.
	\item \texttt{maxfreq} - Highest frequency contributing to the RM.
	\item \texttt{channelwidth} - Median channel width of the spectrum.
	\item \texttt{Nchan} - Number of channels in the spectrum.
	\item \texttt{rmsf\_fwhm} - Full-width at half maximum of the RMSF.
	\item \texttt{noise\_chan} - Median noise per-channel in Stokes $Q$ and $U$.
	\item \texttt{telescope} - Name of telescope. `\texttt{ASKAP}' for all components.
	\item \texttt{int\_time} - Integration time of the observation.
	\item \texttt{epoch} - Median epoch of the observation.
	\item \texttt{leakage} - Instrumental leakage estimate at the position of the component.
	\item \texttt{beamdist} - Angular separation from the centre of nearest tile. \texttt{Separation\_Tile\_Centre} in \citetalias{Hale2021}.
	\item \texttt{catalog} - Name of catalogue. `\texttt{SPICE-RACS-DR1}' for all components.
	\item \texttt{cat\_id} - Gaussian component ID in catalogue. \texttt{Gaussian\_ID} in \citetalias{Hale2021}.
	\item \texttt{complex\_test} - Faraday complexity metric. \texttt{sigma\_add OR Second moment} for all components.
	\item \texttt{dataref} - Data reference. Our CASDA collection, \url{https://data.csiro.au/collection/csiro:58508} for all components.
\end{itemize}

We provide the following additional columns, which are outside of the \textsc{RM-Table} standard:
\begin{itemize}
	\item \texttt{ra\_err} - Error in Right Ascension. \texttt{E\_RA} in \citetalias{Hale2021}.
	\item \texttt{dec\_err} - Error in Declination. \texttt{E\_Dec} in \citetalias{Hale2021}.
	\item \texttt{total\_I\_flux} - Total flux density in Stokes $I$ of the component. \texttt{Total\_Flux\_Gaussian} in \citetalias{Hale2021}.
	\item \texttt{total\_I\_flux\_err} - Error in total flux density in Stokes $I$. \texttt{E\_Total\_Flux\_Gaussian} in \citetalias{Hale2021}.
	\item \texttt{peak\_I\_flux} - Peak flux density of the component in Stokes $I$. \texttt{Peak\_Flux} in \citetalias{Hale2021}.
	\item \texttt{peak\_I\_flux\_err} - Error in peak flux density in Stokes $I$. \texttt{E\_Peak\_Flux} in \citetalias{Hale2021}.
	\item \texttt{maj} - Major axis of the fitted Gaussian in in Stokes $I$. \texttt{Maj} in \citetalias{Hale2021}.
	\item \texttt{maj\_err} - Error in the major axis of the Gaussian fit. \texttt{E\_Maj} in \citetalias{Hale2021}.
	\item \texttt{min} - Minor axis of Gaussian fit. \texttt{Min} in \citetalias{Hale2021}.
	\item \texttt{min\_err} - Error in the minor axis of the Gaussian fit. \texttt{E\_Min} in \citetalias{Hale2021}.
	\item \texttt{pa} - Position angle of the Gaussian fit. \texttt{PA} in \citetalias{Hale2021}.
	\item \texttt{pa\_err} - Error in the position angle of the Gaussian fit. \texttt{E\_PA} in \citetalias{Hale2021}.
	\item \texttt{dc\_maj} - Major axis of the deconvolved Gaussian fit. \texttt{Maj\_DC} in \citetalias{Hale2021}.
	\item \texttt{dc\_maj\_err} - Error in the major axis of the deconvolved Gaussian fit. \texttt{E\_Maj\_DC} in \citetalias{Hale2021}.
	\item \texttt{dc\_min} - Minor axis of deconvolved Gaussian fit
	\item \texttt{dc\_min\_err} - Error in minor axis of deconvolved Gaussian fit. \texttt{E\_Min\_DC} in \citetalias{Hale2021}.
	\item \texttt{dc\_pa} - Position angle of deconvolved Gaussian fit. \texttt{Min\_DC} in \citetalias{Hale2021}
	\item \texttt{dc\_pa\_err} - Error in position angle of deconvolved Gaussian fit. \texttt{E\_Min\_DC} in \citetalias{Hale2021}.
	\item \texttt{N\_Gaus} - Number of Gaussians associated with the source. \texttt{N\_Gaus} in \citetalias{Hale2021}.
	\item \texttt{source\_id} - Total intensity component ID. \texttt{Source\_ID} in \citetalias{Hale2021}.
	\item \texttt{tile\_id} - RACS tile ID. \texttt{Tile\_ID} in \citetalias{Hale2021}.
	\item \texttt{sbid} - ASKAP scheduling block ID. \texttt{SBID} in \citetalias{Hale2021}.
	\item \texttt{start\_time} - Observation start time. \texttt{Obs\_Start\_Time} in \citetalias{Hale2021}.
	\item \texttt{separation\_tile\_centre} - Angular separation from the nearest tile centre, an alias for \texttt{beamdist} above. \texttt{Separation\_Tile\_Centre} in \citetalias{Hale2021}.
	\item \texttt{s\_code} - Source complexity classification (see \citetalias{Hale2021}). \texttt{S\_Code} in \citetalias{Hale2021}.
	\item \texttt{fdf\_noise\_th} - Theoretical noise in the Faraday spectrum.
	\item \texttt{refwave\_sq\_pol} - Reference wavelength squared used for polarisation. Corresponds to \texttt{reffreq\_pol} above.
	\item \texttt{snr\_polint} - Signal-to-noise ratio in polarised intensity.
	\item \texttt{stokesI\_model\_coef} - A comma-separated list of the Stokes $I$ model coefficients corresponding to Equation~\ref{eqn:power}. This format is compatible with the fitting routines in \textsc{RM-Tools}.
	\item \texttt{fdf\_noise\_mad} - Noise in the Faraday spectrum derived using the median absolute deviation.
	\item \texttt{fdf\_noise\_rms} - Noise in the Faraday spectrum derived using the standard deviation.
	\item \texttt{sigma\_add} - $\sigma_\text{add}$ complexity metric (see Equation~\ref{eqn:sigma_add}).
	\item \texttt{sigma\_add\_err} - Error in $\sigma_\text{add}$ complexity metric.
\end{itemize}

Further, we add the following Boolean flag columns (see \S\ref{sec:flags} below):
\begin{itemize}
    \item \texttt{snr\_flag} - Signal-to-noise flag.
    \item \texttt{leakage\_flag} - Leakage flag.
    \item \texttt{channel\_flag} - Number of channels flag.
    \item \texttt{stokesI\_fit\_flag\_is\_negative} - Stokes $I$ model array contains negative values.
    \item \texttt{stokesI\_fit\_flag\_is\_close\_to\_zero} - Stokes $I$ model array is close to 0.
    \item \texttt{stokesI\_fit\_flag\_is\_not\_finite} - Stokes $I$ model array contains non-numerical values.
    \item \texttt{stokesI\_fit\_flag\_is\_not\_normal} - Stokes $I$ model residuals are not normally distributed.
    \item \texttt{stokesI\_fit\_flag} - Overall Stokes $I$ model fit flag.
    \item \texttt{complex\_sigma\_add\_flag} - $\sigma_\text{add}$ complexity flag.
    \item \texttt{complex\_M2\_CC\_flag} - Second moment complexity flag.
    \item \texttt{local\_rm\_flag} - RM is an outlier compared to nearby components.
\end{itemize}

Finally, our work has not yet provided the following columns which are part of \textsc{RM-Table}. As such, these all take on their default or blank values:
\begin{itemize}
	\item \texttt{rm\_width\_err} - Error in the RM width.
	\item \texttt{Ncomp} - Number of RM components.
	\item \texttt{fracpol\_err} - Error in fractional polarisation.
	\item \texttt{stokesQ\_err} - Error in Stokes $Q$ flux density.
	\item \texttt{stokesU\_err} - Error in Stokes $U$ flux density.
	\item \texttt{stokesV} - Stokes $V$ flux density.
	\item \texttt{stokesV\_err} - Error in Stokes $V$.
	\item \texttt{interval} - Interval of time between observations.
	\item \texttt{type} - Source classification.
	\item \texttt{notes} - Notes.
\end{itemize}

\subsubsection{Selection criteria and flags}\label{sec:flags}
We perform our full analysis for all \ncomponents\ components detected in Stokes $I$. Of course, not all components were detected in polarisation, nor are all polarised components suitable for full analysis. We wish to maximise the usage of this catalogue, and therefore make minimal cuts to it. Instead, we include several flags which can then be used, as appropriate, for a variety of science cases. In practice when we decide on a flag value we are making a value judgement between higher completeness or higher correctness. The flags and subsets we provide here are weighted towards higher correctness. Users of the catalogue will certainly be able to extract larger subsets of data with appropriate weighting and understanding of the errors and systematics.

Here we adopt the \textsc{CASA} flag convention, where a value of `\texttt{True}' indicates potentially bad data. Users should take care when using these flags, as this is the inverse to a \textsc{NumPy}-like convention where `\texttt{True}' would indicate that the value should be included.

\paragraph{Component blending}
As we outline in \S\ref{sec:spectra}, the positions in the \citetalias{Hale2021} catalogue contain overlapping components which were separated by Gaussian component fitting, but are blended in our spectral extraction. We provide the flag \texttt{is\_blended\_flag} if a given component is within 1\unit{PSF} of angular separation of any other components in its given source. We recommend using this flag in combination with the \texttt{N\_blended} and \texttt{blend\_ratio} columns to assess the impact of blended components on a given spectrum.

\paragraph{Signal-to-noise}
At low SNR, the value of a derived RM can become unreliable~\citep{George2012, Macquart2012}. We therefore flag all components with a polarised SNR$<8$, above which we consider the RM to be reliable.

\paragraph{Faraday complexity}
As described in \S\ref{sec:complex}, we provide two normalised metrics of complexity: $\sigma_\text{add}$ and $m_{2,CC}/\delta\phi$. We flag a component as being Faraday complex if either $m_{2,CC}/\delta\phi>1$ (\texttt{complex\_M2\_CC\_flag}) or $\sigma_\text{add}/\delta\sigma_\text{add} > 10$ (\texttt{complex\_sigma\_add\_flag}). We refer the reader to \S\ref{sec:complex_rev} for further discussion of these values.

\paragraph{Number of channels}
Flagging spectra during the various stages of processing results in some spectra with a high number of blanked channels. If a spectrum is entirely blanked, we do not perform RM synthesis or any further processing and we do not include it in the final catalogue. For the remaining spectra, we add a flag if more than half of the 288 channels were flagged for the given spectrum. Such components are of reduced quality, and should generally be avoided for analysis.

\paragraph{Stokes $I$ fitting}
We provide several flags for the Stokes $I$ model fitting, as described in \S\ref{sec:spectra}. We evaluate the fitted model at each frequency and if any part of this evaluated model array is \qty{0}{\jansky\per\,beam} or a non-numeric (e.g. `\texttt{nan}' or `\texttt{inf}') value we apply the \texttt{stokesI\_fit\_flag\_is\_negative} or \texttt{stokesI\_fit\_flag\_is\_not\_finite}, respectively. In practice, we find these flags are needed to catch two pathological cases. First, low-signal spectra can cross \qty{0}{\jansky\per PSF}, which can produce a \qty{0}{\jansky\per PSF} or negative model. Second, if a spectrum is particularly poorly behaved, the fitting can diverge, leading to non-numeric values. Further, if any parts of this evaluated model lies within 1 standard deviation (from the data) of \qty{0}{\jansky\per\,beam} we apply \texttt{stokesI\_fit\_flag\_is\_close\_to\_zero}. If any of these flags are \verb|True|, or if the fitting routine fails to fit any form of Equation~\ref{eqn:power}, we set the overall flag \texttt{stokesI\_fit\_flag} to \verb|True|.

In addition to the tests above, we also inspect the residuals of the fitted model. Using \textsc{SciPy}'s \texttt{stats.normaltest} routine we check if the distribution of the residuals are normally distributed. We use the condition that if the hypothesis $p$-value is $<10^{-6}$ we apply \texttt{stokesI\_fit\_flag\_is\_not\_normal}. Under the assumption that residuals ought to be normally distributed, this corresponds to a 1 in $10^6$ chance of producing a false positive.

\paragraph{Local RM}
Even when using the above flags, there remain some components with outlying RM with respect to the ensemble of nearby components. Such components would be undesirable for use in an RM grid experiment as a high intrinsic RM is likely due to source properties and not the foreground medium~\citep[e.g.][]{Pasetto2016}. To identify these components, we use Voronoi binning~\citep[implemented in the \textsc{VorBin} library,][]{Cappellari2003} to group components on the sky into local ensembles. We perform this binning on components where \texttt{snr\_flag}, \texttt{leakage\_flag}, \texttt{channel\_flag}, and \texttt{stokes\_I\_fit\_flag} are all \texttt{False}. Rather than binning using a metric of signal-to-noise, we instead target Voronoi bins with 50 components per bin.

For each bin we apply iterative MADFM sigma-clipping to a threshold of $3\sigma$ about the median RM. We flag components which are identified as outliers by this process. The remaining components which did not go through this process receive a flag of `\texttt{False}'.

\paragraph{Leakage}
As we detail in \ref{app:leakage}, we fit a polynomial envelope to the residual fractional widefield leakage in our catalogue as function of the distance from the nearest field centre. For each component in the catalogue, given its separation from the nearest field centre, if its polarisation fraction places it below our fitted leakage envelope we set the \texttt{leakage\_flag} to \texttt{True}.

\subsubsection{Recommended basic subsets}\label{sec:subsets}
For users of our catalogue in need of a simple subset of the catalogue we recommend three basic subsets. First is the subset where our Stokes $I$ model is reliable (`\texttt{goodI}'), where \texttt{channel\_flag} and \texttt{stokesI\_fit\_flag} are \texttt{False}. Next, the subset where we have a reliable detection of polarised signal (`\texttt{goodL}'), but not necessarily a reliable RM, where \texttt{leakage\_flag} is \texttt{False}, the polarised SNR is $\geq5$, and the component is in \texttt{goodI}. Finally, to obtain a subset with reliable RMs (`\texttt{goodRM}') we recommend components in \texttt{goodL} where \texttt{snr\_flag} is also \texttt{False}. We have designed these flags to be reliable and relatively conservative. We encourage advanced users to explore beyond these subsets, taking into account caveats in these data which we explore in \S\ref{sec:analysis}. For use in a RM-grid, users may also wish to consider removing complex sources and outlying RMs by selecting where \texttt{complex\_flag} and \texttt{local\_rm\_flag} \texttt{False}.

\begin{center}
\centering
\onecolumn
\begin{longtable}{l|ccc}
\caption {The first two rows of the SPICE-RACS DR1 catalogue. We have transposed the table for readability. We define all column names in \S\ref{sec:catalogue}.}\label{tab:catalogue}\\
\toprule
\endfirsthead
\caption*{\textbf{Table \ref{tab:catalogue} Continued}}\\
\endhead
\endfoot
\bottomrule
\endlastfoot
Column & Units & Row 1 & Row 2 \\
\midrule
\texttt{source\_id} & $\mathrm{---}$ & \texttt{RACS\_1146-18A\_1502} & \texttt{RACS\_1146-18A\_1570} \\
\texttt{cat\_id} & $\mathrm{---}$ & \texttt{RACS\_1146-18A\_1681} & \texttt{RACS\_1146-18A\_1757} \\
\texttt{tile\_id} & $\mathrm{---}$ & \texttt{RACS\_1213-18A} & \texttt{RACS\_1213-18A} \\
\texttt{ra} & $\mathrm{deg}$ & \num{1.799e2} & \num{1.798e2} \\
\texttt{ra\_err} & $\mathrm{arcsec}$ & \num{6.000e-2} & \num{2.000e-2} \\
\texttt{dec} & $\mathrm{deg}$ & \num{-2.183e1} & \num{-2.171e1} \\
\texttt{dec\_err} & $\mathrm{arcsec}$ & \num{8.000e-2} & \num{2.000e-2} \\
\texttt{total\_I\_flux} & $\mathrm{Jy}$ & \num{7.486e-2} & \num{2.274e-1} \\
\texttt{total\_I\_flux\_err} & $\mathrm{Jy}$ & \num{5.787e-3} & \num{1.643e-2} \\
\texttt{peak\_I\_flux} & $\mathrm{Jy\,PSF^{-1}}$ & \num{6.297e-2} & \num{2.213e-1} \\
\texttt{peak\_I\_flux\_err} & $\mathrm{Jy\,PSF^{-1}}$ & \num{3.740e-4} & \num{3.520e-4} \\
\texttt{maj\_axis} & $\mathrm{arcsec}$ & \num{2.952e1} & \num{2.548e1} \\
\texttt{maj\_axis\_err} & $\mathrm{arcsec}$ & \num{1.900e-1} & \num{4.000e-2} \\
\texttt{min\_axis} & $\mathrm{arcsec}$ & \num{2.524e1} & \num{2.527e1} \\
\texttt{min\_axis\_err} & $\mathrm{arcsec}$ & \num{1.400e-1} & \num{4.000e-2} \\
\texttt{pa} & $\mathrm{deg}$ & \num{1.740e2} & \num{6.480e1} \\
\texttt{pa\_err} & $\mathrm{deg}$ & \num{4.700e-1} & \num{8.430e0} \\
\texttt{dc\_maj\_axis} & $\mathrm{arcsec}$ & \num{1.566e1} & \num{4.660e0} \\
\texttt{dc\_maj\_axis\_err} & $\mathrm{arcsec}$ & \num{1.900e-1} & \num{4.000e-2} \\
\texttt{dc\_min\_axis} & $\mathrm{arcsec}$ & \num{3.150e0} & \num{3.590e0} \\
\texttt{dc\_min\_axis\_err} & $\mathrm{arcsec}$ & \num{1.400e-1} & \num{4.000e-2} \\
\texttt{dc\_pa} & $\mathrm{deg}$ & \num{1.740e2} & \num{6.480e1} \\
\texttt{dc\_pa\_err} & $\mathrm{deg}$ & \num{4.700e-1} & \num{8.430e0} \\
\texttt{stokesI\_err} & $\mathrm{Jy\,PSF^{-1}}$ & \num{3.620e-4} & \num{3.500e-4} \\
\texttt{beamdist} & $\mathrm{deg}$ & \num{4.317e0} & \num{4.300e0} \\
\texttt{N\_Gaus} & $\mathrm{---}$ & $1$ & $1$ \\
\texttt{sbid} & $\mathrm{---}$ & \texttt{8584} & \texttt{8584} \\
\texttt{start\_time} & $\mathrm{MJD}$ & \num{5.860e4} & \num{5.860e4} \\
\texttt{separation\_tile\_centre} & $\mathrm{deg}$ & \num{4.317e0} & \num{4.300e0} \\
\texttt{s\_code} & $\mathrm{---}$ & \texttt{S} & \texttt{S} \\
\texttt{rm} & $\mathrm{rad\,m^{-2}}$ & \num{-4.275e0} & \num{-5.768e0} \\
\texttt{rm\_err} & $\mathrm{rad\,m^{-2}}$ & \num{1.496e0} & \num{7.031e-1} \\
\texttt{polint} & $\mathrm{Jy\,PSF^{-1}}$ & \num{3.057e-3} & \num{9.228e-3} \\
\texttt{polint\_err} & $\mathrm{Jy\,PSF^{-1}}$ & \num{1.951e-4} & \num{2.362e-4} \\
\texttt{stokesQ} & $\mathrm{Jy\,PSF^{-1}}$ & \num{-3.299e-5} & \num{3.006e-3} \\
\texttt{stokesU} & $\mathrm{Jy\,PSF^{-1}}$ & \num{3.070e-3} & \num{8.732e-3} \\
\texttt{polangle} & $\mathrm{deg}$ & \num{4.531e1} & \num{3.550e1} \\
\texttt{polangle\_err} & $\mathrm{deg}$ & \num{1.820e0} & \num{7.327e-1} \\
\texttt{derot\_polangle} & $\mathrm{deg}$ & \num{7.431e1} & \num{7.148e1} \\
\texttt{derot\_polangle\_err} & $\mathrm{deg}$ & \num{9.872e0} & \num{3.665e0} \\
\texttt{fracpol} & $\mathrm{---}$ & \num{8.612e-2} & \num{4.522e-2} \\
\texttt{reffreq\_pol} & $\mathrm{Hz}$ & \num{8.712e8} & \num{9.086e8} \\
\texttt{reffreq\_beam} & $\mathrm{Hz}$ & \num{8.712e8} & \num{9.086e8} \\
\texttt{reffreq\_I} & $\mathrm{Hz}$ & \num{8.882e8} & \num{8.877e8} \\
\texttt{fdf\_noise\_th} & $\mathrm{Jy\,PSF^{-1}}$ & \num{1.951e-4} & \num{2.362e-4} \\
\texttt{rmsf\_fwhm} & $\mathrm{rad\,m^{-2}}$ & \num{4.709e1} & \num{5.498e1} \\
\texttt{refwave\_sq\_pol} & $\mathrm{m^{2}}$ & \num{1.184e-1} & \num{1.089e-1} \\
\texttt{stokesI} & $\mathrm{Jy\,PSF^{-1}}$ & \num{3.603e-2} & \num{2.058e-1} \\
\texttt{stokesI\_fit\_flag\_is\_negative} & $\mathrm{---}$ & \texttt{False} & \texttt{False} \\
\texttt{stokesI\_fit\_flag\_is\_close\_to\_zero} & $\mathrm{---}$ & \texttt{False} & \texttt{False} \\
\texttt{stokesI\_fit\_flag\_is\_not\_finite} & $\mathrm{---}$ & \texttt{False} & \texttt{False} \\
\texttt{stokesI\_fit\_flag\_is\_not\_normal} & $\mathrm{---}$ & \texttt{False} & \texttt{False} \\
\texttt{stokesI\_chi2\_red} & $\mathrm{---}$ & \num{1.145e0} & \num{4.010e0} \\
\texttt{snr\_polint} & $\mathrm{---}$ & \num{1.574e1} & \num{3.910e1} \\
\texttt{minfreq} & $\mathrm{Hz}$ & \num{7.440e8} & \num{7.440e8} \\
\texttt{maxfreq} & $\mathrm{Hz}$ & \num{1.031e9} & \num{1.031e9} \\
\texttt{channelwidth} & $\mathrm{Hz}$ & \num{1.000e6} & \num{1.000e6} \\
\texttt{Nchan} & $\mathrm{---}$ & $275$ & $270$ \\
\texttt{rm\_width} & $\mathrm{rad\,m^{-2}}$ & \num{1.350e1} & \num{1.126e1} \\
\texttt{stokesI\_model\_coef} & $\mathrm{---}$ & \makecell{\texttt{nan},\\\texttt{nan},\\\texttt{nan},\\\texttt{-0.47615490049552056},\\\texttt{0.03570472872802541}} & \makecell{\texttt{nan},\\\texttt{nan},\\\texttt{nan},\\\texttt{1.5412339631245944},\\\texttt{0.19855502989246643}} \\
\texttt{stokesI\_model\_coef\_err} & $\mathrm{---}$ & \makecell{\texttt{nan},\\\texttt{nan},\\\texttt{nan},\\\texttt{0.14196683816999223},\\\texttt{0.00048307191814766517}} & \makecell{\texttt{nan},\\\texttt{nan},\\\texttt{nan},\\\texttt{0.02702178622671984},\\\texttt{0.0005112067063229497}} \\
\texttt{stokesI\_model\_order} & $\mathrm{---}$ & \num{1.000e0} & \num{1.000e0} \\
\texttt{noise\_chan} & $\mathrm{Jy\,PSF^{-1}}$ & \num{3.305e-3} & \num{4.050e-3} \\
\texttt{fdf\_noise\_mad} & $\mathrm{Jy\,PSF^{-1}}$ & \num{1.436e-4} & \num{2.121e-4} \\
\texttt{fdf\_noise\_rms} & $\mathrm{Jy\,PSF^{-1}}$ & \num{3.128e-4} & \num{4.975e-4} \\
\texttt{beam\_maj} & $\mathrm{deg}$ & \num{6.944e-3} & \num{6.944e-3} \\
\texttt{beam\_min} & $\mathrm{deg}$ & \num{6.944e-3} & \num{6.944e-3} \\
\texttt{beam\_pa} & $\mathrm{deg}$ & \num{0.000e0} & \num{0.000e0} \\
\texttt{sigma\_add} & $\mathrm{---}$ & \num{5.777e-1} & \num{1.258e0} \\
\texttt{sigma\_add\_err} & $\mathrm{---}$ & \num{5.243e0} & \num{9.015e-2} \\
\texttt{snr\_flag} & $\mathrm{---}$ & \texttt{False} & \texttt{False} \\
\texttt{leakage\_flag} & $\mathrm{---}$ & \texttt{False} & \texttt{False} \\
\texttt{channel\_flag} & $\mathrm{---}$ & \texttt{False} & \texttt{False} \\
\texttt{stokesI\_fit\_flag} & $\mathrm{---}$ & \texttt{False} & \texttt{False} \\
\texttt{complex\_sigma\_add\_flag} & $\mathrm{---}$ & \texttt{False} & \texttt{True} \\
\texttt{complex\_M2\_CC\_flag} & $\mathrm{---}$ & \texttt{False} & \texttt{False} \\
\texttt{local\_rm\_flag} & $\mathrm{---}$ & \texttt{False} & \texttt{False} \\
\texttt{is\_blended\_flag} & $\mathrm{---}$ & \texttt{False} & \texttt{False} \\
\texttt{blend\_ratio} & $\mathrm{---}$ & \texttt{nan} & \texttt{nan} \\
\texttt{N\_blended} & $\mathrm{---}$ & $0$ & $0$ \\
\texttt{spectral\_index} & $\mathrm{---}$ & \num{-4.762e-1} & \num{1.541e0} \\
\texttt{spectral\_index\_err} & $\mathrm{---}$ & \num{1.420e-1} & \num{2.702e-2} \\
\texttt{int\_time} & $\mathrm{s}$ & \num{8.958e2} & \num{8.958e2} \\
\texttt{epoch} & $\mathrm{MJD}$ & \num{5.897e4} & \num{5.897e4} \\
\texttt{l} & $\mathrm{deg}$ & \num{2.873e2} & \num{2.872e2} \\
\texttt{b} & $\mathrm{deg}$ & \num{3.947e1} & \num{3.956e1} \\
\texttt{pos\_err} & $\mathrm{arcsec}$ & \num{2.251e1} & \num{2.251e1} \\
\texttt{rm\_method} & $\mathrm{---}$ & \makecell{\texttt{RM Synthesis -},\\\texttt{ Fractional polarization}} & \makecell{\texttt{RM Synthesis -},\\\texttt{ Fractional polarization}} \\
\texttt{telescope} & $\mathrm{---}$ & \texttt{ASKAP} & \texttt{ASKAP} \\
\texttt{pol\_bias} & $\mathrm{---}$ & \texttt{2012PASA...29..214G} & \texttt{2012PASA...29..214G} \\
\texttt{catalog\_name} & $\mathrm{---}$ & \texttt{SPICE-RACS-DR1} & \texttt{SPICE-RACS-DR1} \\
\texttt{ionosphere} & $\mathrm{---}$ & \texttt{FRion} & \texttt{FRion} \\
\texttt{flux\_type} & $\mathrm{---}$ & \texttt{Peak} & \texttt{Peak} \\
\texttt{aperture} & $\mathrm{deg}$ & \num{0.000e0} & \num{0.000e0} \\
\texttt{leakage} & $\mathrm{---}$ & \num{3.182e-2} & \num{3.133e-2} \\
\texttt{complex\_flag} & $\mathrm{---}$ & \texttt{False} & \texttt{True} \\
\texttt{rm\_width\_err} & $\mathrm{rad\,m^{-2}}$ & \texttt{nan} & \texttt{nan} \\
\texttt{complex\_test} & $\mathrm{---}$ & \texttt{sigma\_add OR Second moment} & \texttt{sigma\_add OR Second moment} \\
\texttt{Ncomp} & $\mathrm{---}$ & $1$ & $1$ \\
\texttt{fracpol\_err} & $\mathrm{---}$ & \texttt{nan} & \texttt{nan} \\
\texttt{stokesQ\_err} & $\mathrm{Jy}$ & \texttt{nan} & \texttt{nan} \\
\texttt{stokesU\_err} & $\mathrm{Jy}$ & \texttt{nan} & \texttt{nan} \\
\texttt{stokesV} & $\mathrm{Jy}$ & \texttt{nan} & \texttt{nan} \\
\texttt{stokesV\_err} & $\mathrm{Jy}$ & \texttt{nan} & \texttt{nan} \\
\texttt{obs\_interval} & $\mathrm{d}$ & \texttt{nan} & \texttt{nan} \\
\texttt{dataref} & $\mathrm{---}$ & \makecell{\texttt{https://},\\\texttt{data.csiro.au/},\\\texttt{collection/},\\\texttt{csiro:58508/}} & \makecell{\texttt{https://},\\\texttt{data.csiro.au/},\\\texttt{collection/},\\\texttt{csiro:58508/}} \\
\texttt{type} & $\mathrm{---}$ & $\mathrm{---}$ & $\mathrm{---}$ \\
\texttt{notes} & $\mathrm{---}$ & $\mathrm{---}$ & $\mathrm{---}$ \\
\end{longtable}
\twocolumn
\end{center}

\section{QUALITY ASSESSMENT}\label{sec:analysis}

\subsection{Spectral Noise}\label{sec:noise}

\begin{figure}
	\centering
	\begin{subfigure}[b]{\columnwidth}
		\includegraphics[width=\textwidth]{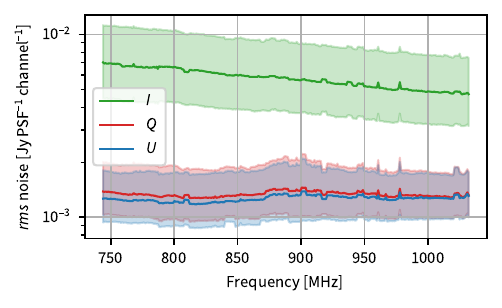}
		\caption{}\label{fig:noise_chan}
	\end{subfigure}
	\begin{subfigure}[b]{\columnwidth}
		\includegraphics[width=\textwidth]{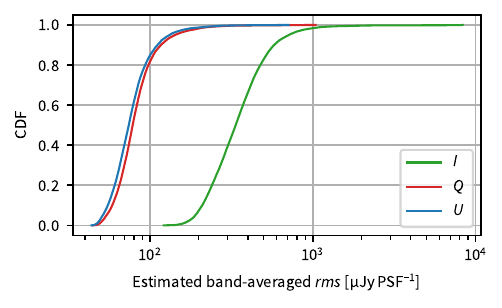}
		\caption{}\label{fig:noise_cdf}
	\end{subfigure}
	\caption{Measured \textit{rms} noise in each Stokes parameter across all observed fields. (\subref{fig:noise_chan}) Noise as a function of frequency. We show the median noise with a solid line, and the $\pm1\sigma$ range as a shaded region. (\subref{fig:noise_cdf}) The cumulative distribution function (CDF) of estimated band-averaged noise for each Stokes parameter. In Stokes $Q$ and $U$ we are approaching the theoretical noise limit, whereas in Stokes $I$ the noise by a factor of 3 to 4 higher. We attribute this to the higher level of artefacts and sidelobes in the Stokes $I$ images.}\label{fig:noise}
\end{figure}

We begin our assessment of our catalogue by inspecting the properties of the noise as a function of frequency, Stokes parameter, and direction on the sky. In Figure~\ref{fig:noise} we show the distribution of our estimated \textit{rms} noise per channel, per Stokes parameter across all components in our region. In Stokes $I$, we find an average $rms$ of $\stokesInoise$\,\unit{\milli\jansky\per PSF\per channel}, which corresponds to a band-averaged value of $\stokesIavnoise$\,\unit{\micro\jansky\per PSF} (assuming noise scaling with the square-root of the number of channels), with a $90^\text{th}$ percentile value of \qty{580}{\micro\jansky\per PSF}. This is marginally higher than values reported in \citetalias{McConnell2020} and \citetalias{Hale2021} for multi-frequency synthesis (MFS) images, where we found a median and $90^\text{th}$-percentiles \textit{rms} noise of \qty{250}{\micro\jansky\per PSF} and \qty{330}{\micro\jansky\per PSF}, respectively. We can attribute this, in part, to local $rms$ variations due to bright sources and declination effects, our larger final PSF (\ang{;;25}, achieved by convolution with a Gaussian kernel) and to our shallower deconvolution threshold. The latter is necessitated by the lower signal-to-noise in each channel image of our frequency cubes against the full-band MFS images. We also note that some observations that do not appear in \citetalias{McConnell2020} are included in \citetalias{Hale2021} catalogue due to their PSF. Some of these fields have a higher $rms$ noise than ones in \citetalias{McConnell2020}, but could be convolved to the common \ang{;;25} resolution.

Looking to Stokes $Q$ and $U$, we find that the $rms$ noise is $\stokesQnoise$\,\unit{\milli\jansky\per PSF\per channel} and $\stokesUnoise$\,\unit{\milli\jansky\per PSF\per channel}, respectively. Converting to an estimated band-averaged value, we find a median noise of $\stokesQavnoise$\,\unit{\micro\jansky\per PSF} and $\stokesUavnoise$\,\unit{\micro\jansky\per PSF} for $Q$ and $U$, respectively. After performing RM-synthesis the median noise in linear polarisation is $\stokesLnoise$\,\unit{\micro\jansky\per PSF}. \textsc{RM-Tools} measures this noise value by taking the MAD of the Faraday spectrum, excluding the main peak. As such, Faraday complexity can induce an increased measured `noise' into this computation. Overall, these values are close to the expected \textit{rms} values for naturally weighted images. Even though we use $-0.5$ robust image weighting, this is applied per-channel. As such, computing an average along the frequency axis after this channel imaging should approach the naturally weighted noise. 

In Figure~\ref{fig:noise_chan} we can see that the $rms$ noise in Stokes $Q$ and $U$ has a weak local maximum around $\sim$\qty{900}{\mega\hertz}. We find no such feature in Stokes $I$, even after removing the apparent spectral dependence. We are able to attribute this feature to observations made earlier in the RACS-low observing campaign. As outlined in Table~\ref{tab:fields}, our subset of fields is comprised roughly of two epochs; which can be selected with SBID greater or lesser than 9000. We find that the the early set of observations (SBID$<9000$) exhibit this local maximum, whereas the $Q,U$ noise in later observations (SBID$>9000$) are approximately flat as function of frequency. We therefore attribute this feature to lower quality bandpass solutions during early RACS-low observations.

\begin{figure*}
	\centering
	\begin{subfigure}[b]{0.49\columnwidth}
		\includegraphics[width=\textwidth]{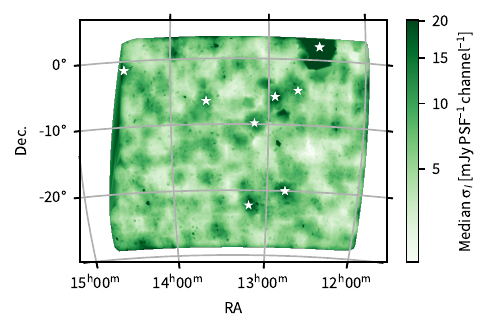}
		\caption{}\label{fig:noise_map_I}
	\end{subfigure}
	\begin{subfigure}[b]{0.49\columnwidth}
		\includegraphics[width=\textwidth]{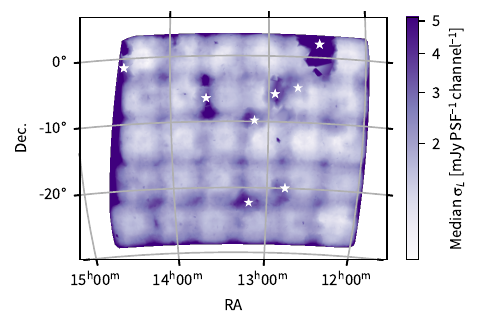}
		\caption{}\label{fig:noise_map_L}
	\end{subfigure}
	\caption{Spatial distribution of \textit{rms} noise in (\protect\subref{fig:noise_map_I}) Stokes $I$ and; (\protect\subref{fig:noise_map_L}) linearly polarised intensity ($L$). White stars indicate the position of components with a Stokes $I$ flux density $>$\qty{3}{\jansky\per PSF}. Our linear mosaicking of adjacent beams and fields (weighted by inverse-variance) produces a spatial pattern in the resulting noise. This effect is particularly noticeable in the $\sigma_L$ distribution, where the boundaries of our 30 square fields are apparent.}\label{fig:noise_map}
\end{figure*}

Finally, we can turn our attention to the spatial distribution of the \textit{rms} noise in $I$ and $L$ across the observed fields, which we show in Figure~\ref{fig:noise_map}. In both maps we see artefacts around the brightest sources in the field. In particular, the worst noise appears in the North-West portion of the field, surrounding 3C273 (\texttt{source\_id}: \texttt{RACS\_1237+00A\_3595}), which has an integrated Stokes I flux density of $\sim$\qty{64}{\jansky} in RACS-low. Looking throughout the rest of the field we can see greater variance in the total intensity noise, indicating the impact of sources to the \textit{rms} noise. By contrast, in the polarised intensity noise the square sensitivity pattern of the beam configuration is clearly apparent. This is to be expected as RACS-low was conducted with a non-interleaved \texttt{square\_6x6} footprint, and a wide (\ang{1.05}) beam pitch. We conclude that, away from the brightest sources, we are approaching the sensitivity limit in Stokes $Q$ and $U$, but are artefact-limited in Stokes $I$. We note that since we are close to the thermal noise limit in Stokes $Q$ and $U$ our noise measurements become sensitive to residual calibration errors, such as the feature we describe above. 

Users of our catalogue should remain aware that the noise in our catalogue is not spatially uniform. Without appropriately weighting for uncertainties, this distribution can cause spurious correlations to appear when assessing bulk properties and statistics within our catalogue. In addition to the flag columns we provide, users should also make appropriate use of the uncertainty columns when undertaking statistical analysis with our catalogue.

\subsection{Linear Polarisation and Rotation Measure}\label{sec:rm}

We now inspect the polarisation properties of SPICE-RACS components after applying our internal flags to select our recommended basic subsets (see \S\ref{sec:subsets}). We summarise the basic statistics of these subsets in Table~\ref{tab:pol_param}, following the approximate format of \citet{Adebahr2022} (their Table 5) for later comparison. Of the \ncomponents\ total components we obtain reliable fits (\texttt{goodI}) to the Stokes $I$ spectrum for \ngoodI\ ($\sim23\%$) components, we detect \npol\ ($\sim8\%$) components in linear polarisation (\texttt{goodL}), and \nrms\ ($\sim5.5\%$) components have a reliable RM (\texttt{goodRM}). Given our observed area of $\sim$\qty{1300}{\degreesq}, our average RM density is $\sim$\qty{4}{\perdegreesq} (the \texttt{goodRM} subset) with $\sim$\qty{6.5}{\perdegreesq} for polarised sources above $5\sigma$ (the \texttt{goodL} subset).

We can compare our recovered source densities to an `expected' value by extrapolating from \citet{Rudnick2014, Rudnick2014_Err}. \citet{Rudnick2014} provide a power-law scaling relation with polarised source density $\propto \sigma_L^{-0.6}$. Assuming that beam and wavelength-dependent depolarisation are negligible to first order, we find an expected spatial density of \qty{9(4)}{\perdegreesq} and \qty{7(3)}{\perdegreesq} at $5\sigma$ and $8\sigma$, respectively, for SPICE-RACS. As we detail in \ref{app:leakage}, our residual widefield leakage is $\geq1\%$ across our survey area, whereas \citet{Rudnick2014} report residual leakage of $\sim0.4\%$. If we are able to further improve our residual leakage in future releases, we can therefore expect to approach these values.

\begin{figure*}
    \centering
    \includegraphics{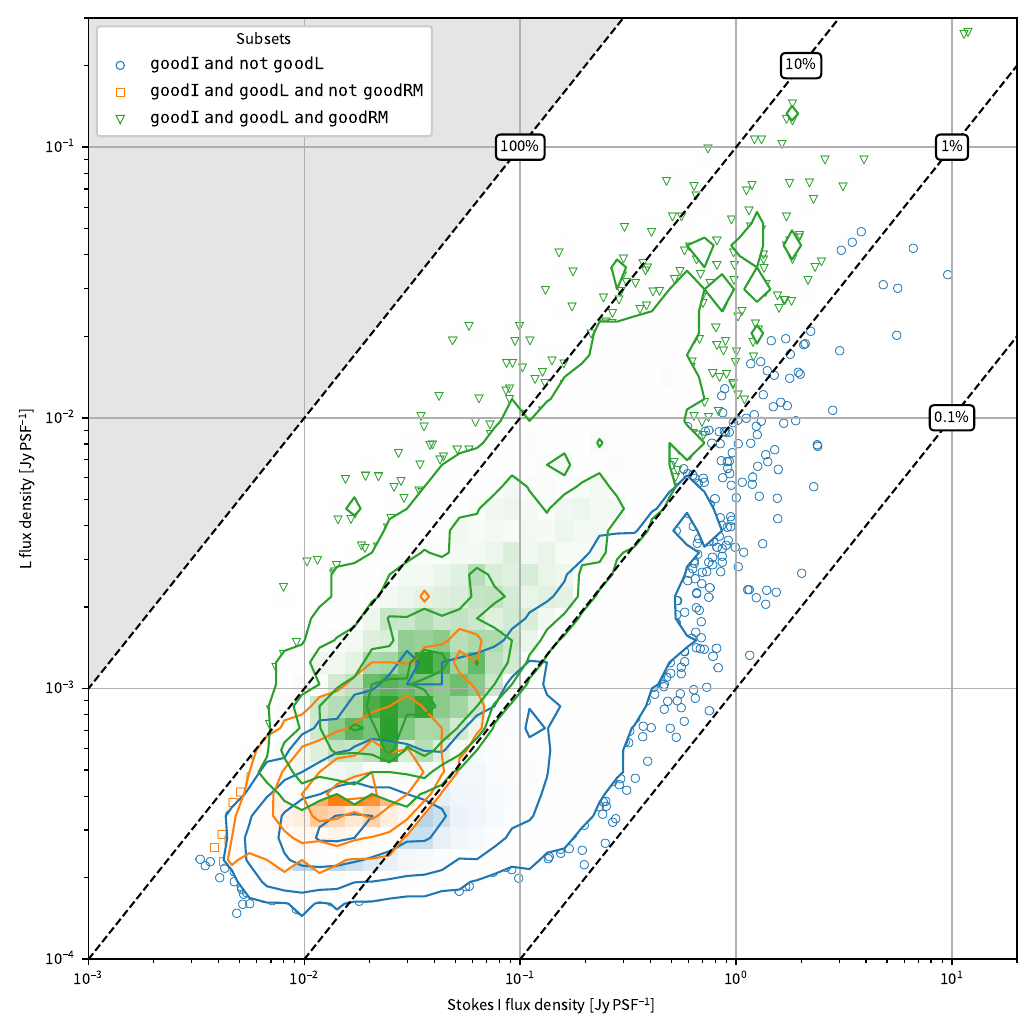}
    \caption{Linearly polarised flux density ($L$) against Stokes $I$ flux density for SPICE-RACS DR1. Each coloured/shaped marker corresponds to our basic subsets as defined in \S\ref{sec:subsets}: The blue circles represent components for which we have a reliable fit to the Stokes $I$ spectrum, but the linear polarisation may be spurious, the orange squares represent components that have a reliable linear polarisation detection but a potentially unreliable RM, and the green triangles represent components with a reliable RM. The dashed black lines show contours of constant fractional polarisation, and the grey shaded region is the area of $>100\%$ fractional polarisation. Where the scatter points become over-dense we show the density of points as a 2D histogram. The contour levels of the histogram are at the $2^\text{nd}$, 16$^\text{th}$, 50$^\text{th}$, 84$^\text{th}$, and 98$^\text{th}$ percentiles.}
    \label{fig:flux_flux}
\end{figure*}
\begin{figure*}
    \centering
    \includegraphics{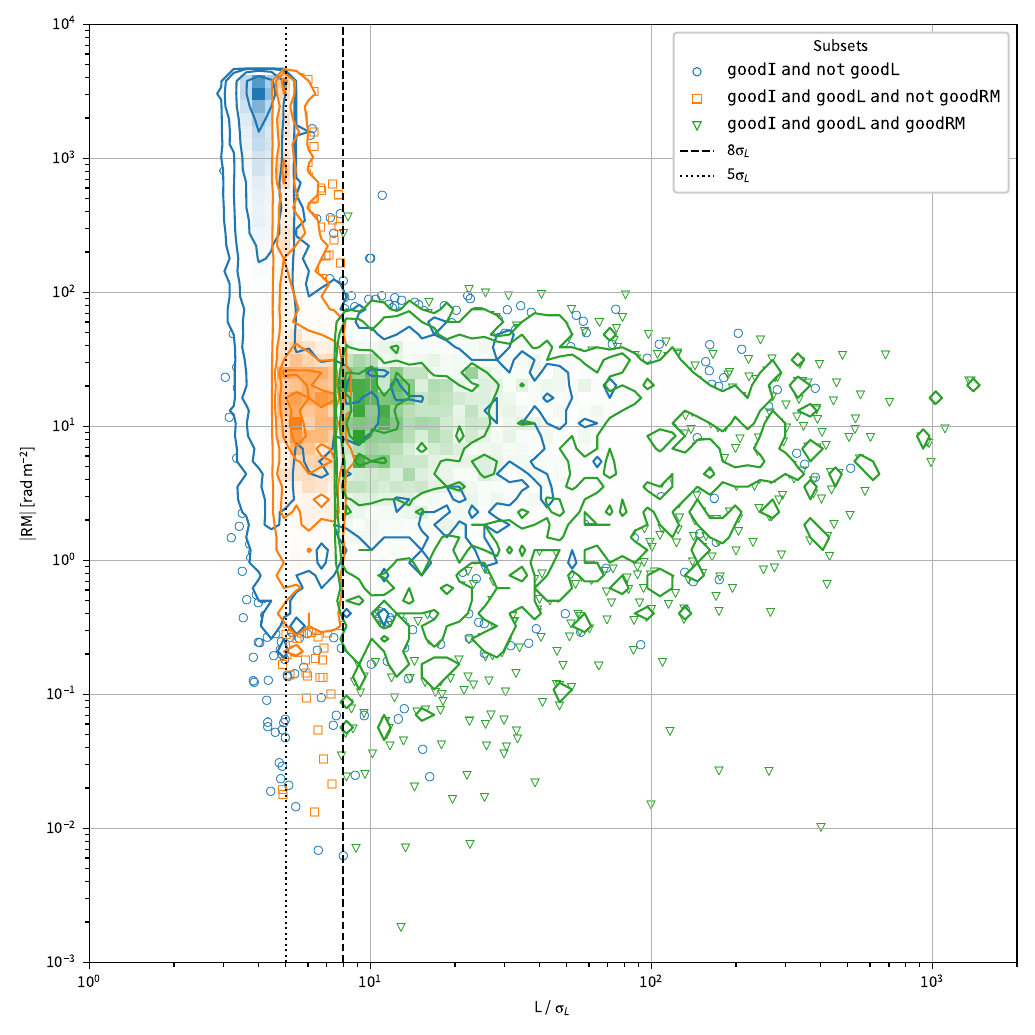}
    \caption{The absolute value of the rotation measure ($|\text{RM}|$) as a function of the linearly polarised signal-to-noise ratio ($L/\sigma_L$). Each coloured/shaped marker corresponds to our basic subsets as defined in \S\ref{sec:subsets} as per Figure~\ref{fig:flux_flux}. The dashed and dotted vertical lines represent the $8\sigma$ and $5\sigma$ levels, respectively. We note that to be included in the \texttt{goodRM} and \texttt{goodL} subsets a component must have $L/\sigma_L$ $\geq8$ and $\geq5$, respectively (see \S\ref{sec:subsets} for further details). Where the scatter points become over-dense we show the density of points as a 2D histogram. The contour levels of the histogram are at the $2^\text{nd}$, 16$^\text{th}$, 50$^\text{th}$, 84$^\text{th}$, and 98$^\text{th}$ percentiles.}
    \label{fig:flux_rm}
\end{figure*}

In Figure~\ref{fig:flux_flux} we show the distribution of the (bias-corrected) linearly polarised flux density against the total flux density, split by our basic subsets. The distribution of reliable polarised sources bears overall resemblance to previous large radio polarimetric surveys \citep[e.g.][]{Hales2014, Anderson2021, OSullivan2023}. Notably, however, due to our residual leakage we have no components with $\leq1\%$ fractional polarisation in our \texttt{goodL} or \texttt{goodRM} subsets. Looking now at Figure~\ref{fig:flux_rm}, we show the distribution of the absolute RM value against the polarised SNR for the same subsets. At SNRs less than $8\sigma$ we can see a large spike in high absolute value RMs, which are certainly spurious. This gives us confidence in our recommended thresholds for the \texttt{goodRM} basic subset. However, we also see that the bulk population of RMs, which we can take to be real, appear to extend below the $8\sigma$ threshold. We conclude that there are therefore many more scientifically useful RMs below the $8\sigma$ $L/\sigma_L$ level that can be extracted with appropriate and careful weighting of the uncertainties and systematics.

Using the classification from \citetalias{Hale2021} (see their \S5.2.1 and Eqn. 1), we further subdivide our catalogue into unresolved and extended components. For our DR1 subset, we find the same fraction of components ($\sim40\%$) are unresolved in total intensity at \ang{;;25} as in the all-sky catalogue. For components detected in linear polarisation (both \texttt{goodL} and \texttt{goodRM}), we find $\sim20\%$ are unresolved. Looking at the fractional polarisation across our subsets, we find the median linear polarisation is around $\sim3\%$ for both \texttt{goodL} and \texttt{goodRM} and for unresolved and extended components. In this comparison we note, however, that our linear polarisation analysis uses pencil-beam spectral extraction compared to the source-finding used in \citetalias{Hale2021}.

\begin{table*}
\caption{Summary of polarization statistics for components in SPICE-RACS \citep[following the format of][]{Adebahr2022}. Each row corresponds to the subsets in \S\ref{sec:flags} (\texttt{goodI}, \texttt{goodL}, and \texttt{goodRM}) and \texttt{I} denotes all the components detected in Stokes $I$. We define columns as follows: `$N$' represents a number count of a given subset, `$F$' a fraction, and `$L/I$' the average linear polarisation fraction.  The subscripts $S$ and $E$ denote the subset of sources that are unresolved and resolved, respectively. The error ranges given represent the 16$^\text{th}$ and 84$^\text{th}$ percentiles of the population distribution.\label{tab:pol_param}}
\begin{tabular}{c|ccccccccc}
\toprule
Subset & $N$ & $F$ & $N_S$ & $N_E$ & $F_S$ & $F_E$ & $L/I$ & $L/I_S$ & $L/I_E$ \\
 &  & $\mathrm{\%}$ &  &  & $\mathrm{\%}$ & $\mathrm{\%}$ & $\mathrm{\%}$ & $\mathrm{\%}$ & $\mathrm{\%}$ \\
\midrule
All & 105912 & 100.0 & 41925 & 63987 & 39.6 & 60.4 & -- & -- & -- \\
$\texttt{goodI}$ & 24680 & 23.3 & 9195 & 15485 & 37.3 & 62.7 & -- & -- & -- \\
$\texttt{goodL}$ & 9092 & 8.58 & 2057 & 7035 & 22.6 & 77.4 & $3.0 \substack{+2.5\\-1.2}$ & $2.7 \substack{+1.9\\-1.1}$ & $3.1 \substack{+2.7\\-1.3}$ \\
$\texttt{goodRM}$ & 5818 & 5.49 & 1104 & 4714 & 19.0 & 81.0 & $3.4 \substack{+3.0\\-1.6}$ & $2.9 \substack{+2.4\\-1.2}$ & $3.5 \substack{+3.1\\-1.6}$ \\
\bottomrule
\end{tabular}
\end{table*}

\begin{figure*}
	\centering
	\begin{subfigure}[b]{\columnwidth}
	    \includegraphics[width=\textwidth]{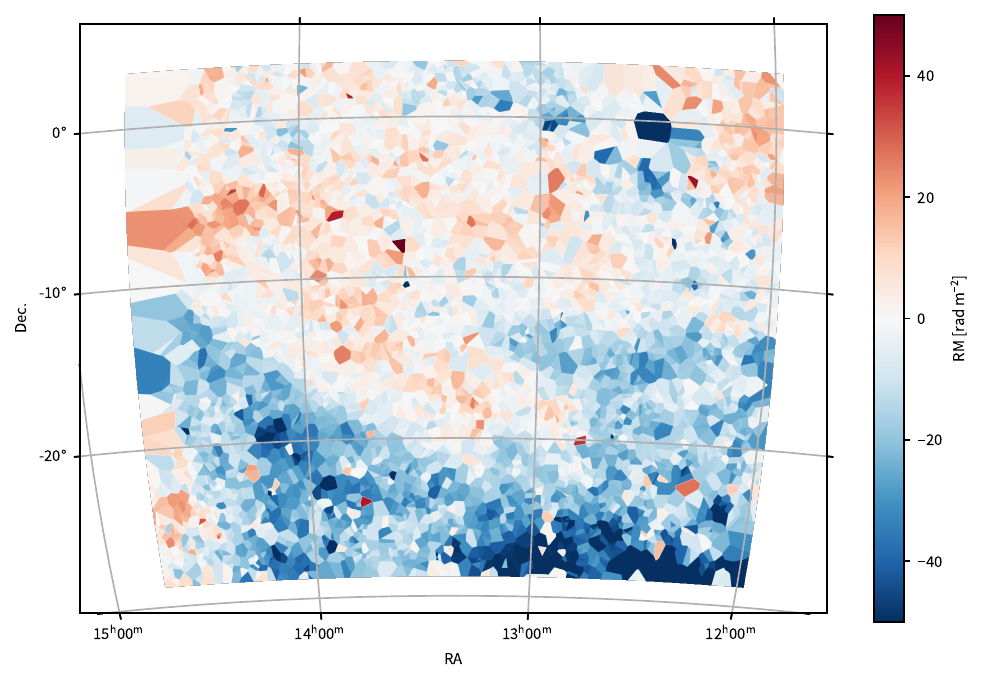}
	    \caption{SPICE-RACS (our work)}\label{fig:spice_rmgrid}
	\end{subfigure}

	\begin{subfigure}[b]{0.49\columnwidth}
		    \includegraphics[width=\textwidth]{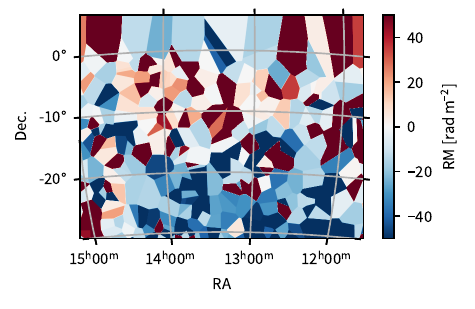}
	    \caption{\citetalias{Schnitzeler2019}~\citep{Schnitzeler2019}}\label{fig:spass_rmgrid}
	\end{subfigure}
	\begin{subfigure}[b]{0.49\columnwidth}
		    \includegraphics[width=\textwidth]{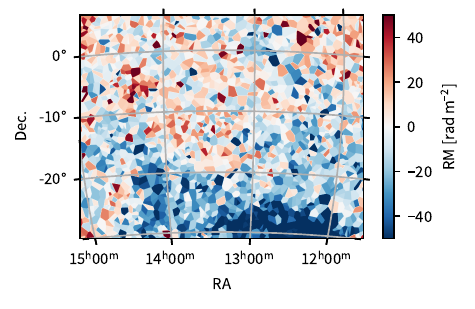}
	    \caption{\citetalias{Taylor2009}~\citep{Taylor2009}}\label{fig:nvss_rmgrid}
	\end{subfigure}
	\caption{Spatial distribution of rotation measures (RM) across the sky from (\subref{fig:spice_rmgrid}) SPICS-RACS-DR1 (\texttt{goodRM} subset, see \S\ref{sec:subsets}), (\subref{fig:spass_rmgrid}) \citetalias{Schnitzeler2019}, and (\subref{fig:nvss_rmgrid}) \citetalias{Taylor2009} using nearest-neighbour interpolation.}\label{fig:rmgrid}
\end{figure*}

\begin{figure}
	\centering
    \begin{subfigure}[b]{\columnwidth}
    	\includegraphics[width=\textwidth]{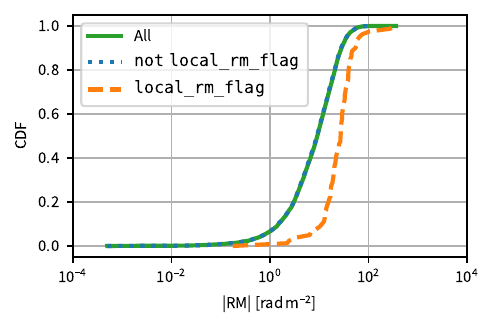}
        \caption{}\label{fig:abs_rm_cdf}
	\end{subfigure}

    \begin{subfigure}[b]{\columnwidth}
    	\includegraphics[width=\textwidth]{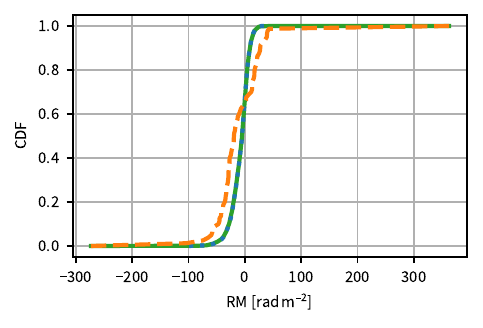}
         \caption{}\label{fig:rm_cdf}
	\end{subfigure}
	\caption{The cumulative distribution (CDF) of (\subref{fig:abs_rm_cdf}) absolute RM (log-scale) and (\subref{fig:rm_cdf}) RM with and without applying the \texttt{local\_rm\_flag}. Here we select RMs from the \texttt{goodRM} subset (see \S\ref{sec:subsets}).}\label{fig:rm_cdfs}
\end{figure}

We show the spatial distribution of RMs from the \texttt{goodRM} subset in Figure~\ref{fig:rmgrid}. As previously mentioned, we will provide a more detailed description and analysis of the structure present in these RMs in a forthcoming paper. We show the distribution of both absolute RM and RM in Figure~\ref{fig:rm_cdfs} and with and without the \verb|local_rm_flag|. Before applying the \verb|local_rm_flag|, the RMs span from \qty{\minRM(\minRMerr)}{\radian\per\metre\squared} to \qty{\maxRM(\maxRMerr)}{\radian\per\metre\squared}. The median absolute RM is about \qty{\medianabsRM}{\radian\per\metre\squared} with a standard deviation of \qty{\stdabsRM}{\radian\per\metre\squared}. Within this set there are only \nrmgthundred\ components with an absolute RM greater than \qty{100}{\radian\per\metre\squared}. Applying the \verb|local_rm_flag|, we find \nlocalrmflag\ components are identified as being outliers with respect to their local RM value. After excluding these values, the overall RM distribution remains mostly unchanged (as shown in Figure~\ref{fig:rm_cdfs}), with most of flagged RMs having slightly higher absolute RM ($>$\qty{10}{\radian\per\metre\squared}) than the total distribution. We leave investigation of the properties, and potential sources, of these outlying RMs to future work.

We now assess our RMs by crossmatching against historical polarisation catalogues. We crossmatch against the \citet{VanEck2023} consolidated catalogue (v1.1.0) using the STILTS~\citep{STILTS} \texttt{tmatch2} routine with a maximum sky separation of \ang{;;60} (the beam-width of \citetalias{Taylor2009}, which comprises the majority of the catalogue). From our total \ncomponents\ components, we find \ncross\ matches in the consolidated catalogue; \ncrossgood\ of these have reliable RMs in SPICE-RACS-DR1 with remaining \ncrossbad\ being flagged. In Table~\ref{tab:rm_crossmatch_bad} we show which flags were applied to the set of \ncrossbad\ flagged components and which original survey they were matched with. Across each of the matched surveys the majority of components were flagged for being below our fractional polarisation leakage threshold (\verb|leakage_flag|). After \verb|stokesI_fit_flag|, the most common flag is \verb|snr_flag|. In future releases, we plan to improve both our leakage threshold and our signal-to-noise near bright components which will help to reduce the number of true polarised components flagged in SPICE-RACS.

\begin{table*}
	\caption{The \ncross\ SPICE-RACS-DR1 components that crossmatch with historical surveys. For each survey we give the number and percentage (in parentheses) of components flagged by a given criterion (see \S\ref{sec:flags}). In the final column we give the counts of unflagged components.}\label{tab:rm_crossmatch_bad}
	\begin{tabular}{lcccccc}
\toprule
                        Survey &   snr\_flag & leakage\_flag & channel\_flag & stokesI\_fit\_flag &  Total flagged &  Total unflagged \\
\midrule
                          NVSS & 61 (15.6\%) &  361 (92.1\%) &     2 (0.5\%) &        18 (4.6\%) &            392 &             1021 \\
                   S-PASS/ATCA & 15 (31.2\%) &   38 (79.2\%) &     0 (0.0\%) &        8 (16.7\%) &             48 &               28 \\
          Farnes et al. (2014) &   0 (0.0\%) &  18 (100.0\%) &     0 (0.0\%) &         0 (0.0\%) &             18 &               18 \\
          Tabara et al. (1980) &   0 (0.0\%) &   8 (100.0\%) &     0 (0.0\%) &         0 (0.0\%) &              8 &                5 \\
                       POGS-II &  2 (66.7\%) &    2 (66.7\%) &     0 (0.0\%) &        1 (33.3\%) &              3 &                4 \\
             Kim et al. (2016) &   0 (0.0\%) &   2 (100.0\%) &     0 (0.0\%) &         0 (0.0\%) &              2 &                1 \\
                        POGS-I &   0 (0.0\%) &     0 (0.0\%) &     0 (0.0\%) &         0 (0.0\%) &              0 &                2 \\
Simard-Normandin et al. (1981) & 1 (100.0\%) &     0 (0.0\%) &     0 (0.0\%) &       1 (100.0\%) &              1 &                1 \\
          Battye et al. (2011) & 1 (100.0\%) &     0 (0.0\%) &     0 (0.0\%) &       1 (100.0\%) &              1 &                0 \\
                           All & 80 (16.9\%) &  429 (90.7\%) &     2 (0.4\%) &        29 (6.1\%) &            473 &             1080 \\
\bottomrule
\end{tabular}

\end{table*}

\begin{figure*}
	\centering
	\includegraphics[width=\textwidth]{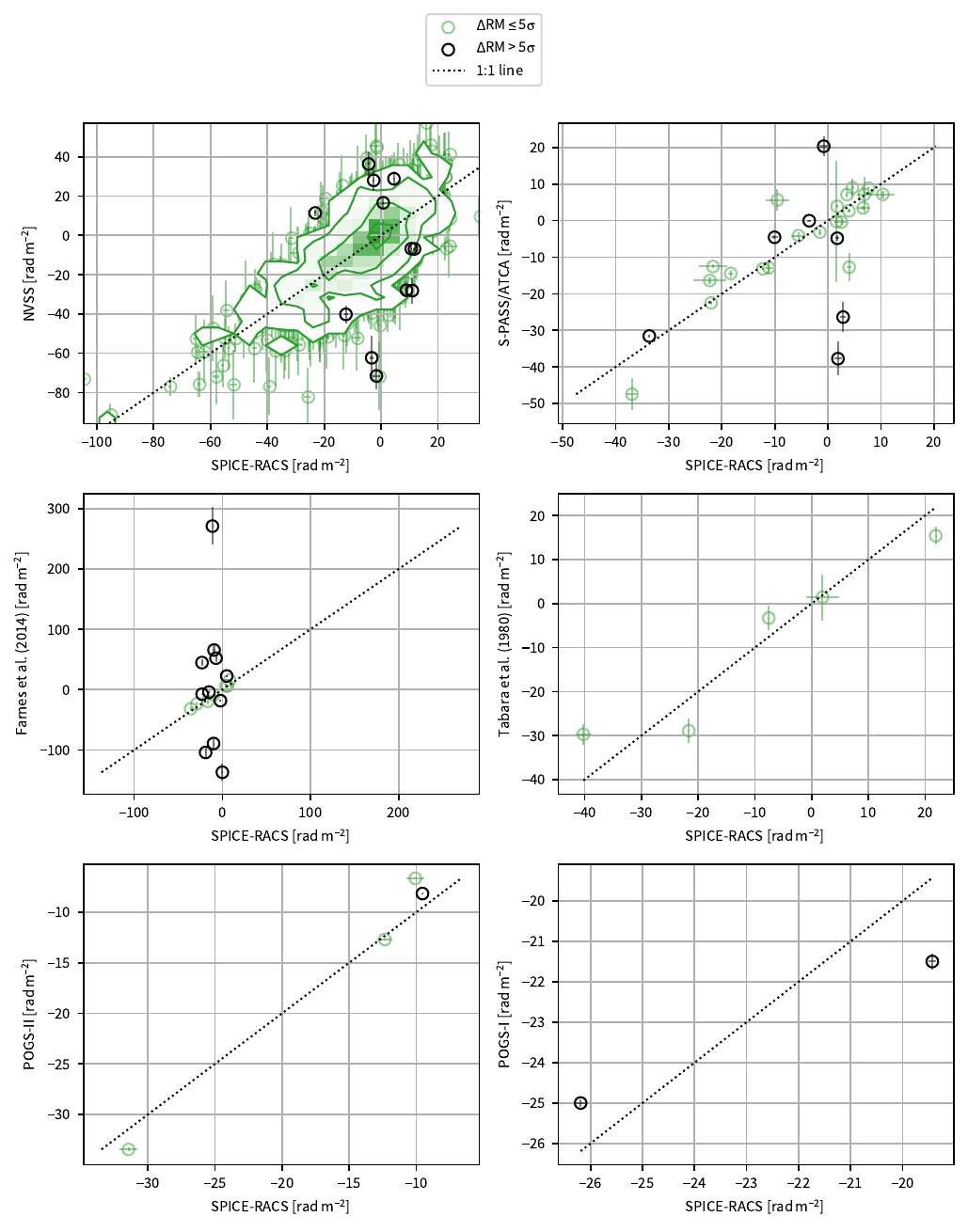}
	\caption{Rotation measures of SPICE-RACS-DR1 against other surveys. The SPICE-RACS RMs are drawn from the \texttt{goodRM} subset~(see \ref{sec:subsets}). We show points that have a RM difference ($\Delta\text{RM}$) of less than $5\sigma$ in green, and the remaining outlier points in black. Due to the high number of matched components, we show the inlying points from \citetalias{Taylor2009} as a density plot. The contour levels of the 2D histogram are at the $2^\text{nd}$, 16$^\text{th}$, 50$^\text{th}$, 84$^\text{th}$, and 98$^\text{th}$ percentiles. We note that the \citet{Farnes2014} RMs have a reference frequency of $\sim$\qty{5}{\giga\hertz}, and also showed similar scatter when compared to the \citetalias{Taylor2009} RMs.}\label{fig:cross}
\end{figure*}

Looking now to the crossmatched components with reliable SPICE-RACS RMs (the \texttt{goodRM} subset), we compare the RMs from all surveys with at least two matched components in Figure~\ref{fig:cross}. The included surveys are \citetalias{Taylor2009}, \citetalias{Schnitzeler2019}, \citet{Farnes2014}, \citet{Tabara1980}, POGS-II~\citep{Riseley2020}, and POGS-I~\citep{Riseley2018}. Out of these surveys, \citetalias{Taylor2009}, \citetalias{Schnitzeler2019}, and \citet{Farnes2014} each have $\geq10$ matched components. For each of these larger surveys, 90\% of the matched RMs are within 2.6, 9.2, and 13 standard deviations, respectively. We note that \citet{Farnes2014} derived their broadband results from high-frequency data; they estimate $\sim$\qty{5}{\giga\hertz} as the approximate reference frequency. \citet{Farnes2014} also find a similar scatter as we see in Figure~\ref{fig:cross} when they compare their RMs to \citetalias{Taylor2009} (see their Figure 6). We conclude that the large discrepancy between SPICE-RACS and \citet{Farnes2014} is due to the properties of the latter catalogue, and not spurious results from SPICE-RACS DR1.

\begin{figure*}
	\centering
	\begin{subfigure}[b]{0.49\columnwidth}
		\includegraphics[width=\textwidth]{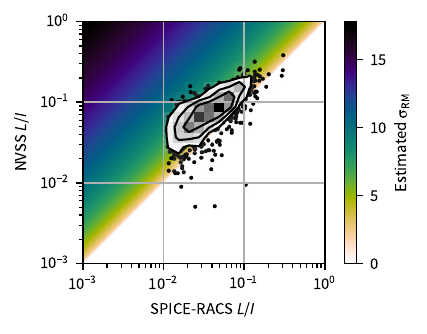}
		\caption{}\label{fig:cross_fracpol_dens}
	\end{subfigure}
	\begin{subfigure}[b]{0.49\columnwidth}
		\includegraphics[width=\textwidth]{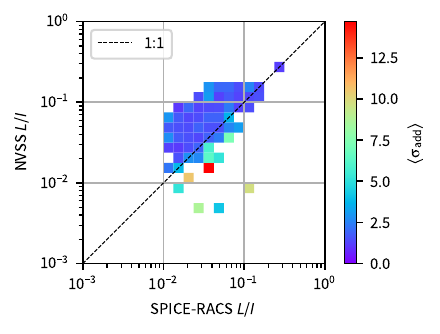}
		\caption{}\label{fig:cross_fracpol_sigma}
	\end{subfigure}
	\caption{The fractional linear polarisation ($L/I$) of SPICE-RACS components cross-matched with components from \citetalias{Taylor2009}. In (\subref{fig:cross_fracpol_dens}) we show the scatter points as a 2D density histogram where the data are over-dense. The contour levels of the 2D histogram are at the $2^\text{nd}$, 16$^\text{th}$, 50$^\text{th}$, 84$^\text{th}$, and 98$^\text{th}$ percentiles. Here the background colour scale is the estimated $\sigma_\text{RM}$ (see Eqn.~\ref{eqn:sigma_rm}) for a given fractional polarisation ratio. In (\subref{fig:cross_fracpol_sigma}) we bin collections of matched components in a 2D histogram and compute the median complexity metric $\left<\sigma_\text{add}\right>$ within each bin. }\label{fig:fractions}
\end{figure*}

In Figure~\ref{fig:fractions} we compare linear polarisation from \citetalias{Taylor2009} components with those that cross-match in our catalogue. Since \citetalias{Taylor2009} was derived from observations at $\sim$\qty{1.4}{\giga\hertz} we expect that most of our lower-frequency observations should be depolarised with respect to the higher frequency data~\citep{Sokoloff1998}. If we assume that components are depolarising due to a random, external dispersion in RM described by $\sigma_\text{RM}$ \citep[see e.g.][Eqns. 41 and 2, respectively]{Sokoloff1998, OSullivan2018}, then the fractional polarisation is given by
\begin{equation}
    \frac{L}{I}(\lambda) \propto e^{-2\sigma^2_\text{RM}\lambda^4}.
\end{equation}
We note that this $\sigma_\text{RM}$ is referring to a dispersion \emph{within} a given spectrum, and should not be confused with RM dispersion between sources~\citep[e.g.][]{Schnitzeler2010}. Given observations at two reference wavelengths ($\lambda_1$, $\lambda_2$), $\sigma_\text{RM}$ can be estimated as
\begin{equation}
    \sigma_\text{RM} = \sqrt{ \frac{\ln{ \left[\frac{(L/I)_1}{(L/I)_2}\right]}}{2\left(\lambda_2^4 - \lambda_1^4\right)}}.\label{eqn:sigma_rm}
\end{equation}
For our set of components which match with \citetalias{Taylor2009}, we find average polarisation fractions of \medfracspicexnvssspice\ and \medfracspicexnvssnvss\ for SPICE-RACS and \citetalias{Taylor2009}, respectively. Converting to $\sigma_\text{RM}$ using Equation~\ref{eqn:sigma_rm} we find an average estimated value of \medsigmarm\,\unit{\radian\per\metre\squared}. For comparison, \citet{OSullivan2018} reported a value of \qty{14.1(1.8)}{\radian\per\metre\squared} from $QU$-fitting spectra between \qtyrange[range-units = single]{1}{2}{\giga\hertz}. Given our lower frequency band, we expect to probe an overall sample of more Faraday simple sources than \citet{OSullivan2018}. We derive this expectation from the fact that complex polarised sources are known to depolarise as a function $\lambda^2$~\citep{Burn1966,Tribble1991,Sokoloff1998}. To first order, we can therefore expect to find a a higher average fractional polarisation at higher frequency and vice versa. Therefore, after taking into account both spectral index and observational sensitivity, sources detected at low frequencies are almost all Faraday simple~\citep[e.g.][]{OSullivan2023}. Conversely, higher frequency observations~\citep[e.g.][]{Anderson2015,OSullivan2018} are able to detect the population of Faraday complex sources.

There is also a small subset of components for which the SPICE-RACS fractional polarisation is greater than \citetalias{Taylor2009}. Inspecting the Faraday complexity metrics, we find that $\sigma_\text{add}$ is \sigmaaddcrossgt\ where SPICE-RACS has a higher fractional polarisation than \citetalias{Taylor2009}, compared to \sigmaaddcrosslt\ for the remainder. \citetalias{Taylor2009} suffers from bandwidth depolarisation for $|\text{RM}|>100$\,\unit{\radian\per\metre\squared}~\citep{Taylor2009,Ma2019}, however almost all of our detected RMs are below this threshold. Further, we find no correlation between absolute RM and complexity, nor these outlying components. We can therefore attribute the subset of sources with higher fractional polarisation to repolarisation effects~\citep[e.g.][]{Anderson2015}. Further investigation of depolarisation/repolarisation requires deeper analsysis (e.g. $QU$-fitting), which we leave for future work.


From the matched RMs, we identify outlying points as those beyond $5\sigma$ in the difference in RM ($\Delta\text{RM}$). We investigate two potential components that can explain the outlying points. First, since each of these surveys were conducted at different frequencies and with different bandwidths, components which are Faraday complex can produce different measured RMs in the different surveys. We provide a detailed assessment of our complexity metrics below. Secondly, as we found above, polarisation leakage remains a significant systematic effect in our observations. Crucially, if we have underestimated the amount of instrumental leakage in the position of a given component, our measured RM will be contaminated. In the case of leakage from Stokes $I$ into $Q$ and $U$, the smooth Stokes $I$ spectrum will impart a false RM signal at \qty{0}{\radian\per\metre\squared}.

\begin{figure*}
	\centering
	\begin{subfigure}[b]{0.49\columnwidth}
		\includegraphics[width=\textwidth]{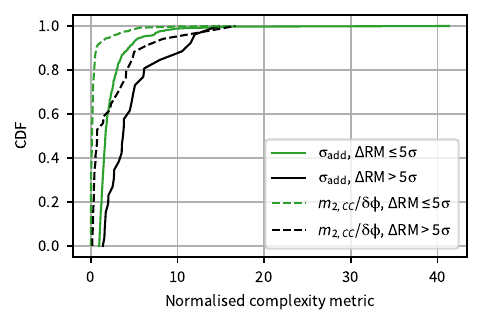}
		\caption{}\label{fig:complex_cdf}
	\end{subfigure}
	\begin{subfigure}[b]{0.49\columnwidth}
		\includegraphics[width=\textwidth]{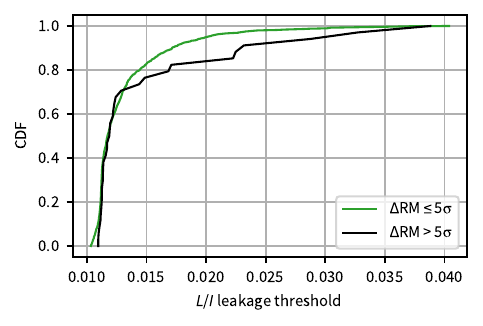}
		\caption{}\label{fig:leak_cdf}
	\end{subfigure}
	\caption{Distributions of inlying (green) and outlying (black) RMs (as shown in Figure~\ref{fig:cross}) as a function of (\subref{fig:complex_cdf}) normalised Faraday complexity and (\subref{fig:leak_cdf}) leakage threshold.}\label{fig:outliers}
\end{figure*}

\begin{figure}
	\centering
	\includegraphics[width=\textwidth]{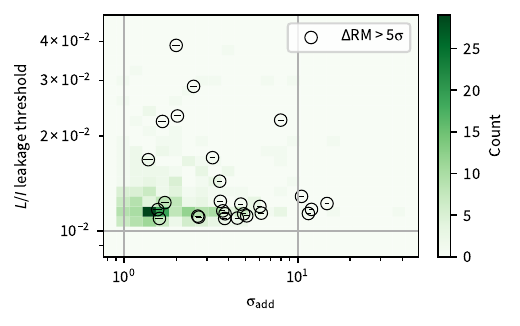}
	\caption{Scatter between our $\sigma_\text{add}$ complexity metric and leakage threshold for inlying (green) and outlying (black) points (as shown in Figure~\ref{fig:cross}). Due to the density of inlying points we show the data as a 2D density histogram.}\label{fig:cross_leak_sigma}
\end{figure}

In Figure~\ref{fig:outliers} we show the distributions of both our complexity metrics, $\sigma_\text{add}$ and $m_{2,\text{CC}}/\delta\phi$, and our leakage threshold for both inlying and outlying components. We find that outlying components are statistically more complex. For the inlying set $\sigma_\text{add}=\insigma$ and $m_{2,\text{CC}}/\delta\phi=\inmcc$, whereas for the outlying set $\sigma_\text{add}=\outsigma$ and $m_{2,\text{CC}}/\delta\phi=\outmcc$. Looking at our leakage threshold distributions, we find that the upper percentile portion of the outlier set tends to be higher than the inlier set; with the 97.6$^\text{th}$ percentile being \outleakhi\% and \inleakhi\%, respectively. Stated another way, if an outlying component is towards the edge of a field, with a higher leakage threshold, the outlier is more likely to be closer to the edge of the field with respect to an inlying component. Finally, we confirm that complexity and the leakage threshold are not correlated for inlying components, which we show in Figure~\ref{fig:cross_leak_sigma}. For outlying components there is not a general correlation, however almost all of the outliers that have a high leakage value are also classed as complex. We discuss potential sources of this in \S\ref{sec:complex_rev}.

\begin{figure*}
	\centering
	\begin{subfigure}[b]{0.49\columnwidth}
		\includegraphics[width=\textwidth]{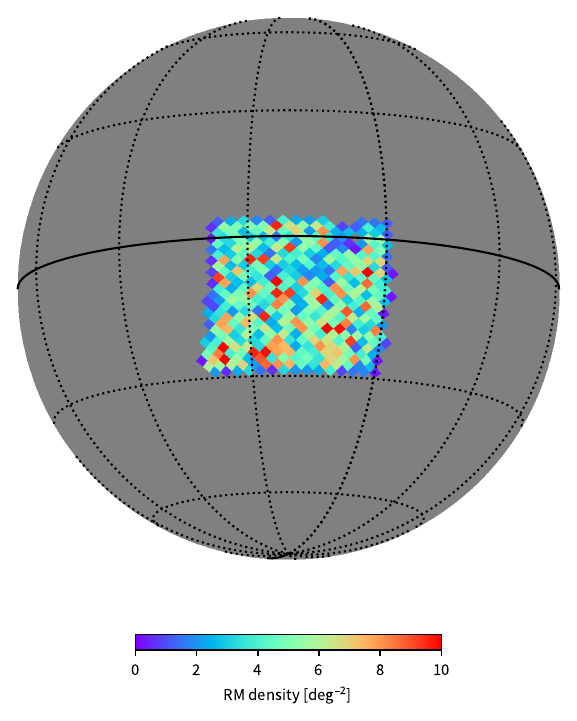}
		\caption{SPICE-RACS DR1 (our work)}\label{fig:density_spice}
	\end{subfigure}
	\begin{subfigure}[b]{0.49\columnwidth}
		\includegraphics[width=\textwidth]{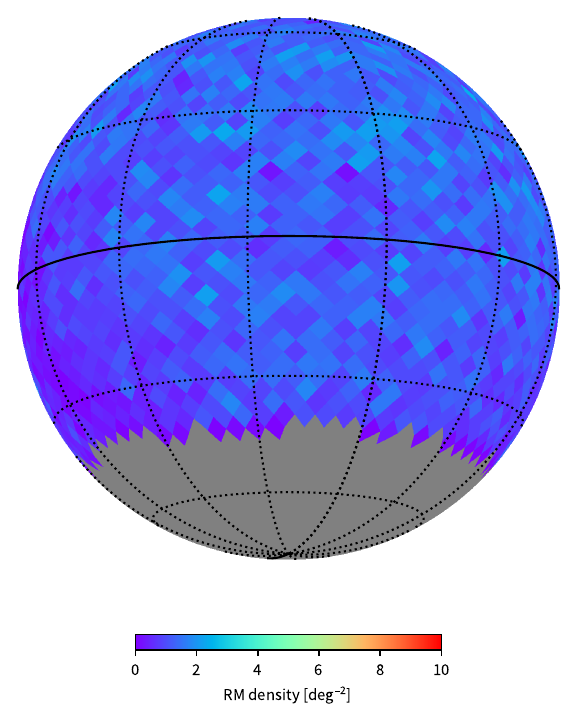}
		\caption{\citetalias{Taylor2009}~\citep{Taylor2009}}\label{fig:density_nvss}
	\end{subfigure}
	\begin{subfigure}[b]{0.49\columnwidth}
		\includegraphics[width=\textwidth]{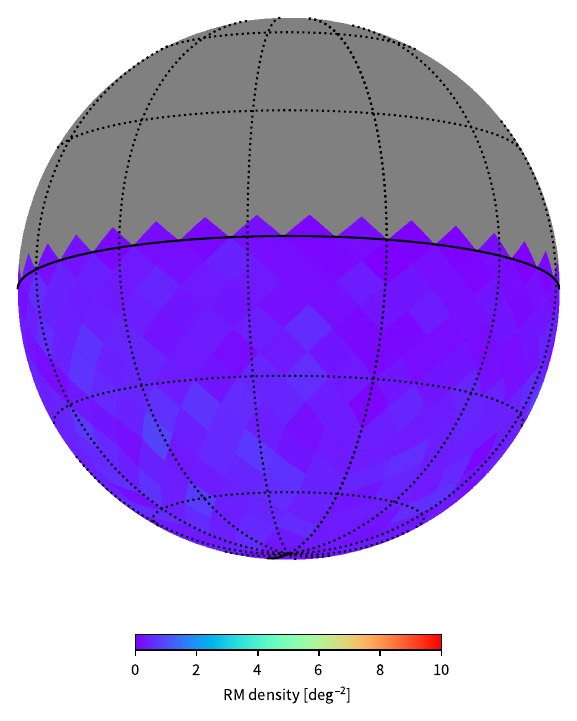}
		\caption{\citetalias{Schnitzeler2019}~\citep{Schnitzeler2019}}\label{fig:density_spass}
	\end{subfigure}
	\caption{RM density of SPICE-RACS compared to other large area surveys. Here we show the sky using an orthographic projection in equatorial coordinates, centred on the SPICE-RACS DR1 area. We use HEALPix grids with $N_\text{side}$ of 32, 16, and 8 for panels (\subref{fig:density_spice}), (\subref{fig:density_nvss}), (\subref{fig:density_spass}), respectively.}\label{fig:density}
\end{figure*}

Finally, we can assess our RM density and compare with historical surveys. We construct three coarse HEALPix\footnote{\url{http://healpix.sf.net}} grids with $N_\text{side}$ parameters 32, 16, and 8, corresponding to pixel resolutions of $\sim\ang{;110}$, $\sim\ang{;220}$, and $\sim\ang{;440}$, respectively. We then count the density of RMs from SPICE-RACS-DR1 (\texttt{goodRM} subset), \citetalias{Taylor2009}, and \citetalias{Schnitzeler2019} on the $N_\text{side}$ 32, 16, 8 grids, respectively. We show these density maps in Figure \ref{fig:density}. Our SPICE-RACS RM grid covers a total sky area of \qty{1306}{\degreesq}, with an RM density of $4.2\substack{+2.7\\-2.1}\,$\unit{\perdegreesq}. The lowest RM densities occur on the edge of our processed region, where our sensitivity decreases and leakage increases, as well as in the area affected by artefacts around 3C273. For comparison, we find RM densities of $1.1\substack{+0.4\\-0.4}\,$\unit{\perdegreesq} and $0.3\substack{+0.2\\-0.1}\,$\unit{\perdegreesq} for \citetalias{Taylor2009} and \citetalias{Schnitzeler2019}, respectively.

\subsection{Faraday Complexity}\label{sec:complex_rev}

As we describe in \S\ref{sec:complex} and \S\ref{sec:catalogue}, we provide two normalised complexity metrics, $\sigma_\text{add}$ and $m_{2,\text{CC}}/\delta\phi$. Having investigated the $\sigma_\text{add}$ parameter, and its reported error $\delta\sigma_\text{add}$, we find that the metric is susceptible to two failure cases; each applying when $\sigma_\text{add}/\delta\sigma_\text{add}$ is small. First, if a single channel (or small subset of channels) are affected by a noise or RFI spike, the resulting distribution (which is computed internally within \textsc{RM-Tools}) of $\sigma_{\text{add},Q}$ or $\sigma_{\text{add},U}$ is heavily skewed to larger values. This causes an over-reported value in the $\delta\sigma_\text{add}$. Further, since we compute $\sigma_\text{add}$ as the quadrature sum of $\sigma_{\text{add},Q}$ and $\sigma_{\text{add},U}$ (Equation~\ref{eqn:sigma_add}), themselves each being positive definite values, we will have a Ricean-like bias at low signal-to-noise. We therefore only consider $\sigma_\text{add}$ a reliable metric where $\sigma_\text{add}/\delta\sigma_\text{add}>10$ (where \texttt{complex\_sigma\_add\_flag} is \verb|True|). Where we do classify $\sigma_\text{add}$ as reliable, we indeed find that increasing values of $\sigma_\text{add}$ correspond to greater complexity in the spectra (see~\ref{app:complex_spectra}).

\begin{figure}
	\centering
	\includegraphics[width=\textwidth]{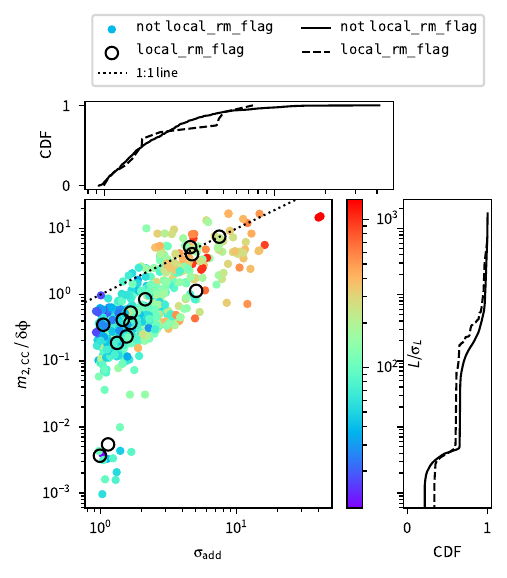}
	\caption{A comparison of the normalised Faraday complexity metrics described in \S\ref{sec:complex}. Here we only show points for which $\sigma_\text{add}/\delta\sigma_\text{add}>10$, which is why the $m_{2,CC}$ CDF does not reach 0. The scatter plot shows the correlation between these metrics, with points coloured by the components' polarised signal-to-noise ratio ($L/\sigma_L$). Black points indicate components that have been flagged as having RMs which are outlying from the local ensemble of SPICE-RACS RMs. For each metric, we also show the distribution of the metric values for both inlying and outlying components. We note that the $m_{2,CC}$ values (vertical axis) appear quantised due to the discretisation and numerical precision of the Faraday depth axis and the placement of \textsc{clean} components.}\label{fig:complexity}
\end{figure}

In Figure~\ref{fig:complexity} we show the distributions and correlation between these two metrics for all SPICE-RACS components with reliable RMs. Of the \nrms\ components in the \texttt{goodRM} subset, we find \ncomplex\ components that are classified as complex. We see that the $\sigma_\text{add}$ metric is more sensitive than $m_{2,\text{CC}}/\delta\phi$, with $\sigma_\text{add}$ flagging \ncomplexsigma\ components as complex and $m_{2,\text{CC}}/\delta\phi$ flagging \ncomplexmcc. Further, $m_{2,\text{CC}}/\delta\phi$ only flags \ncomplexmccnotsigma\ components that were not flagged by $\sigma_\text{add}$; whereas $\sigma_\text{add}$ flags \ncomplexsigmanotmcc\ components that were not flagged by $m_{2,\text{CC}}/\delta\phi$.

For components that are classed as complex, we see a good correlation between the two metrics. We check whether the components that were flagged as complex were also flagged as outlying from their local ensemble RMs, but we find no significant deviation from the distribution of inlying points. We do, however, find that the complexity correlates with polarised signal-to-noise ($L/\sigma_L$). To better understand this correlation we directly inspect the spectra themselves.

\begin{figure*}
	\centering
	\includegraphics[width=\textwidth]{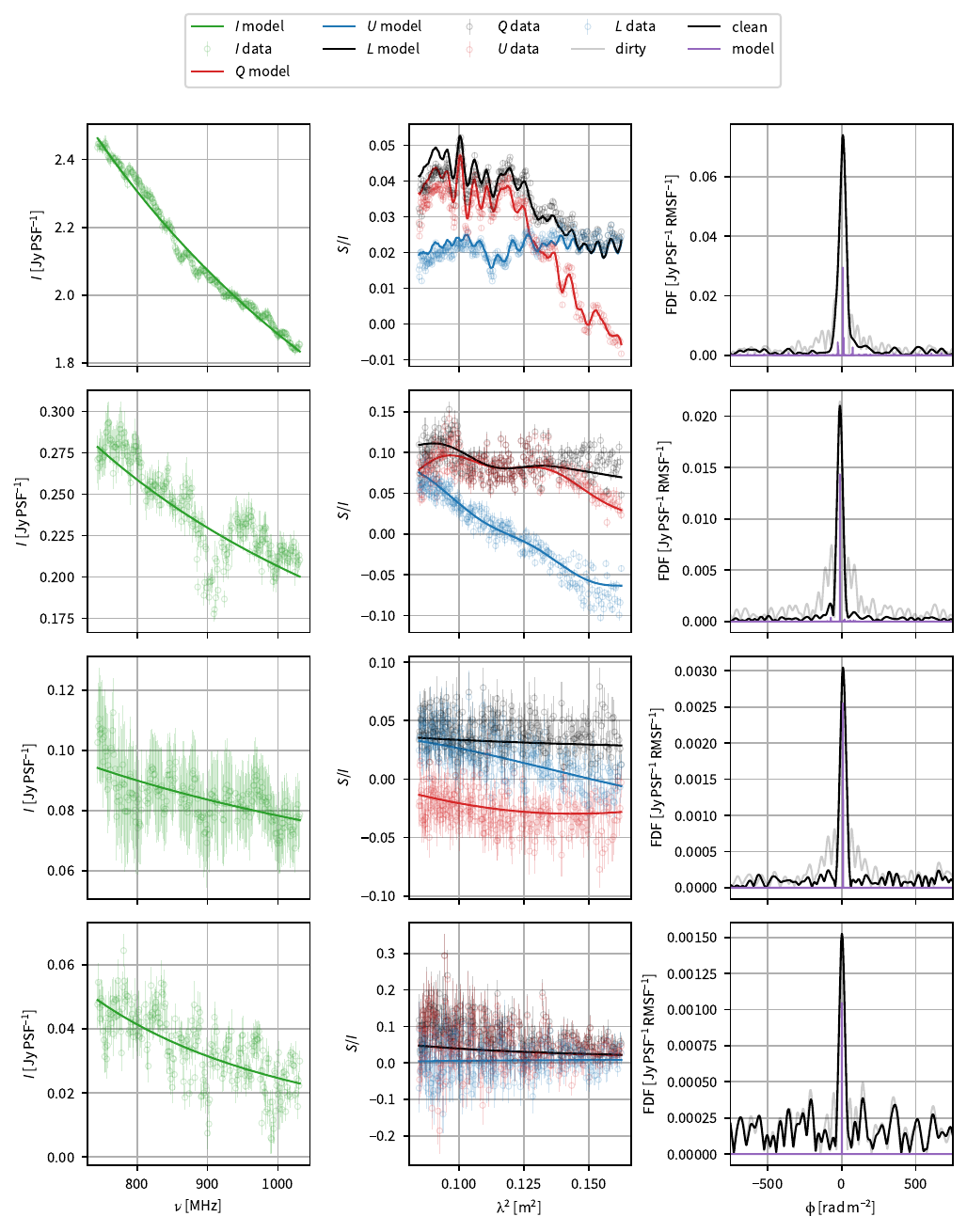}
	\caption{Example spectra of four sources from SPICE-RACS nearest to the 99.9$^\text{th}$, 97.7$^\text{th}$, 84$^\text{th}$, and 50$^\text{th}$ percentile in polarised signal-to-noise from top to bottom. In the left column we show the Stokes $I$ data and our fitted power-law model. In the middle column we show the fractional Stokes $Q$ and $U$, and polarised intensity $L$. Here the scatter points are the observed data divided by the Stokes $I$ model, and the solid lines is the \textsc{RM-clean} model transformed to $\lambda^2$ space. In the right-hand column we show the dirty, clean, and model Faraday dispersion functions (FDFs) in polarised intensity.}\label{fig:spectra}
\end{figure*}

In Figure~\ref{fig:spectra} we show the Stokes $I$ spectra as a function of frequency, the $L$, $Q$, $U$ spectra as a function of $\lambda^2$, and the clean, dirty, and model Faraday spectra. Here we show four representative spectra nearest to the 50$^\text{th}$, 84$^\text{th}$, 97.7$^\text{th}$, and 99.9$^\text{th}$ percentile in $L/\sigma_L$. Looking at the brightest Stokes $I$ spectrum we can see both a power-law structure as well as a sinusoidal ripple across the band.

ASKAP has a known ripple across its band with a period of $\sim$\qty{25}{\mega\hertz}, which has been attributed to a standing wave effect between the dish surface and the receiver~\citep{Sault2015}. We will leave a deep investigation of characterising and mitigating this effect to future work. We do note however, that the period of the ripple may vary in concert with other observational parameters such as elevation. Further, the ripple in our data is seen after the application of the bandpass calibration. Therefore, residual errors from the bandpass are also superimposed. For the purposes of this investigation we will only consider a constant \qty{25}{\mega\hertz} ripple.

We consider the two ways a \qty{25}{\mega\hertz} ripple may present itself in our observations. As described in \citet{Sault2015}, the standing wave ripple presents itself in the instrumental gains. For a given component which emits in all Stokes parameters, this ripple can present itself in the data \emph{multiplicatively} with the true emission from the sky. Secondly, in case of a bright Stokes $I$ component, leakage of the ripple in Stokes $I$ can be introduced \emph{additively} to the other Stokes parameters.

\begin{figure}
	\centering
	\begin{subfigure}[b]{\columnwidth}
	\includegraphics[width=\textwidth]{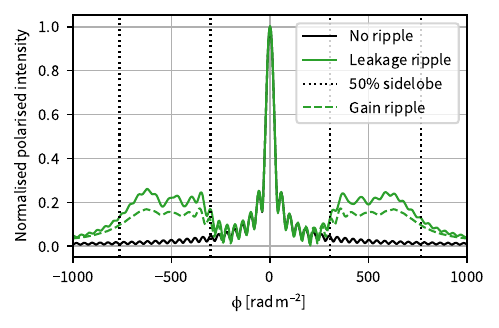}
	\caption{}\label{fig:ripple_rmsf}
	\end{subfigure}
	\begin{subfigure}[b]{\columnwidth}
	\includegraphics[width=\textwidth]{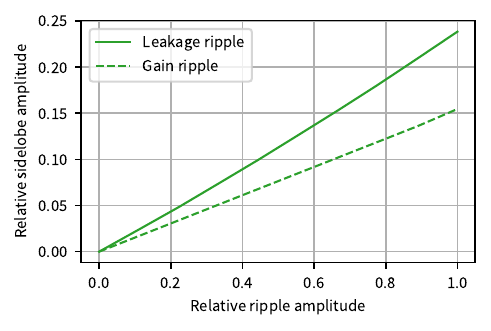}
	\caption{}\label{fig:ripple_sidelobes}
	\end{subfigure}
	\caption{The effect of the \qty{25}{\mega\hertz} standing-wave ripple on an ideal Faraday depth spectrum. (\subref{fig:ripple_rmsf}) The Faraday spectrum for an ideal, noiseless component (black) in the presence of a leakage ripple (solid green) and a gain ripple (dashed green). (\subref{fig:ripple_sidelobes}) The maximum peak of the leakage ripple in the Faraday spectrum relative to a unitary polarised signal as a function of the ripple amplitude.}\label{fig:ripple}
\end{figure}

We will now consider how the two cases that can introduce the standing wave ripple into Stokes $Q$ and $U$ may affect our Faraday spectra. First, it is important to consider that the ripple is periodic in frequency, and will therefore distribute across a broad range of Faraday depths. In Figure~\ref{fig:ripple} we show the noiseless Faraday spectrum, with $\text{RM}=0$, in the presence of both a gain ripple and leakage ripple. In each case we have generated a unit polarised signal with the amplitude of the ripple also being unity. In the case of a gain ripple, the RMSF is effectively given higher sidelobes. These sidelobes translate in Faraday depth with the true Faraday depth of the component. In the case of leakage, the response has no dependence on the Faraday depth of the source; the true spectrum is superposed with the response from the ripple. For both the leakage ripple and $\text{RM}=0$ gain ripple case, we find the leakage response spans \qtyrange[range-units = single]{302}{764}{\radian\per\metre\squared} in $|\phi|$ at the 50\% level. Lastly, we find that the leakage has a marginally higher amplitude in the Faraday spectrum than a gain ripple. For a ripple with unit amplitude, the maximum amplitude of the Faraday spectrum response is 24\% and 15\% for the leakage and gain ripples, respectively. These responses decrease linearly with the ripple amplitude. In conclusion, we urge caution in the interpretation of high signal-to-noise spectra. Whilst they are not included in our catalogue, care should be taken in analysis of secondary components in our \textsc{clean} Faraday spectra. In particular, secondary RM components found in the range $300<|\phi|<800$\,\unit{\radian\per\metre\squared} should be treated with a reasonable level of scepticism.

\subsection{Spectral Indices}
Using our fitted models to the Stokes $I$ spectra we are able to assess the distribution of spectral indices. Here we are interested in the spectra of all components, regardless of their polarisation. We identify components with a reliable fit in our catalogue by indexing where both the \verb|stokesI_fit_flag| and \verb|channel_flag| are false (see \S\ref{sec:flags}). Of the \ncomponents\ components we analyse, we successfully fit \ngoodI\ spectra in Stokes $I$. Since we image each \qty{1}{MHz} independently, this number is in line with our expectations. If we assume that the $rms$ noise increases in proportion to the square root of the number of channels, a $5\sigma$ detection using the full band is only $0.3\sigma$ in a single channel; with only $\sim10\%$ of our sample of components having $5\sigma$ per channel. Further, we find that \verb|goodI| subset is indeed drawn from the high SNR Stokes $I$ components.

Of the \verb|goodI| subset, \nalpha\ components have a reliable fit to the spectral index, with the remaining \nflat\ spectra only being fit with a flat intensity model. We make three notes of caution in interpreting our spectral indices. First, they are derived from, at most, \qty{288}{\mega\hertz} of bandwidth and are therefore not as robust as broad-band spectral indices. Secondly, our applied holographic primary beams may differ slightly from the true primary beams. If the error in our holographic beams is chromatic, this will also affect our spectral indices. Finally, since we both apply a constant \qty{100}{m} $uv$-cut to the visibilities and extract spectra from the peak pixel, our spectral indices may be unreliable for extended sources.

\begin{figure}
    \centering
    \includegraphics{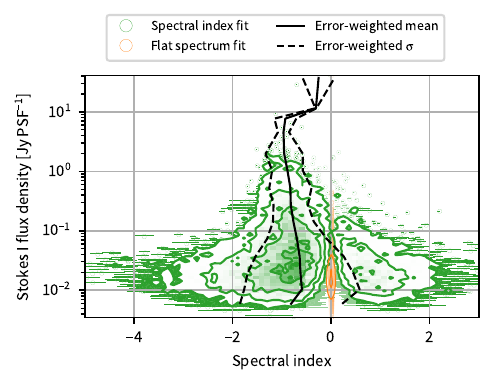}
    \caption{The spectral index of Stokes $I$ spectra against flux density from our fitted spectra (taken from the peak pixel). In green we show the measurements from each component, including errors in flux density and spectral index. In orange we show the components which have a flat spectral index fit; for all these components we take the spectral index to be $\alpha=0$. Where the points are over-dense, we show the distribution as a density plot. The contour levels of the 2D histogram are at the $2^\text{nd}$, 16$^\text{th}$, 50$^\text{th}$, 84$^\text{th}$, and 98$^\text{th}$ percentiles. In black we show the error-weighted mean spectral index in flux density bins, along with the error-weighted standard deviation.}
    \label{fig:spectral_index}
\end{figure}

We find an error-weighted average spectral index of \avgalpha (here the errors represent the error-weighted ensemble standard deviation). We show the distribution of our fitted spectral indices, as a function of flux density, in Figure~\ref{fig:spectral_index}. Above about \qty{1}{\jansky\per PSF} we have relatively few components, making bulk statistics unreliable. Below \qty{1}{\jansky\per PSF}, however, we see that most bright components are in line with the bulk average index of $\sim-0.8$, but tend to flatten with decreasing flux density to $-0.4\pm0.9$. Our bulk spectral indices, and trend to flatten with decreasing flux density, are consistent with previous spectral index measurements at $\sim$\qty{1}{\giga\hertz} \citep[e.g.][]{Gasperin2018}; giving us confidence in our in-band results.

\section{DATA ACCESS}\label{sec:access}

All of our data final products can be accessed through the CSIRO ASKAP Science Data Archive \citep[CASDA;][]{casda,Huynh2020}. These data products are held together in a CASDA `collection' at \url{https://data.csiro.au/collection/csiro:58508}. Due to the large number and overall size of our data products we have split our data products into two sets. The first set contains the data from our \texttt{goodI} subset (see \S\ref{sec:subsets}), referring to \ngoodI\ components. The second is a superset of the first containing information on all \ncomponents\ components we analyse here. The same atomic data products and formats are available in each set, however accessibility differs between the two. The second, larger set is provided as a simple CSIRO Data Access Portal (DAP) deposit at \url{https://data.csiro.au/collection/csiro:58409}. Within this deposit the data products have been consolidated for bulk storage and retrieval. The first, smaller set is provided as a CASDA collection; providing greater granularity and searchability of the data products. Namely, these data can be retrieved through CASDA's `Observation Search\footnote{\url{https://data.csiro.au/domain/casdaObservation}}' and `Skymap Search\footnote{\url{https://data.csiro.au/domain/casdaSkymap}}'.

Our final data products are as follows:
\begin{itemize}
    \item \verb|spice-racs.dr1.corrected.{cut,full}.xml| \\
    Our component catalogue, as described in \S\ref{sec:catalogue}, in VOTable \citep[v1.3,][]{VOTable13} format for the first (\verb|cut|) and second (\verb|full|) sets of components described above.
    \item \path{RACS_DR1_Sources_spice-racs.dr1.corrected.{cut,full}.xml} \\
    A subset of the `Source' catalogue described in \citetalias{Hale2021} to match the fields in the SPICE-RACS DR1 subset. This catalogue corresponds to the cutout cubelets we present.
    \item \verb|{source_id}.cutout.{image,weights}.*.fits| \\
    Image and weights cube cutout in Stokes $I$, $Q$, and $U$, as described in \S\ref{sec:cutout}, cutout around a RACS-low source identified by `\verb|source_id|'. The data dimensions are J2000 right ascension and declination, Stokes parameter, and frequency.
    \item \verb|{cat_id}_polspec.fits| \\
    Polarisation spectra extracted on the position of the RACS-low component identified by `\verb|cat_id|', as described in \S\ref{sec:spectra}, in \textsc{PolSpectra}\footnote{\url{https://github.com/CIRADA-Tools/PolSpectra}}~\citep{VanEck2023} format. Each table contains a single row, corresponding to the component. We enumerate the columns below.
    \item \verb|spice_racs_dr1_polspec_{cut,full}.tar| \\
    A tarball containing all of the \textsc{PolSpectra} for the first (\verb|cut|) and second (\verb|full|) sets of components described above.
\end{itemize}

\subsection{PolSpectra Columns}

The \textsc{PolSpectra} standard defines the following mandatory columns:
\begin{itemize}
    \item \verb|source_number| \\
    Simple source identifier (incrementing integer number).
    \item \verb|ra| \\
    As per \S\ref{sec:catalogue}.
    \item \verb|dec| \\
    As per \S\ref{sec:catalogue}.
    \item \verb|l| \\
    As per \S\ref{sec:catalogue}.
    \item \verb|b| \\
    As per \S\ref{sec:catalogue}.
    \item \verb|freq| \\
    Frequency array in \unit{\hertz}.
    \item \verb|stokesI| \\
    Stokes $I$ flux density array in \unit{\jansky\per PSF}.
    \item \verb|stokesI_error| \\
    Errors on Stokes $I$ flux density array in \unit{\jansky\per PSF}.
    \item \verb|stokesQ| \\
    Stokes $Q$ flux density array in \unit{\jansky\per PSF}.
    \item \verb|stokesQ_error| \\
    Errors on Stokes $Q$ flux density array in \unit{\jansky\per PSF}.
    \item \verb|stokesU| \\
    Stokes $Q$ flux density array in \unit{\jansky\per PSF}.
    \item \verb|stokesU_error| \\
    Errors on Stokes $U$ flux density array in \unit{\jansky\per PSF}.
    \item \verb|beam_maj| \\
    As per \S\ref{sec:catalogue}.
    \item \verb|beam_min| \\
    As per \S\ref{sec:catalogue}.
    \item \verb|beam_pa| \\
    As per \S\ref{sec:catalogue}.
    \item \verb|Nchan| \\
    As per \S\ref{sec:catalogue}.
\end{itemize}

In addition, we also provide the following columns:
\begin{itemize}
    \item \verb|cat_id| \\
    As per \S\ref{sec:catalogue}.
    \item \verb|telescope| \\
    As per \S\ref{sec:catalogue}.
    \item \verb|epoch| \\
    As per \S\ref{sec:catalogue}.
    \item \verb|integration_time| \\
    As per \S\ref{sec:catalogue}.
    \item \verb|leakage| \\
    As per \S\ref{sec:catalogue}.
    \item \verb|channel_width| \\
    As per \verb|channelwidth| in \S\ref{sec:catalogue}.
    \item \verb|flux_type| \\
    As per \S\ref{sec:catalogue}.
    \item \verb|faraday_depth| \\
    The Faraday depth array in \unit{\radian\per\metre\squared}.
    \item \verb|faraday_depth_long| \\
    Double-length Faraday depth array (in \unit{\radian\per\metre\squared}), matching the RMSF for use in \textsc{RM-clean}
    \item \verb|FDF_Q_dirty| \\
    The real part of the dirty Faraday spectrum array in \\\unit{\jansky\per PSF \per RMSF}.
    \item \verb|FDF_U_dirty| \\
    The imaginary part of the dirty Faraday spectrum array in \unit{\jansky\per PSF \per RMSF}.
    \item \verb|FDF_Q_clean| \\
    The real part of the \textsc{clean} Faraday spectrum array in \unit{\jansky\per PSF \per RMSF}.
    \item \verb|FDF_U_clean| \\
    The imaginary part of the \textsc{clean} Faraday spectrum array in \unit{\jansky\per PSF \per RMSF}.
    \item \verb|FDF_Q_model| \\
    The real part of the \textsc{clean} model of the Faraday spectrum array in \unit{\jansky\per PSF \per RMSF}.
    \item \verb|FDF_U_model| \\
    The imaginary part of the \textsc{clean} model of the Faraday spectrum array in \unit{\jansky\per PSF \per RMSF}.
    \item \verb|RMSF_Q| \\
    The real part of the RMSF array in normalised units.
    \item \verb|RMSF_U| \\
    The imaginary part of the RMSF array in normalised units.
\end{itemize}

\section{FUTURE AND OUTLOOK}\label{sec:discussion}
The data we present and publicly release here is the first of several we plan to make as part of the SPICE-RACS project. Here we have only processed 30 RACS-low fields; which represents $\sim3\%$ of the overall survey. The ASKAP observatory has now conducted a further three RACS epochs. These surveys have now covered the mid (band 2, \qtyrange[range-units = single]{1296}{1440}{\mega\hertz}) and high (band 3, \qtyrange[range-units = single]{1524}{1812}{\mega\hertz}) frequency ASKAP bands, and another epoch of the low (band 1, \qtyrange[range-units = single]{744}{1032}{\mega\hertz}) band. As well as providing much wider combined bandwidth, there have been a number of improvements to scheduling and operations, as well as online and offline processing. These are detailed in the RACS-mid paper~\citep[Paper~IV][]{Duchesne2023}. Notably, the second epoch of RACS-low (RACS-low2) has been observed with a higher Northern declination limit of $\delta\sim$\ang{50}.

For future releases of SPICE-RACS, we will initially focus on producing an all-sky polarisation catalogue using RACS-low2. These data provide both the largest $\lambda^2$ and areal coverage, with the best data quality. For our purposes, RACS-low2 supersedes RACS-low. There may be interest in the processing of RACS-low in addition to RACS-low2 for variability studies, but we will leave this work to a later time. Extrapolating from the RM density we find here, we can expect to catalogue $>10^5$ RMs across the Southern sky. This will be the largest RM catalogue produced to date, with an increase in RM areal density of at least 5 and 25 times over \citetalias{Taylor2009} and \citetalias{Schnitzeler2019}, respectively. This catalogue will only be superseded after a substantial fraction ($\sim20\%$) of the POSSUM survey is completed.

In producing an all-sky polarisation catalogue there are a number of improvements we hope to make over our first data release. Off-axis leakage remains a key limiting systematic in our catalogue presented here. The ASKAP observatory now regularly performs holographic observations following each primary beam-forming observation. As such, we now have an independent characterisation of the primary beams, including widefield leakage, for a portion of RACS-mid, and the entirety of RACS-high and RACS-low2. We also expect the sensitivity and resolution of RACS-low2 to be improved over RACS-low. These are primarily driven by improved scheduling, with RACS observations being scheduled within \qty{1}{\hour} of the meridian, and additional processing steps such as peeling of bright sources outside the field of view. Here we have also explored the impact of a spectral ripple on SPICE-RACS data. Together with the observatory, we will continue to characterise the impact of this ripple and work to nullify its effects in future catalogues.

Following the production of an all-sky band 1 polarisation catalogue, we will work to fold in the RACS-mid and high observations. Along with the sensitivity provided by increased bandwidth, including these data will improve our ability to detect and classify Faraday complex components. To first order, a combined low and mid-SPICE-RACS would have a Faraday depth resolution, maximum Faraday depth, and maximum Faraday depth scale of $\delta\phi\approx$ \qty{30}{\radian\per\metre\squared} or $\mathcal{W_\text{max}}\approx$ \qty{12}{\radian\per\metre\squared}, $\phi_\text{max} \approx$\qty{6700}{\radian\per\metre\squared}, and $\phi_{\text{max-scale}}\approx$ \qty{72}{\radian\per\metre\squared}, respectively (see Equations~\ref{eqn:delta_phi}, \ref{eqn:phi_max}, and \ref{eqn:phi_max_scale}). Similarly, a full low, mid, and high-SPICE-RACS would have $\delta\phi\approx$ \qty{26}{\radian\per\metre\squared}, $\phi_\text{max}\approx$ \qty{1e4}{\radian\per\metre\squared}, and $\phi_{\text{max-scale}}\allowbreak\approx$  \qty{113}{\radian\per\metre\squared}. By using RACS-low2 in this combination, we can produce a survey covering the entire Southern sky up to a declination of $\sim+\ang{50}$, with an expected $rms$ noise $\sim$\qty{40}{\micro\jansky\per PSF}. Care will need to be taken to handle differences in $uv$-coverage, as well as the performance of \textsc{RM-clean} with large frequency gaps. 

If the technical challenges of combining these data can be overcome, the benefits will be significant. The wide areal coverage of SPICE-RACS will provide the opportunity for ultra-broadband studies though with both low (e.g. LoTSS, POGS) and high frequency (e.g. Apertif, VLASS) radio polarisation surveys. SPICE-RACS is also the ideal pilot survey for followup observations with MeerKAT or the Australia Telescope Compact Array (c/o QUOCKA Heald et al. in prep.). Whilst POSSUM will provide much deeper observations over a large area of the Southern sky, the bandwidth provided by a combined RACS-low-mid-high survey will be unmatched across large areas until the era of the SKA.  

In a forthcoming paper, we will use the SPICE-RACS-DR1 catalogue to derive an RM grid behind the nearby \ion{H}{II} region of the Spica Nebula. Using these data, we will explore the magneto-ionic properties of the Galactic ISM towards this region.

\section{CONCLUSION}\label{sec:conclusions}
Here we have described the first data release (DR1) of Spectra and Polarisation in Cutouts of Extragalactic components from RACS (SPICE-RACS); the project to produce linear polarisation results from the Rapid ASKAP Continuum Survey~\citep[RACS][]{McConnell2020}.

We have processed 30 fields from the first epoch of RACS-low in Stokes $I$, $Q$, and $U$. These data cover \qtyrange[range-units = single]{744}{1032}{\mega\hertz} at \qty{1}{\mega\hertz} spectral resolution. Using the total intensity catalogue of \citet{Hale2021}, we have produced cutout images and spectra of \ncomponents\ radio components over an area of $\sim$\qty{1300}{\degreesq}. The angular resolution of these images is \ang{;;25}, and we measure an \textit{rms} noise of \qty{80}{\micro\jansky\per PSF} in Stokes $Q$ and $U$ across the full bandwidth. We have corrected our images for ionospheric Faraday rotation, primary beam attenuation, and on- and off-axis leakage.

From these spectra, we have produced a spectro-polarimetric catalogue of all \ncomponents\ radio components in the RM-Table~\citep{VanEck2023} standard. \nrms\ of these components have a reliable derived rotation measure (RM); corresponding to an areal density of $\sim$\qty{4}{\perdegreesq}. After cross-matching with the \citetalias{Taylor2009} catalogue, we find 90\% of the matched components are within $2.6\sigma$ in RM. Further, we provide metrics and classification of Faraday complexity, and find that \ncomplex\ components exhibit detectable Faraday complexity. We urge caution in their interpretation, however, in the face of the uncorrected standing wave ripple in ASKAP data. Finally, we have derived in-band spectral indices in determining our RMs. We find that \nalpha\ components have a reliable fitted spectral index, with an average value of $-0.8\pm0.4$ and trend to flatter spectral indices with lower flux density. Of the components with a fitted spectral index, $\sim60\%$ are within $1\sigma$ of the average $-0.8$ value.

The work we present here lays the foundation for the all-Southern-sky RM catalogue we will produce from RACS. We make our images, spectra, and catalogue publicly available through the ASKAP Science Data Archive \citep[CASDA;][]{casda,Huynh2020}\footnote{\url{https://data.csiro.au/collection/csiro:58508}}. Our processing pipeline repository is also made available, and open-source, on GitHub\footnote{\url{https://github.com/AlecThomson/arrakis} v1.0.0.}.

\begin{acknowledgement}
We thank the anonymous referee for their careful and insightful review of this work.

This scientific work uses data obtained from Inyarrimanha Ilgari Bundara / the Murchison Radio-astronomy Observatory. We acknowledge the Wajarri Yamaji People as the Traditional Owners and native title holders of the Observatory site. CSIRO’s ASKAP radio telescope is part of the Australia Telescope National Facility \footnote{\url{https://ror.org/05qajvd42}}. Operation of ASKAP is funded by the Australian Government with support from the National Collaborative Research Infrastructure Strategy. ASKAP uses the resources of the Pawsey Supercomputing Research Centre. Establishment of ASKAP, Inyarrimanha Ilgari Bundara, the CSIRO Murchison Radio-astronomy Observatory and the Pawsey Supercomputing Research Centre are initiatives of the Australian Government, with support from the Government of Western Australia and the Science and Industry Endowment Fund.

The acknowledgements were compiled using the Astronomy Acknowledgement Generator. Some of the results in this paper have been derived using the \textsc{healpy} and HEALPix packages~\citep{Gorski2005}. This research made use of: NASA's Astrophysics Data System; \textsc{spectralcube}, a library for astronomical spectral data cubes; \textsc{Astropy}, a community-developed core Python package for Astronomy \citep{2018AJ....156..123A, 2013A&A...558A..33A}; \textsc{SciPy} \citep{Virtanen_2020}; \textsc{ds9}, a tool for data visualization supported by the Chandra X-ray Science Center (CXC) and the High Energy Astrophysics Science Archive Center (HEASARC) with support from the JWST Mission office at the Space Telescope Science Institute for 3D visualization; \textsc{NumPy} \citep{harris2020array}; \textsc{matplotlib}, a Python library for publication-quality graphics \citep{Hunter:2007}. This work made use of the \textsc{IPython} package \citep{PER-GRA:2007}; \textsc{pandas} \citep{McKinney_2010, McKinney_2011}; TOPCAT, an interactive graphical viewer and editor for tabular data \citep{2005ASPC..347...29T}

C.F.~acknowledges funding provided by the Australian Research Council (Future Fellowship FT180100495 and Discovery Projects DP230102280), and the Australia-Germany Joint Research Cooperation Scheme (UA-DAAD). C.F.~further acknowledges high-performance computing resources provided by the Leibniz Rechenzentrum and the Gauss Centre for Supercomputing (grants~pr32lo, pr48pi and GCS Large-scale project~10391), the Australian National Computational Infrastructure (grant~ek9) and the Pawsey Supercomputing Centre (project~pawsey0810) in the framework of the National Computational Merit Allocation Scheme and the ANU Merit Allocation Scheme. CLH acknowledges funding from the Leverhulme Trust through an Early Career Research Fellowship. NM-G is the recipient of an Australian Research Council Australian Laureate Fellowship (project number FL210100039) funded by the Australian Government. MH acknowledges funding from the European Research Council (ERC) under the European Union's Horizon 2020 research and innovation programme (grant agreement No 772663). This work has been supported by funding from the Australian Research Council, the Natural Sciences and Engineering Research Council of Canada, the Canada Research Chairs Program, and the Canada Foundation for Innovation.

\section*{Data Availability}
The data that support the findings of this study are openly available in the CSIRO ASKAP Science Data Archive and the CSIRO Data Access Portal at \url{https://doi.org/10.25919/w37t-nw98} and \url{https://data.csiro.au/collection/csiro:58409}. All software is publicly available on GitHub, with the primary processing software at \url{https://github.com/AlecThomson/arrakis} and supplementary scripts at \url{https://github.com/AlecThomson/spica}.
\end{acknowledgement}


\bibliography{correct}

\appendix

\section{Widefield leakage}\label{app:leakage}

Here we detail our determination and evaluation of widefield leakage using field sources. We compute the position of a source within a beam in the instrument frame as:
\begin{align*}
    \ell = \rho \sin(\theta),\\
    m = \rho \cos(\theta),
\end{align*}
where $\rho$ is the offset from a given beam centre, and $\theta$ is the position angle. We then follow the procedure outlined in \S\ref{sec:off_leak}; namely, spectra extraction, flagging, and model-fitting. We test several model-fitting routines, but we find that simple least-squares is the most robust as a function of frequency. In practice we see other routines, such as variance-weighted least squares or Markov-chain-Monte-Carlo, are more easily biased by true polarised sources. In future work, we hope to provide a more robust rejection of such sources from our fitting, which will allow us to take advantage of different fitting routines.

Having derived a best-fitting Zernike polynomial, we evaluate the leakage surfaces on a dense grid spanning the largest separation of the field sources from the beam centre. We then regrid and interpolate these grids to match those from the holographic observations of the Stokes $I$ primary beams; producing a hypercube of Stokes $I$ response, and $Q$, $U$ leakage as a function of frequency and beam. We show these surfaces at the central frequency of \qty{888}{\MHz} in Figure~\ref{fig:primary_beams}.

\begin{figure*}
	\begin{center}
	\begin{subfigure}[b]{\columnwidth}
		\includegraphics[width=\textwidth]{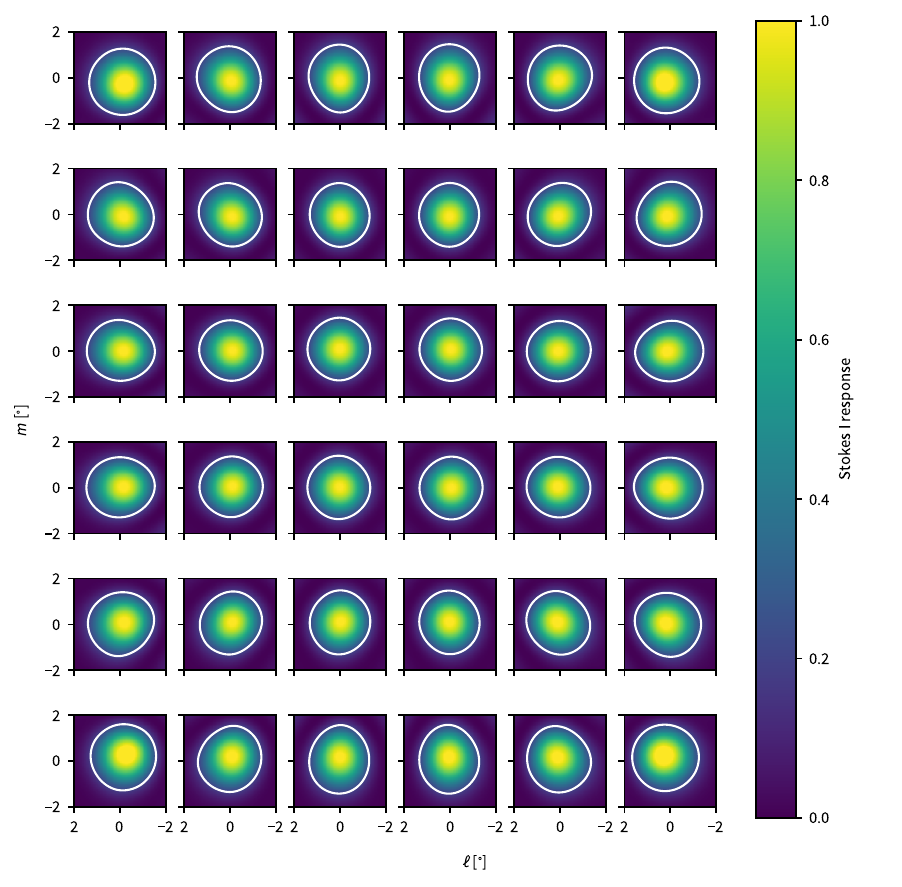}
		\caption{Stokes $I$ response from holography.}
	\end{subfigure}
	\end{center}
\end{figure*}
\begin{figure*}\ContinuedFloat
	\begin{center}
	\begin{subfigure}[b]{\columnwidth}
		\includegraphics[width=\textwidth]{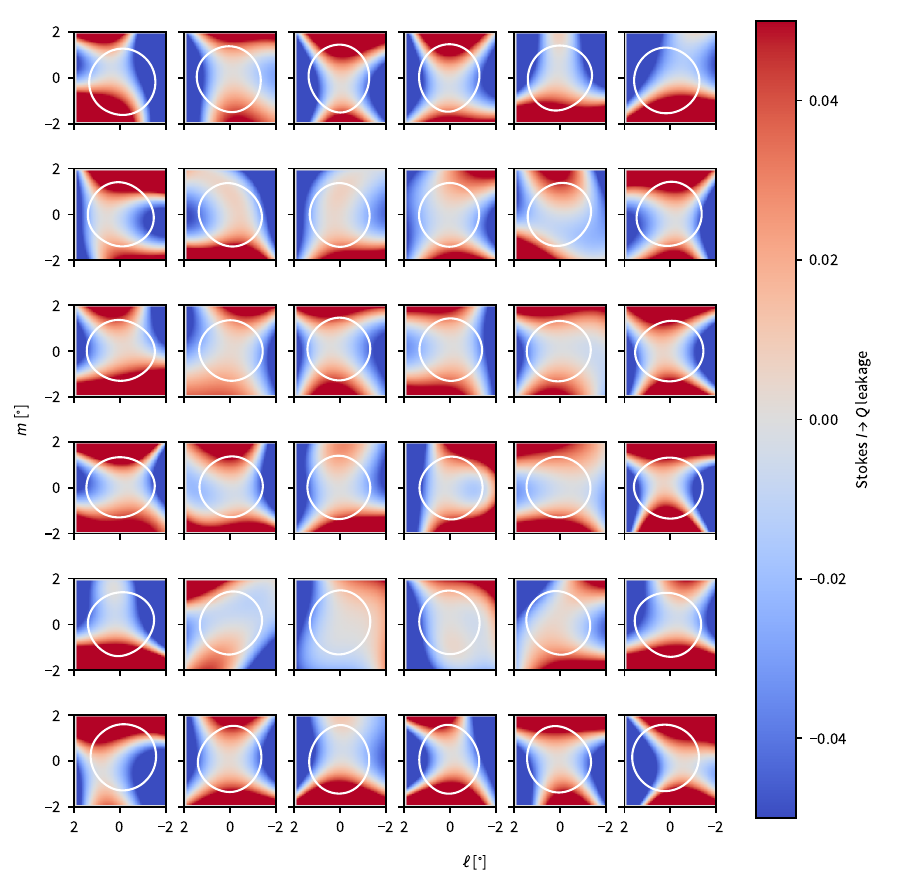}
		\caption{Stokes $I$ to $Q$ leakage fitted from field sources.}
	\end{subfigure}
	\end{center}
\end{figure*}
\begin{figure*}\ContinuedFloat
	\begin{center}
	\begin{subfigure}[b]{\columnwidth}
		\includegraphics[width=\textwidth]{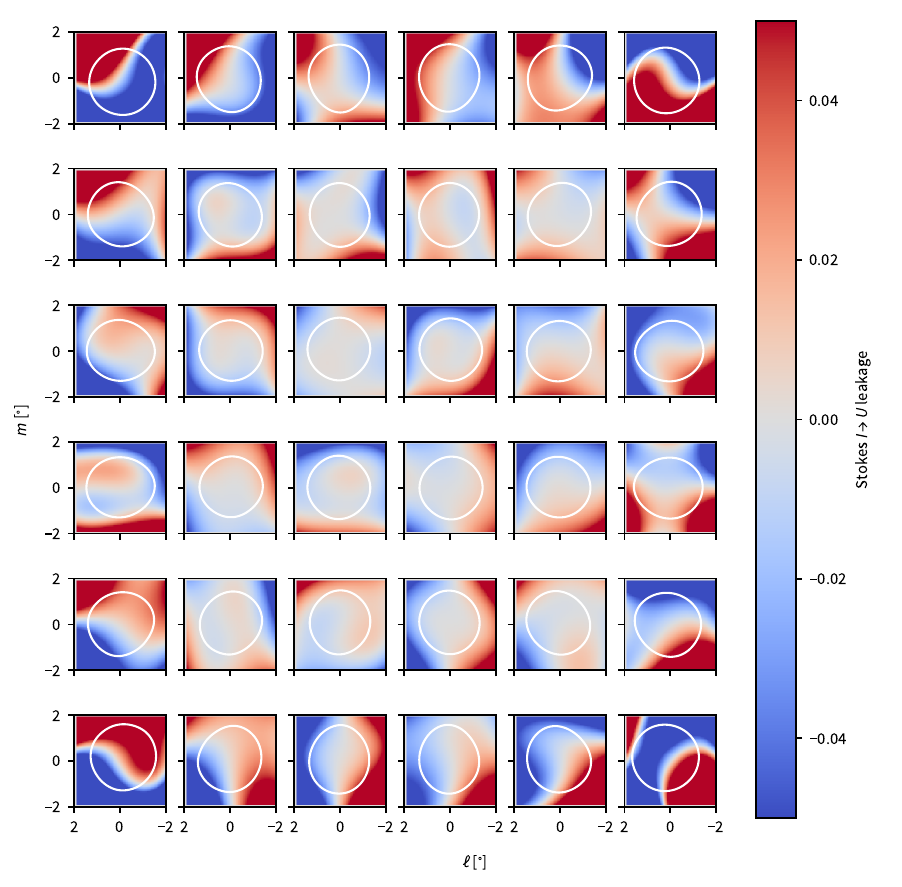}
		\caption{Stokes $I$ to $U$ leakage fitted from field sources.}
	\end{subfigure}
		\caption{Models of the primary beam \qty{888}{\MHz} from observations with beam weight SB8247.}\label{fig:primary_beams}
	\end{center}
\end{figure*}

\begin{figure*}
	\begin{subfigure}[b]{0.49\columnwidth}
		\includegraphics[width=\textwidth]{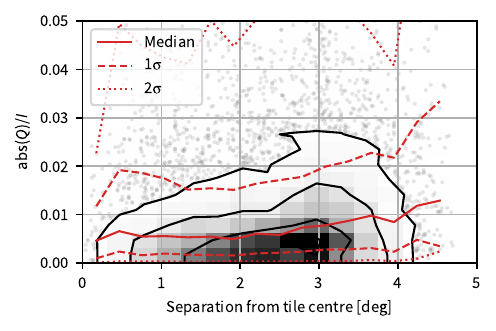}
		\caption{}\label{fig:leakage_res_q}
	\end{subfigure}
	\begin{subfigure}[b]{0.49\columnwidth}
		\includegraphics[width=\textwidth]{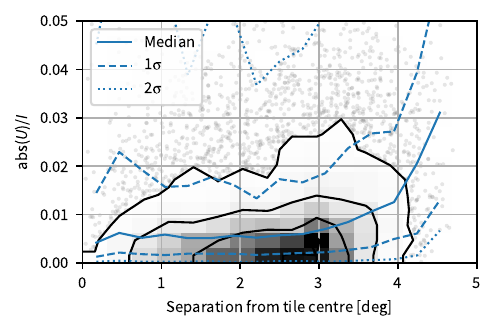}
		\caption{}\label{fig:leakage_res_u}
	\end{subfigure}
	\begin{center}
	\begin{subfigure}[b]{\columnwidth}
		\includegraphics[width=\textwidth]{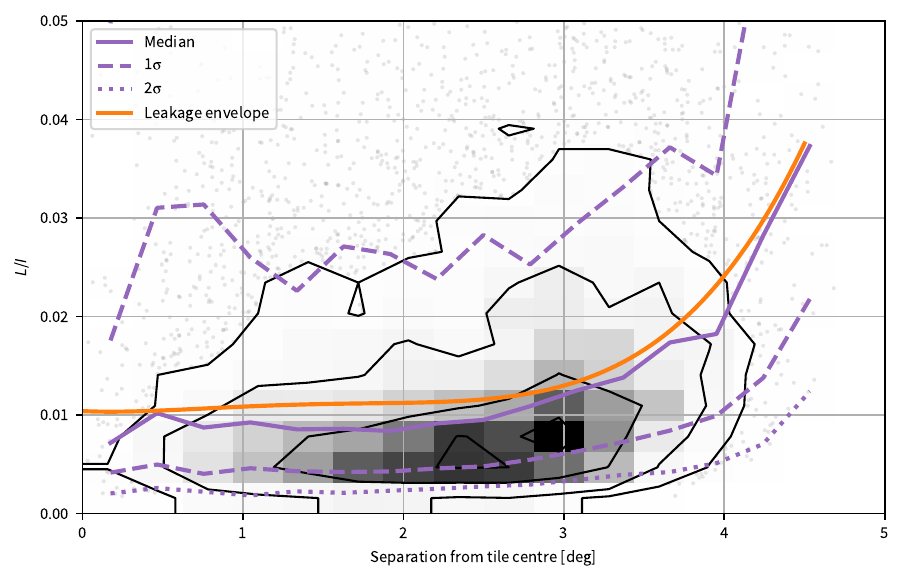}
		\caption{}\label{fig:leakage_res_l}
	\end{subfigure}
		\caption{Residual leakage from Stokes $I$ into; (\subref{fig:leakage_res_q}) Stokes $Q$; (\subref{fig:leakage_res_u}) Stokes $U$; and (\subref{fig:leakage_res_l}) polarised intensity ($L$). In each panel we show the leakage distribution of bright ($100\sigma$) Stokes $I$ sources as a function of separation from the centre of the nearest observed tile. The solid, dashed, and dotted lines show the median, $\pm1\sigma$, and $\pm2\sigma$ levels, respectively, in angular separation bins. The orange curve we show in (\subref{fig:leakage_res_l}) defines leakage value in the catalogue.  Where the scatter points are over-dense we show the data as a 2D density histogram. The contour levels of the 2D histogram are at the $2^\text{nd}$, 16$^\text{th}$, 50$^\text{th}$, 84$^\text{th}$, and 98$^\text{th}$ percentiles.}\label{fig:leakage_res}
	\end{center}
\end{figure*}

After applying our primary beam attenuation and leakage correction using \texttt{LINMOS} we can evaluate the residual leakage in the catalogue. After mosaicking adjacent beams together into a tile we know that the resulting leakage surface will be smoothly varying and increasing in amplitude towards the edge of each tile. In our evaluation we reduce the two-dimensional surface to one dimension, and inspect the Stokes $Q$ and $U$ and polarised intensity ($L$) flux densities as a function of separation from the tile centre; which we show in Figure~\ref{fig:leakage_res}. For both Stokes $Q$ and $U$ we find that the median fraction is less than $1\%$ out to \ang{3} separation from the nearest tile centre. In Stokes $Q$, this then rises to about $1.5\%$ at the tile edges (separation $>\ang{4}$). In Stokes $U$ the performance is worse, with the residual leakage rising over $3\%$ at the tile edges. This results in a maximum leakage in $L$ on the order of about $4\%$. To quote a leakage level in the catalogue we fit a polynomial to the binned median of $L/I$ as a function of tile centre separation, offset by a small (0.2\%) value to avoid under-quoting the leakage. For each component in our catalogue, we evaluate this fitted polynomial and quote this value in the \texttt{leakage} column.

\section{Noise estimation robustness}\label{app:noise}
As we describe in \S\ref{sec:spectra}, we estimate the $rms$ noise in the local vicinity of a component by computing the corrected MADFM within an annulus about the centre of our image cube cutout. We set inner and outer radii of this annulus are 1 and 3.1\,PSFs, respectively. Despite using the robust MADFM, our noise measurement may be over-estimated if a radio component falls within our annulus. To test the effect of this, we simulate an unresolved component falling completely within our noise annulus.

We begin by constructing a $100\times100\times288$ (RA$\times$Dec.$\times$ $\nu$) pixel cube of random noise; representing a single cubelet. We draw the noise from a Gaussian distribution with a mean of 0 and a standard deviation of $\sigma=\sqrt{288}\times$\qty{80}{\micro\jansky\per beam}. We note though, that we will normalise by this $\sigma$ later. We also take our pixels to be \ang{;;2.5}, as in the real cutouts.
 
\begin{figure}
    \centering
    \includegraphics{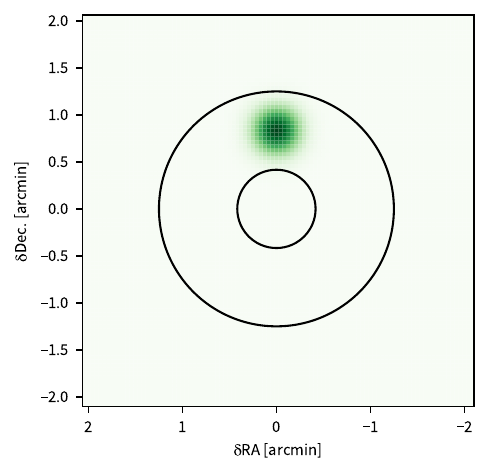}
    \caption{A simulated Gaussian component (in the green colour scale), injected within our noise-estimation annulus (black lines).}
    \label{fig:noise_error_model}
\end{figure}

\begin{figure}
    \centering
    \includegraphics{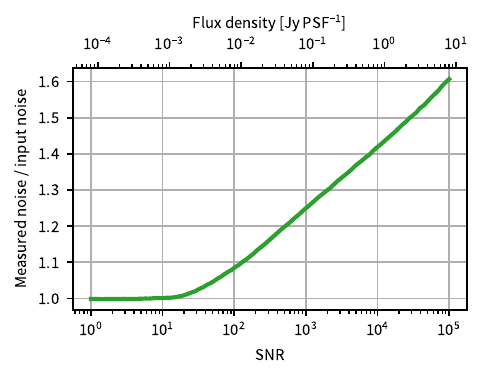}
    \caption{Simulating the effect of a Gaussian component without our noise estimation. We show the ratio of recovered noise within our noise annulus to the true input noise as a function of the input signal-to-noise ratio (SNR). We convert input SNR to a band-averaged flux density using an noise value of \qty{80}{\micro\jansky\per beam}.}
    \label{fig:noise_error_snr}
\end{figure}

We now generate a two-dimensional Gaussian model with a FWHM of \ang{;;25}, matching our PSF, and place it within the our noise estimation annulus; we show this model in Figure~\ref{fig:noise_error_model}. We scale the model using a sign-to-noise value which we sample over 100 logarithmically-space bins in the range 1 to 100000. After scaling the Gaussian, we inject it into the noise cube and run it through our noise estimation procedure. In Figure~\ref{fig:noise_error_snr} we show the measured noise divided by the true input $rms$ noise as a function of the SNR of the injected Gaussian. We see that even for a component as bright as \qty{1}{\jansky\per beam} the overestimation of the noise is less than 50\%. In reality, an adjacent component this bright would likely introduce artefacts and sidelobes into all nearby sources. We conclude that our noise estimation approach is sufficiently robust for our purposes.

\section{Faraday Complex Spectra}\label{app:complex_spectra}

\begin{figure*}
    \centering
    \includegraphics{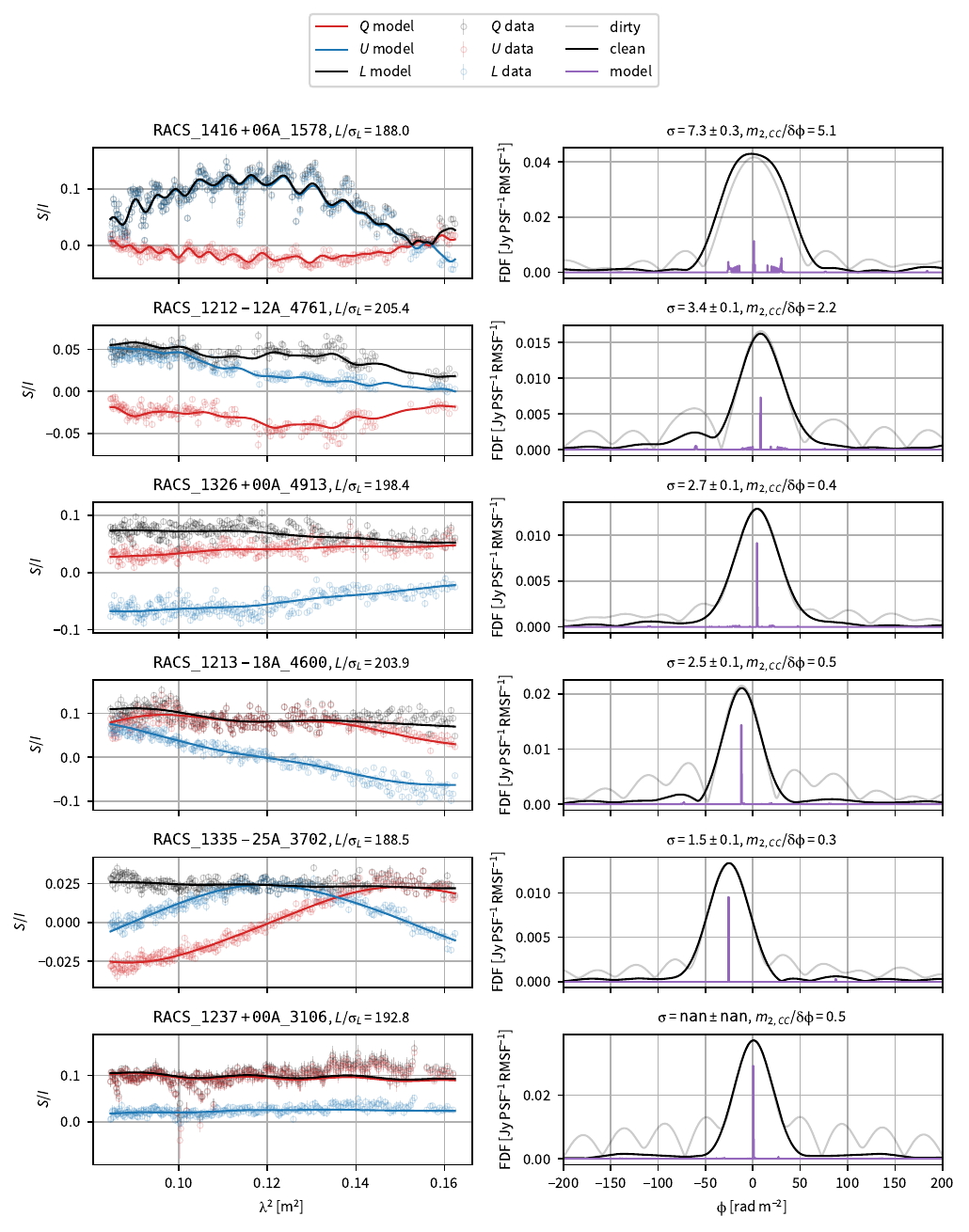}
    \caption{Example SPICE-RACS spectra with approximately constant polarised SNR ($L/\sigma_L$) over a range of normalised complexity metrics ($\sigma_\text{add}$ and $m_{2,CC}/\delta\phi$). The spectra are sorted by $\sigma_\text{add}$. In the left column we show the fractional Stokes $Q/I$ and $U/I$ as a function of $\lambda^2$, with the labels showing the Gaussian ID and the polarised SNR. In the right column we show the Faraday spectra and we display the normalised complexity metrics in the labels. Where $\sigma_\text{add}/\delta\sigma_\text{add} < 10$ we set the value and error to \texttt{nan} (see discussion in \S\ref{sec:complex_rev}).}
    \label{fig:complex_spectra}
\end{figure*}

As a demonstration of our Faraday complexity metrics we perform a simple search of our catalogue to identify a small set of example spectra with approximately similar linearly polarised SNR ($L/\sigma_L$) with a range in our normalised complexity metrics $\sigma_\text{add}$ and $m_{2,CC}/\delta\phi$ (see \S\ref{sec:complex}). Our search narrowed down 6 example spectra with an SNR of $\sim200$ in polarisation. We show these spectra, ordered by $\sigma_\text{add}$, in Figure~\ref{fig:complex_spectra}.

\end{document}